\pdfoutput=1
\documentclass[aps, prb, reprint, superscriptaddress, eqsecnum, longbibliography]{revtex4-2} 
\usepackage{graphicx, hyperref}
\hypersetup{
colorlinks = true,
allcolors = blue
}
\usepackage{amsmath, amssymb, bm}
\usepackage{braket}
\usepackage[dvipsnames]{xcolor}
\usepackage{comment}
\usepackage{manfnt}
\usepackage{lineno}
\usepackage{here}

\allowdisplaybreaks

\begin{document}

\title{Electrical and thermal magnetotransport and the Wiedemann-Franz law in semimetals with electron-electron scattering}
\author{Keigo Takahashi} 
\affiliation{Department of Physics, University of Tokyo, 7-3-1 Hongo, Bunkyo, Tokyo 113-0033, Japan}
\author{Hiroyasu Matsuura}
\affiliation{National Institute of Advanced Industrial Science and Technology (AIST), 1-1-1 Umezono, Tsukuba, Ibaraki 305-8568, Japan}
\author{Hideaki Maebashi}
\affiliation{National Institute of Advanced Industrial Science and Technology (AIST), 1-1-1 Umezono, Tsukuba, Ibaraki 305-8568, Japan}
\author{Masao Ogata}
\affiliation{Department of Physics, University of Tokyo, 7-3-1 Hongo, Bunkyo, Tokyo 113-0033, Japan}
\affiliation{National Institute of Advanced Industrial Science and Technology (AIST), 1-1-1 Umezono, Tsukuba, Ibaraki 305-8568, Japan}
\date{\today}

\begin{abstract}
We study the electrical and thermal transport properties and the violation of the Wiedemann-Franz (WF) law of two-carrier semimetals using exact treatments of the Boltzmann equation with the impurity and electron-electron scatterings in a magnetic field. For comparison, we also study those in the case of Baber scattering: a single-carrier system with an impurity scattering and phenomenological momentum-dissipative electron-electron scattering. In both systems, the longitudinal and transverse WF laws, $L = L_{\text{H}} = L_{0}= \pi^2k_B^2/3e^2$, hold at zero temperature, where the Lorenz ratio $L$ and the Hall Lorenz ratio $L_{\text{H}}$ are ratios of thermal conductivity $\kappa_{\mu\nu}$ to electrical conductivity $\sigma_{\mu\nu}$ divided by temperature. However, the electron-electron scattering makes Lorenz ratios deviate from $L_{0}$ with increasing temperature. To describe the WF law in a magnetic field, we introduce another set of Lorenz ratios, $\widetilde{L}$ and $\widetilde{L}_{\text{H}}$, defined as the ratios of the resistivity and the Hall coefficient to their thermal counterparts. The WF laws for them, $\widetilde{L} = \widetilde{L}_{\text{H}} = L_{0}$, and their violation are helpful for the discussion of $L$ and $L_{\text{H}}$. For Baber scattering, our exact result shows $L_{\text{H}}/L_{0} \sim (L/L_{0})^2$ in a weak magnetic field. In semimetals, the violations of the WF laws are significant, reflecting the different temperature dependence between the electrical and thermal resistivities in a magnetic field. This is because the momentum conservation of the electron-electron scattering has a completely different effect on electrical and thermal magnetotransport. We sort out these behaviors using $\widetilde{L}$ and $\widetilde{L}_{\text{H}}$. We also provide a relaxation time approximation, which is useful for comparing theory and experiment.
\end{abstract}

\maketitle

\section{Introduction}
The Wiedemann-Franz (WF) law connects electrical and electronic thermal transport in metals \cite{Ziman1972, AshcroftMermin, Smith_Jensen1989, Ziman2001}. The WF law is expressed as 
\begin{align}
\frac{\kappa_{xx}}{T\sigma_{xx}} = L_{0} = \frac{\pi^2k_{B}^2}{3e^2}, \label{eq:WF_law}
\end{align}
where $\kappa_{xx}$ ($\sigma_{xx}$) is thermal (electrical) conductivity, $T$ is temperature, $k_{B}$ is the Boltzmann constant, and $e < 0$ is the charge of an electron. The Lorenz ratio $L_{}$ is given by
\begin{align}
L_{} = \frac{\kappa_{xx}}{T\sigma_{xx}}. \label{eq:def_Lorenz_ratio}
\end{align}
However, the WF law is broken in several cases, and the violation of the WF law plays a key role in understanding materials \cite{Ziman1972, AshcroftMermin, Smith_Jensen1989, Ziman2001}. A typical case is a downward violation, $L < L_{0}$, caused by inelastic scatterings, which lead to stronger relaxation in thermal transport than in electrical transport. Violations of the WF law in materials such as a type-II Weyl semimetal $\text{WP}_{2}$ \cite{Gooth2018, Jaoui2018} or a heavy-fermion anti-ferromagnet $\text{CeRhIn}_{5}$ \cite{Paglione2005} are considered to be driven by the inelastic electron-electron scattering since these systems have electrical resistivity proportional to $T^2$ and thermal conductivity proportional to $T^{-1}$ at low-temperatures. 

Theoretical efforts have been made on the violation of the WF law by the electron-electron scattering \cite{Herring1967, Herring1967erratum, Bennett1969, Schriempf1969, Kaveh1984, Schulz1995, Principi2015, Lucas2018, Li2018, Zarenia2020, Lee2021, Takahashi2023}. In particular, motivated by the experiments of $\text{WP}_{2}$ \cite{Gooth2018, Jaoui2018}, the Lorenz ratio has been studied in a two-carrier model of the compensated metal, which has equal numbers of electrons and holes \cite{Li2018, Lee2021} using exact transport coefficients of Fermi liquids \cite{Abrikosov1957, Abrikosov1959, Brooker1968, Jensen1968, Smith1969, Jensen1969, Bennett1969, Sykes1970, Ah-Sam1971, Brooker1972, Egilsson1977, Oliva1982, Anderson1987, Smith_Jensen1989, Golosov1995, Golosov1998, Pethick2009, Li2018, Lee2020, Lee2021}. In such multi-band systems, the interband normal electron-electron scattering, which relaxes a relative motion between carriers, can be a major relaxation process leading to $T^2$ electrical resistivity \cite{Baber1937}. In particular, $T^2$ electrical resistivity arises even without other momentum dissipative processes in the compensated system, where only the relative motion contributes to the longitudinal electrical transport \cite{Kukkonen1976, Gantmakher1978, Maldague1979}. The ambipolar thermal conduction in the compensated system \cite{Zarenia2020} and the thermal and thermoelectric transport properties, including the uncompensated cases where the numbers of electrons and holes are different, have been discussed as well \cite{Takahashi2023}.

Magnetotransport allows us to study materials from different perspectives through the Hall and thermal Hall effects, where the transverse electric and thermal currents arise in response to the longitudinal electric field and temperature gradient, respectively, in the presence of a magnetic field \cite{Ziman2001}. In particular, the violation of transverse WF law in a magnetic field, quantified by the Hall Lorenz ratio $L_{\text{H}}$, also reflects the inelastic nature of scatterings \cite{Long1971, Zhang2000, Li2002, Matusiak2005, Onose2008, Shiomi2009, Matusiak2009, Shiomi2010, Matusiak2015, Grissonnanche2016, Lucas2018, Nguyen2020, Huang2021, Tulipman2023, Tu2023}. The Hall Lorenz ratio $L_{\text{H}}$ is defined as
\begin{align}
L_{\text{H}} = \frac{\kappa_{xy}}{T\sigma_{xy}},
\end{align}
where $\kappa_{xy}$ ($\sigma_{xy}$) is transverse thermal (electrical) conductivity. Then the transverse WF law is given by $L_{\text{H}} = L_{0}$. The advantage of investigating the transverse WF law lies in the fact that the thermal Hall effect is often dominated by electrons. In contrast, in the case of longitudinal transport, electronic thermal transport is often masked by thermal transport of phonons or other degrees of freedom.

The WF law can be studied from the perspective of the resistivity tensor using another set of Lorenz ratios $\widetilde{L}_{}$ and $\widetilde{L}_{\text{H}}$ \cite{Amundsen1969, Stephan1972, Fletcher1973, Fletcher1977, Sugihara1980}.
The former is defined in terms of the electrical and thermal resistivities. The latter is defined in terms of the Hall coefficient as $\widetilde{L}_{\text{H}} = R_{\text{H}}/K_{\text{H}}$, where $R_{\text{H}}$ is the Hall coefficient and $K_{\text{H}}$ is a thermal counterpart of the Hall coefficient 
\footnote{
$R_{\text{RL}} = K_{\text{H}}/T$ is often referred to as Righi-Leduc coefficient \cite{Amundsen1969, Stephan1972, Fletcher1973, Fletcher1977, Sugihara1980}. It should be noted that Righi-Leduc coefficient may be differently defined as $R_{\text{RL}} = 1/B \cdot \kappa_{xy}/\kappa_{xx}$  \cite{Bridgman1924, Ziman2001}. The WF law for $R_{\text{RL}}$ expressed as $R_{\text{RL}} = \sigma R_{\text{H}}$. This is satisfied with $L = L_{\text{H}} = L_{0}$ and this has been studied as well \cite{Bridgman1924}.}.
The formal definitions are given later. The WF laws for these ratios, $\widetilde{L} = L_{0}$ and $\widetilde{L}_{\text{H}} = L_{0}$, are satisfied for elastic scatterings. The WF law for $\widetilde{L}_{\text{H}}$ behaves differently from that for $L_{\text{H}}$. For example, $\widetilde{L}_{\text{H}} = L_0$ is satisfied 
but $L_{\text{H}} \neq L_0$ for the single-band metal in the relaxation time approximation (RTA), where electric and thermal transport have different relaxation times \cite{Sondheimer1948, Fletcher1973}. This is because $R_{\text{H}}$ and $K_{\text{H}}$ are insensitive to relaxation times.

Semimetals often show large magnetoresistance \cite{Ziman2001, Hurd1972, Gantmakher1978, Ali2014, Wang2017, Kumar2017} and thus the electrical and thermal resistivities in a magnetic field have been interesting as well.

However, the effects of the electron-electron scattering on magnetotransport in metals and semimetals are not entirely understood, in particular beyond the RTA. The thermal Hall effect and Hall Lorenz ratio have been discussed in a single-carrier metal with the electron-electron scattering using the RTA, taking into account the momentum conservation \cite{Lucas2018, Lee2020magneto}. 
The Hall effect and magnetoresistance in the two-band semimetals with the interband electron-electron scattering have been discussed on the basis of a two-carrier kinetic model, or equivalently, the RTA \cite{Gantmakher1978, Kukkonen1979, Entin2013}. 
 
The aim of the paper is to elucidate the effect of the electron-electron scatterings on the transport properties of metals and semimetals in a magnetic field at low temperatures using the exact treatment of the Boltzmann equation beyond the RTA. We focus on the resistivity, the Hall coefficient, and its thermal counterpart, and the WF law.

\begin{figure}[tbp]
\begin{center}
\rotatebox{0}{\includegraphics[angle=0,width=1\linewidth]{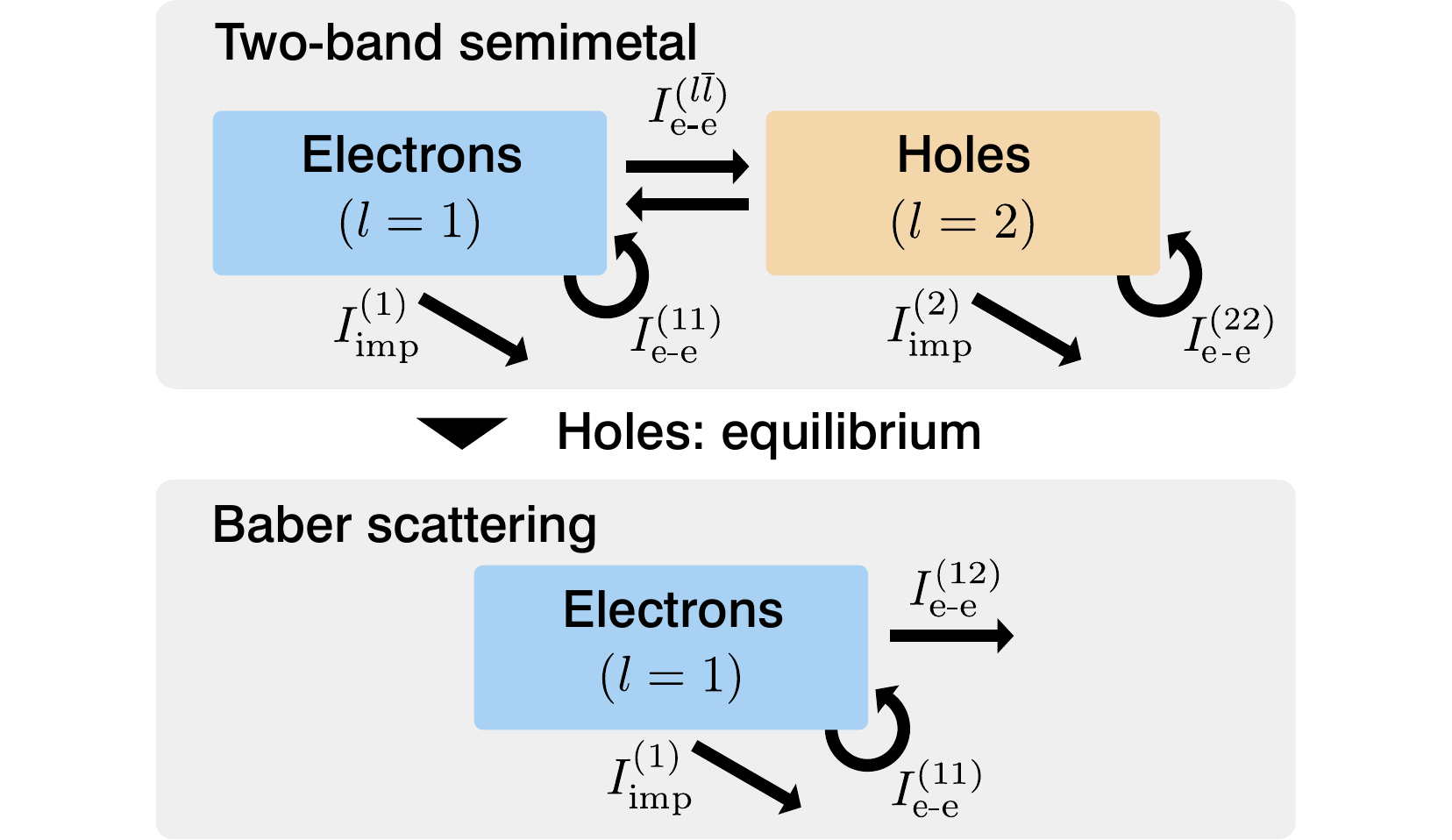}}
\caption{The schematic of two cases in this study: the two-band semimetal and Baber scattering. The black arrows represent how momentum is transferred. $I_{\text{e-e}}^{(ll')}$ indicates the electron-electron scattering between bands $l$ and $l'$ ($l = 1$ means electrons and $l = 2$ means holes) and $I_{\text{imp}}^{(l)}$ indicates the impurity scattering of the band $l$. They are introduced in Sec. \ref{Sec:model_Boltzmann}.}
\label{Fig:Baber_image}
\end{center}
\end{figure}

We study the electrical and thermal transport coefficients of two-band semimetals, considering impurity scattering and intra- and interband electron-electron scatterings in a magnetic field. As a reference and a special case, we study an effective single-carrier system with Baber scattering in which analytical calculations are available.
We first present the Boltzmann equation for the two-band model with momentum-conserving intra- and interband electron-electron scatterings. Then, we obtain an effective single-carrier model as a limiting case: Baber scattering, in which one of the carriers is in equilibrium due to some momentum dissipative relaxation \cite{Baber1937, Bennett1969, Schriempf1969, Ah-Sam1971, Lin2015}. The relation between the two cases is depicted schematically in Fig.~\ref{Fig:Baber_image}. On solving the Boltzmann equation, we use the treatment of the electron-electron scattering originally developed by Abrikosov and Khalatnikov which allows us to treat the electron-electron scatterings exactly at low temperatures $k_BT \ll \varepsilon_{\text{F}}$ \cite{Abrikosov1957, Abrikosov1959, Brooker1968, Jensen1968, Smith1969, Jensen1969, Bennett1969, Sykes1970, Ah-Sam1971, Brooker1972, Egilsson1977, Oliva1982, Anderson1987, Smith_Jensen1989, Golosov1995, Golosov1998, Pethick2009, Li2018, Lee2020, Lee2021}.

For the case of Baber scattering, we will present analytic formulae of the electrical and thermal conductivities where the impurity scattering, the electron-electron scatterings, and the effect of a magnetic field are taken into account. We quantify a small but non-zero magnetoresistance and temperature dependence of $R_{\text{H}}$ and $K_{\text{H}}$. These features originate from the energy dependence of the distribution function, which the RTA cannot describe, even though conductivities by the RTA with properly chosen relaxation times can reproduce those by the exact solutions within an accuracy of order unity, as we will show. Then, we show that the WF laws for $\widetilde{L}_{}$ and $\widetilde{L}_{\text{H}}$ are weakly violated. $L_{\text{H}}/L_{0} \sim (L/L_0)^2$ holds approximately in a weak magnetic field.

For the two-band system, we will use the variational method, which gives numerically exact solutions with sufficient convergence regarding trial functions. We discuss the transport properties, focusing on the fact that the momentum conservation of the electron-electron scattering plays a key role in electric transport, but not in thermal transport. The electrical resistivity increases with temperature almost monotonically, whereas the thermal resistivity shows non-monotonic temperature dependence for a large magnetic field. We discuss the WF law for $\widetilde{L}_{}$, which can be violated upwardly and downwardly depending on the applied magnetic field and the carrier number. We argue that the momentum conservation enhances the violation. 
$R_{\text{H}}$ and $K_{\text{H}}$ reflect the different effects of the momentum conservation as well, and the WF law for $\widetilde{L}_{\text{H}} = R_{\text{H}}/K_{\text{H}}$ is severely violated. This has an important effect on the violation of the transverse WF law through $L_{\text{H}} \simeq L^2/\widetilde{L}_{\text{H}}$ in a weak magnetic field, along with the reduction of $L$.

\begin{figure}[H]
\begin{center}
\rotatebox{0}{\includegraphics[angle=0,width=1\linewidth]{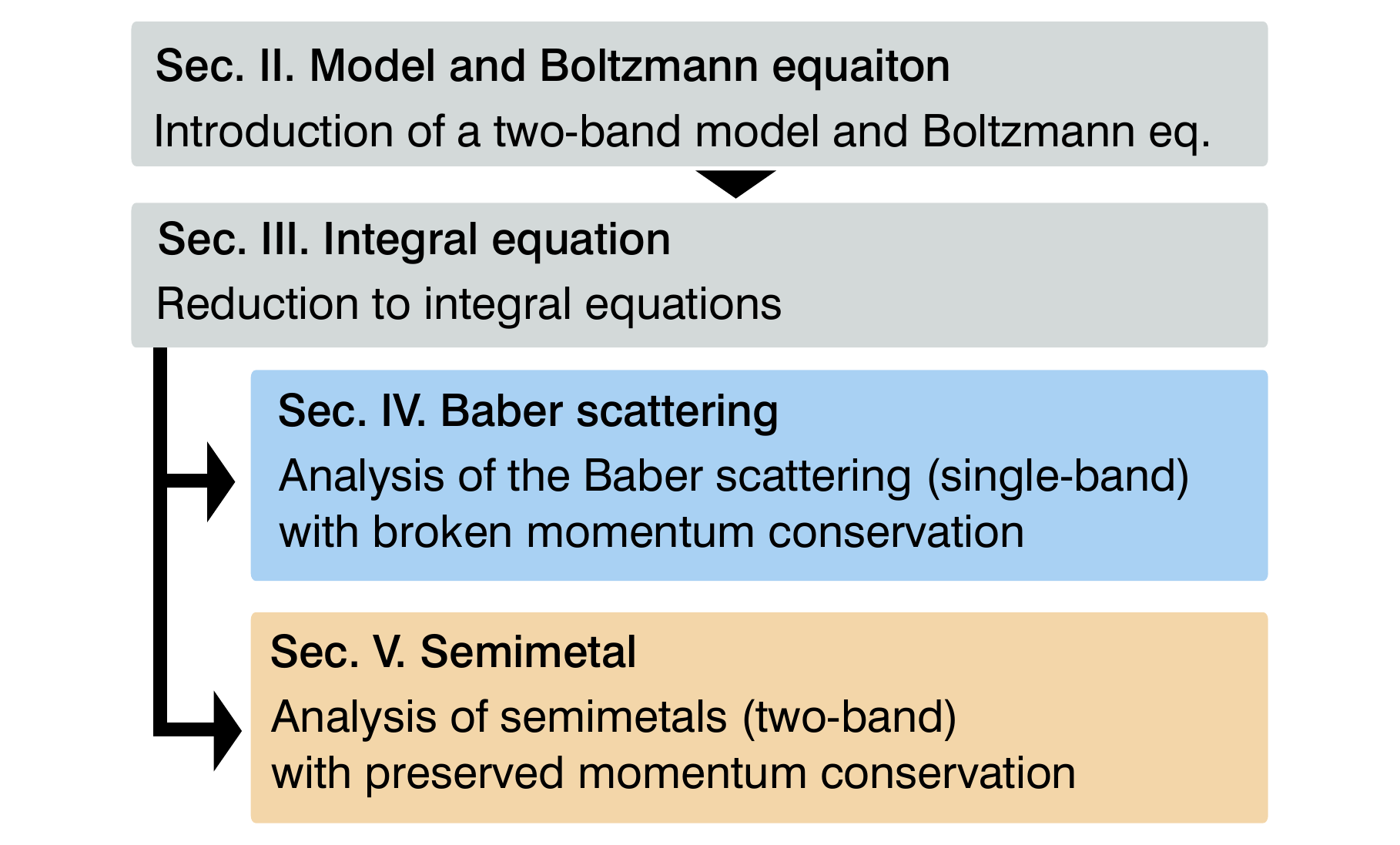}}
\caption{An overview of the paper.}
\label{Fig:overview}
\end{center}
\end{figure}

This paper is organized as follows. In Sec.~\ref{Sec:model_Boltzmann}, the model and the Boltzmann equation are introduced. In Sec.~\ref{Sec:integral_eq}, we present integral equations derived from the Boltzmann equations. In Sec.~\ref{Sec:Baber}, we introduce Baber scattering as a limiting case. We give a set of eigenfunctions for the integral equation for the single-band case. Then, we obtain an analytic formula of transport coefficients in the case of Baber scattering and discuss the transport properties. In Sec.~\ref{Sec:Semimetal}, we discuss the transport properties of the two-band semimetals. The conclusion is given in Sec.~\ref{Sec:Conclusion}. 
An overview from Sec.~\ref{Sec:model_Boltzmann} to Sec.~\ref{Sec:Semimetal} is shown in Fig.~\ref{Fig:overview}.
Appendices provide detailed calculations including the RTA.

\section{Model and Boltzmann equation}\label{Sec:model_Boltzmann}
\subsection{Model}
\begin{figure}[H]
\begin{center}
\rotatebox{0}{\includegraphics[angle=0,width=1\linewidth]{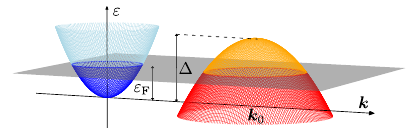}}
\caption{An image of the two-band model \cite{Takahashi2023}.}
\label{Fig:dispersion}
\end{center}
\end{figure}

We study the electrical and thermal transport properties of two-band semimetals in three dimensions with impurity and intra- and interband electron-electron scatterings. We study an effective single-carrier model with Baber scattering as well. The transport equation for the latter is obtained from that for the former by neglecting one carrier as will be discussed in Sec.~\ref{Sec:integral_eq}. As in Refs.~\cite{Li2018, Lee2021, Takahashi2023}, we consider parabolic bands as shown in Fig.~\ref{Fig:dispersion} whose band dispersions are given by 
\begin{align}
\varepsilon_{1,\bm{k}}=& \frac{\hbar^2}{2m_1} \bm{k}^2,\label{eq:dispersion_electrons} \\
\varepsilon_{2,\bm{k}}=& \Delta - \frac{\hbar^2}{2m_2} (\bm{k} - \bm{k}_0)^2, \label{eq:dispersion_holes} 
\end{align}
where $\Delta$ represents the overlap of the bands. Electrons $(l=1)$ and holes $(l=2)$ are assumed to be well separated in the momentum space by $\bm{k}_0$. Hereafter, the momentum of holes is measured from $\bm{k}_0$. The carrier number of band $l$ is given by $n_{l} = k^{3}_{\text{F},l}/3\pi^2$ with $k_{\text{F},l}$ being the Fermi wavenumber of band $l$. We express the ratio of the two Fermi wavenumbers as $\chi = k_{\text{F},2}/k_{\text{F},1}$. Then, the carrier numbers satisfy $n_{2} = \chi^3 n_{1}$. 
The Fermi energy is given by $\varepsilon_{\text{F}} = \mu = m_2\Delta/(\chi^2 m_1 + m_2)$. We only consider low temperatures where the temperature dependence of the chemical potential is negligible.

\subsection{Linearized Boltzmann equation}
Let us introduce the linearized Boltzmann equation for a band $l$ with an external field along $x$-axis \cite{Ziman2001},
\begin{align}
&v^{(l)}_{\bm{k};x} F^{(l)}_{\text{ext}}\left(- \frac{\partial f_0(\varepsilon_{l,\bm{k}})}{\partial \varepsilon_{l,\bm{k}}} \right) \nonumber \\
&= I^{(l)}_{\text{imp}}[\Phi] + I^{(ll)}_{\text{e-e}}[\Phi] + I^{(l\overline{l})}_{\text{e-e}}[\Phi] + M^{(l)}[\Phi], \label{eq:Boltzmann_eq} 
\end{align}
where $f_0(\varepsilon) = (e^{\beta (\varepsilon - \mu)} + 1)^{-1}$ is the Fermi-Dirac distribution function with $\beta = (k_BT)^{-1}$ and $\overline{l} = 3 - l$ denotes the other band to $l$. The external field is given by
\begin{align}
F^{(l)}_{\text{ext}} = 
\left\{
\begin{array}{ll}
eE & (\text{electric field}) \\
\xi_{l,\bm{k}} \left( - \nabla T/T \right) & (\text{temperature gradient}) 
\end{array},
\right.
\end{align}
where $\xi_{l,\bm{k}} = \varepsilon_{l,\bm{k}} - \mu$ and $v^{(l)}_{\bm{k};\nu}~(\nu = x,y,z)$ is $\nu$ component of the group velocity of the band $l$ given by $\bm{v}^{(l)}_{\bm{k}} = \hbar^{-1} \nabla_{\bm{k}} \varepsilon_{l,\bm{k}} = \eta_{l} \hbar \bm{k}/m_{l}$ with $\eta_{1} = 1$ and $\eta_{2} = - 1$.
The first three terms on the right-hand side of Eq.~(\ref{eq:Boltzmann_eq}) are scattering terms, and the last term describes the effect of a magnetic field. The dimensionless function $\Phi^{(l)}(\hat{\bm{k}},\xi)$, which is related to the non-equilibrium part of the distribution function $\delta f^{(l)}(\bm{k}) = f^{(l)}(\bm{k}) - f_0(\varepsilon_{l,\bm{k}})$, is defined as
\begin{align}
\delta f^{(l)}(\bm{k}) = \frac{1}{\beta}\left(- \frac{\partial f_0(\varepsilon_{l,\bm{k}})}{\partial \varepsilon_{l,\bm{k}}} \right) \Phi^{(l)}(\hat{\bm{k}},\xi_{l,\bm{k}}),
\end{align}
where $\hat{\bm{k}} =\bm{k}/k$.

The first term on the right-hand side, $I^{(l)}_{\text{imp}}[\Phi]$, represents the impurity scattering and is given by
\begin{align}
I^{(l)}_{\text{imp}}[\Phi] = \frac{1}{\beta} \left(- \frac{\partial f_0(\varepsilon_{l,\bm{k}})}{\partial \varepsilon_{l,\bm{k}}} \right) \frac{1}{\tau^{(l)}_{\text{imp}}} \Phi^{(l)} (\hat{\bm{k}},\xi_{l,\bm{k}}), \label{eq:imp_scattering}
\end{align}
where the relaxation time $\tau^{(l)}_{\text{imp}}$ is temperature independent.

\begin{figure}[tbp]
\begin{center}
\rotatebox{0}{\includegraphics[angle=0,width= 0.8 \linewidth]{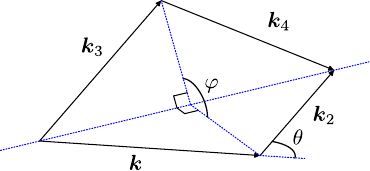}}
\caption{A schematic of the angles $\theta$ and $\varphi$ with a possible combination of momenta fixed on the Fermi surfaces satisfying the momentum conservation $\bm{k} + \bm{k}_2 = \bm{k}_3 + \bm{k}_4$.}
\label{Fig:angles}
\end{center}
\end{figure}

The second term, $I^{(ll)}_{\text{e-e}}[\Phi]$, is the intraband electron-electron scattering and the third term, $I^{(l\overline{l})}_{\text{e-e}}[\Phi]$, is the interband electron-electron scattering. These are given by
\begin{widetext}
\begin{align}
I^{(ll')}_{\text{e-e}}[\Phi] = 
& \frac{2}{V^3} \sum_{\bm{k}_{2}, \bm{k}_{3}, \bm{k}_{4}} \frac{1}{e^{\beta\xi_{l,\bm{k}}} + 1} \frac{1}{e^{\beta\xi_{l',\bm{k}_2}} + 1} \frac{1}{1 + e^{-\beta\xi_{l,\bm{k}_3}}} \frac{1}{1 + e^{-\beta\xi_{l',\bm{k}_4}}} \delta(\xi_{l,\bm{k}} + \xi_{l',\bm{k}_2} - \xi_{l,\bm{k}_3} - \xi_{l',\bm{k}_4}) (2\pi)^3 \delta(\bm{k} + \bm{k}_2 - \bm{k}_3 - \bm{k}_4) \nonumber \\
& \times 
W^{(ll')}(\bm{k},\bm{k}_2;\bm{k}_3,\bm{k}_4) 
(\Phi^{(l)}(\hat{\bm{k}},\xi_{l,\bm{k}}) + \Phi^{(l')}(\hat{\bm{k}}_{2},\xi_{l',\bm{k}_2}) - \Phi^{(l)}(\hat{\bm{k}}_{3},\xi_{l,\bm{k}_3}) - \Phi^{(l')}(\hat{\bm{k}}_{4},\xi_{l',\bm{k}_4})), \label{eq:el_el_scattering_original}
\end{align}
where $W^{(ll')}$ represents the intraband ($l = l'$) and interband ($l \neq l'$) electron-electron scattering probability.
By focusing on the Fermi surfaces, the scattering term becomes \cite{Oliva1982, Anderson1987, Li2018, Lee2021}
\begin{align}
I^{(ll')}_{\text{e-e}}[\Phi] = 
&\frac{m_{l} m_{l'}^2}{8\pi^4 \hbar^{6}} \int_{-\infty}^{\infty} d\xi_{2} d\xi_{3} d\xi_{4} \frac{1}{e^{\beta\xi_{l,\bm{k}}} + 1} \frac{1}{e^{\beta\xi_{2}} + 1} \frac{1}{1 + e^{-\beta\xi_{3}}} \frac{1}{1 + e^{-\beta\xi_{4}}} \delta(\xi_{l,\bm{k}} + \xi_2 - \xi_3 - \xi_4) \nonumber \\
& \times \int \frac{d\Omega}{4\pi} \int \frac{d\varphi_{2}}{2\pi} \frac{W^{(ll')}(\theta,\varphi)}{R^{(ll')}(\theta)} (\Phi^{(l)}(\hat{\bm{k}},\xi_{l,\bm{k}}) + \Phi^{(l')}(\hat{\bm{k}}_{2 },\xi_{2}) - \Phi^{(l)}(\hat{\bm{k}}_{3},\xi_{3}) - \Phi^{(l')}(\hat{\bm{k}}_{4},\xi_{4})), \label{eq:el_el_scattering}
\end{align}
\begin{align}
R^{(ll')}(\theta) = \frac{\sqrt{k_{\text{F},l}^2 + k_{\text{F},l'}^2 + 2k_{\text{F},l}k_{\text{F},l'} \cos \theta}}{2k_{\text{F},l'}},
\end{align}
\end{widetext}
which is an extension of the collision integral originally considered for the single-carrier system by Abrikosov and Khalatnikov \cite{Abrikosov1957, Abrikosov1959}. Note that we exclude the possibility of Umklapp scattering. We can parametrize $W^{(ll')}$ with the two angles $\theta$ and $\varphi$ (depicted in Fig.~\ref{Fig:angles}) since we fix the momenta on the Fermi surfaces and consider the isotropic system.
$\theta$ is the angle between $\bm{k}$ and $\bm{k}_2$ and $\varphi$ is the angle between the two planes, one of which is spanned by $\bm{k}$ and $\bm{k}_2$ and the other is spanned by $\bm{k}_3$ and $\bm{k}_4$. For the angular integral, $d\Omega = \sin \theta d\theta d\varphi$ and  $\varphi_{2}$ is the azimuth angle of $\bm{k}_2$ relative to $\bm{k}$. $R^{(ll')}(\theta)$ represents some geometrical factor \cite{Anderson1987}. In particular, $R^{(ll')}(\theta) = \cos(\theta/2)$ when $k_{\text{F},l} = k_{\text{F},l'}$.

Finally, the effect of a magnetic field along the $z$-axis $M^{(l)}[\Phi]$ is expressed as \cite{Ziman2001, Schulz1995, Lee2020magneto}, 
\begin{align}
M^{(l)}[\Phi] = \frac{e}{\hbar} \cdot \frac{1}{\beta} \left( - \frac{\partial f_{0}(\varepsilon_{l,\bm{k}})}{\partial \varepsilon_{l,\bm{k}}} \right) ( \bm{v}^{(l)}_{\bm{k}} \times \bm{B}) \cdot \nabla_{\bm{k}} \Phi^{(l)} (\hat{\bm{k}},\xi_{l,\bm{k}}), \label{eq:magnetic_field_operator}
\end{align}
where $\bm{B} = (0,0,B)$. Note that the above semiclassical equations do not take into account the effect of Landau quantization, which is out of the scope of the present study.

\subsection{Transport coefficients and various kinds of Lorenz ratios}
The electric current $\bm{j}$ and thermal current $\bm{j}_{q}$ are given by 
\begin{align}
\bm{j} =& \frac{2e}{V} \sum_{l,\bm{k}} \bm{v}^{(l)}_{\bm{k}} \frac{1}{\beta} \left(- \frac{\partial f_0(\varepsilon_{l,\bm{k}})}{\partial \varepsilon_{l,\bm{k}}} \right) \Phi^{(l)}(\hat{\bm{k}},\xi_{l,\bm{k}}), \label{eq:def_el_current} \\
\bm{j}_{q} =& \frac{2}{V} \sum_{l,\bm{k}} \xi_{l,\bm{k}} \bm{v}^{(l)}_{\bm{k}} \frac{1}{\beta} \left(- \frac{\partial f_0(\varepsilon_{l,\bm{k}})}{\partial \varepsilon_{l,\bm{k}}} \right) \Phi^{(l)}(\hat{\bm{k}},\xi_{l,\bm{k}}), \label{eq:def_th_current} 
\end{align}
 where 2 comes from the spin degeneracy. 

Responses of the electric and thermal currents to the electric field $\bm{E}$ and the temperature gradient $\nabla T$ are summarized as follows:
\begin{align}
\left( 
\begin{array}{c}
\bm{j} \\
\bm{j}_{q} \\
\end{array}
\right) =
\left( 
\begin{array}{cc}
\hat{L}_{11} & \hat{L}_{12}\\
\hat{L}_{21} & \hat{L}_{22}
\end{array}
\right)\left( 
\begin{array}{c}
\bm{E} \\
- \nabla T/T \\
\end{array}
\right).
\end{align}
The Onsager relations for transport coefficients are given by ${}^{t}\hat{L}_{11}(-\bm{B}) = \hat{L}_{11}(\bm{B})$, ${}^{t}\hat{L}_{22}(-\bm{B}) = \hat{L}_{22}(\bm{B})$, and ${}^{t}\hat{L}_{12}(-\bm{B}) = \hat{L}_{21}(\bm{B})$ \cite{Ziman2001}. The electrical and thermal conductivity tensors are given by 
\begin{align}
\hat{\sigma} =& \hat{L}_{11}, \label{eq:def_sigma} \\
\hat{\kappa} =& \frac{1}{T} \left[\hat{L}_{22} - \hat{L}_{21}\hat{L}_{11}^{-1}\hat{L}_{12}\right],  \label{eq:def_kappa}
\end{align}
where the second term of $\hat{\kappa}$ originates from the open circuit condition $\bm{j} = 0$. Note that $\sigma_{xy} = -\sigma_{yx}$ and $\kappa_{xy} = -\kappa_{yx}$. As discussed in detail in Sec.~\ref{Sec:integral_eq}, we neglect $- \hat{L}_{21}\hat{L}_{11}^{-1}\hat{L}_{12}/T$ in $\hat{\kappa}$ at low temperatures. 

The electrical resistivity $\rho$ and thermal resistivity $\rho_{\text{th}}$ are given by
\begin{align}
\rho = \frac{\sigma_{xx}}{\sigma_{xx}^2 + \sigma_{xy}^2},~\rho_{\text{th}} = \frac{\kappa_{xx}}{\kappa_{xx}^2 + \kappa_{xy}^2}.
\end{align}
In the following, instead of $\rho_{\text{th}}$, we discuss $WT$ defined by
\begin{align}
WT = TL_0\rho_{\text{th}} = L_0 \cdot \frac{\kappa_{xx}/T}{(\kappa_{xx}/T)^2 + (\kappa_{xy}/T)^2},
\end{align}
which has the same dimension as $\rho$.
In the presence of the magnetic field, corresponding to the Hall coefficient,
\begin{align}
R_{\text{H}} = \frac{1}{B} \cdot \frac{\sigma_{xy}}{\sigma_{xx}^2 + \sigma_{xy}^2}, \label{eq:def_R_H}
\end{align}
we define $K_{\text{H}}$ as 
\begin{align}
K_{\text{H}} = \frac{1}{B} \cdot \frac{\kappa_{xy}/T}{(\kappa_{xx}/T)^2 + (\kappa_{xy}/T)^2}. \label{eq:def_K_H}
\end{align}

Finally, we introduce several kinds of Lorenz ratios, which will be used to compare electric and thermal transport in terms of resistivity and to simplify analyses of $L_{\text{H}}$. In analogy to $L$ and $L_{\text{H}}$, we define
\begin{align}
\widetilde{L}_{} = \frac{\rho}{T \rho_{\text{th}}} = \frac{L_{0}\rho}{WT},~\widetilde{L}_{\text{H}} = \frac{R_{\text{H}}}{K_{\text{H}}},
\end{align}
which are the Lorenz ratios defined in terms of the resistivity tensor. They always satisfy 
\begin{align}
\frac{L_{\text{H}}}{L_{}} = \frac{\widetilde{L}_{}}{\widetilde{L}_{\text{H}}}.
\end{align}
It is easy to show that other types of the WF law, $\widetilde{L} = L_{0}$ and $\widetilde{L}_{\text{H}} = L_{0}$, hold when only the elastic scattering exists.
In the weak magnetic field limit, we have $\widetilde{L}_{} \to {L}_{}~(B \to 0)$ and 
\begin{align}
L_{\text{H}} = \frac{K_{\text{H}}}{R_{\text{H}}} L_{}^2 = \frac{L_{}^2}{\widetilde{L}_{\text{H}}}~(B \to 0). \label{eq:Hall_Lorenz_weak_mag}
\end{align}
This shows that, if $K_{\text{H}}$ behaves in the same way as $R_{\text{H}}$, the Hall Lorenz ratio is proportional to the square of the Lorenz ratio, i.e., $L_{\text{H}}/L_{0} \simeq (L_{}/L_{0})^2$ \cite{Long1971, Zhang2000}.

\section{Integral equation} \label{Sec:integral_eq}
In this section, we present the integral equations for the energy-dependent parts of distribution functions and their relations to the electrical and thermal conductivities. First, we expand and parametrize the distribution functions. Then, we give the integral equations for the two-band system. 

We expand the distribution functions as
\begin{align}
\Phi^{(l)}(\hat{\bm{k}},\xi_{l,\bm{k}}) = \sum_{n,m} Y_{n,m}(\theta_1,\phi_1) \widetilde{\Phi}^{(l)}_{n,m}(\beta \xi_{l,\bm{k}}), \label{eq:distr_expansion_spherical}
\end{align}
where $Y_{n,m}(\theta_1,\phi_1)$ are real spherical harmonics, $\theta_1$ is the polar angle, and $\phi_1$ is the azimuth angle of $\bm{k}$. In particular, $Y_{1,1}(\theta_1,\phi_1) = \sqrt{3/4\pi} \sin \theta_1 \cos \phi_1 = \sqrt{3/4\pi} \cdot k_x/k$ and $Y_{1,-1}(\theta_1,\phi_1) = \sqrt{3/4\pi} \sin \theta_1 \sin \phi_1 = \sqrt{3/4\pi} \cdot k_y/k$. 

We substitute Eq.~({\ref{eq:distr_expansion_spherical}}) into $I^{(ll')}_{\text{e-e}}[\Phi]$ [Eq.~({\ref{eq:el_el_scattering}})] to obtain the integral equation for the degree of freedom of energy following the treatment for the electron-electron scatterings \cite{Abrikosov1957, Abrikosov1959, Brooker1968, Jensen1968, Smith1969, Jensen1969, Bennett1969, Sykes1970, Ah-Sam1971, Brooker1972, Egilsson1977, Oliva1982, Anderson1987, Smith_Jensen1989, Golosov1995, Golosov1998, Pethick2009, Li2018, Lee2020, Lee2021}. The details are shown in Appendix \ref{App:derivation_integral_eq}. 

Because the left-hand side of the Boltzmann equation [Eq.~(\ref{eq:Boltzmann_eq})] is proportional to $k_{x} \propto Y_{1,1}$, in the present case, we only have to consider the modes $(n,m) = (1,\pm1)$. Note that $Y_{1,-1}$ is also involved since the magnetic field $M^{(l)}[\Phi]$ connects $Y_{1,-1}$ with $Y_{1,1}$. The obtained integral equations for $\widetilde{\Phi}^{(l)}_{1,\pm 1}(u)$ with $u = \beta \xi_{l,\bm{k}}$ are shown in Eqs.~(\ref{eq:int_eq_x_axis}) and (\ref{eq:int_eq_y_axis}) in Appendix \ref{App:derivation_integral_eq}.

Furthermore, we consider that $\widetilde{\Phi}^{(l)}_{1, \pm 1}(u)$ is an even function in terms of $u$ when we consider the case of electrical transport, $F^{(l)}_{\text{ext}} = eE$, while it is an odd function in the case of thermal transport $F^{(l)}_{\text{ext}} = \xi_{l,\bm{k}}(- \nabla T/T)$ \cite{Abrikosov1957, Abrikosov1959}. Therefore, it is natural to parametrize the energy dependence of distribution functions as 
\begin{widetext}
\begin{align}
\widetilde{\Phi}^{(l)}_{1,1}(u) =& \eta_{l} \sqrt{\frac{4\pi}{3}} v_{\text{F},l} \tau_{\text{e-e}}^{(l)} \cosh(u/2)  \left[  \beta eE \varphi^{(l)}_{\sigma;x}(u) + \left( -\frac{\nabla T}{T} \right) \varphi^{(l)}_{\kappa;x}(u) \right], \label{eq:distr_x_parametrize} \\
\widetilde{\Phi}^{(l)}_{1,-1}(u) =& \eta_{l} \sqrt{\frac{4\pi}{3}} v_{\text{F},l} \tau_{\text{e-e}}^{(l)} \cosh(u/2)  \left[  \beta eE \varphi^{(l)}_{\sigma;y}(u) + \left( -\frac{\nabla T}{T} \right) \varphi^{(l)}_{\kappa;y}(u) \right], \label{eq:distr_y_parametrize}
\end{align}
\end{widetext}
where $v_{\text{F},l} = \hbar k_{\text{F},l}/m_{l}$. $\varphi^{(l)}_{\sigma;x}(u)$ and $\varphi^{(l)}_{\sigma;y}(u)$ are even functions in terms of $u$, while $\varphi^{(l)}_{\kappa;x}(u)$ and $\varphi^{(l)}_{\kappa;y}(u)$ are odd functions. Here, it turned out to be convenient to include $\tau_{\text{e-e}}^{(l)}$ in Eqs.~(\ref{eq:distr_x_parametrize}) and (\ref{eq:distr_y_parametrize}), which is a characteristic time of electron-electron scattering defined as 
\begin{align}
\frac{1}{\tau_{\text{e-e}}^{(l)}} = \frac{1}{\tau_{\text{e-e}}^{(l1)}} + \frac{1}{\tau_{\text{e-e}}^{(l2)}}, \label{eq:sum_rel_time_ee}
\end{align}
with 
\begin{align}
\frac{1}{\tau_{\text{e-e}}^{(ll')}} = \frac{m_{l} m_{l'}^2 (k_BT)^2}{8\pi^4 \hbar^{6}} \int \frac{d\Omega}{4\pi} \frac{W^{(ll')}(\theta,\varphi)}{R^{(ll')}(\theta)}. \label{eq:def_rel_time_ee}
\end{align} 
We note that the lifetime of electrons and holes on the Fermi surfaces is given by $2/\pi^2 \cdot \tau_{\text{e-e}}^{(l)}$ \cite{Pethick2009} (see also Appendix \ref{App:derivation_integral_eq}). 
Furthermore, by examining the integral equations for $\widetilde{\Phi}^{(l)}_{1,\pm 1}(u)$ in Eqs.~(\ref{eq:int_eq_x_axis}) and (\ref{eq:int_eq_y_axis}), we find that it is convenient to introduce a new distribution function as
\begin{align}
\varphi^{(l)}_{X}(u) := \varphi^{(l)}_{X;x}(u) + i\varphi^{(l)}_{X;y}(u)~(X = \sigma~\text{or}~\kappa).
\end{align}
Then, we reach the following coupled integral equations for $\varphi_{X}^{(l)}(u)$:
\begin{widetext}
\begin{align}
F_{X}(x) =& (\zeta_{1}^2 \pi^2 + x^2) \varphi_{X}^{(1)}(x) - 2\int_{-\infty}^{\infty} \mathcal{G}(x - u)\left( \lambda^{(1)}_{X} \varphi_{X}^{(1)}(u) - \frac{ \tau_{\text{e-e}}^{(2)}}{ \tau_{\text{e-e}}^{(1)}}\beta^{(1)}_{X} \varphi_{X}^{(2)}(u) \right) du, \label{eq:int_eq_electrons} \\
F_{X}(x) =& (\zeta_{2}^2 \pi^2 + x^2) \varphi_{X}^{(2)}(x) - 2\int_{-\infty}^{\infty} \mathcal{G}(x - u)\left( \lambda^{(2)}_{X} \varphi_{X}^{(2)}(u) - \frac{ \tau_{\text{e-e}}^{(1)}}{ \tau_{\text{e-e}}^{(2)}} \beta^{(2)}_{X} \varphi_{X}^{(1)}(u) \right) du, \label{eq:int_eq_holes} 
\end{align}
\end{widetext}
where $F_{X}$ and $\mathcal{G}$ are defined by
\begin{align}
F_{\sigma}(x) =& \frac{2}{\cosh(x/2)}, \\
F_{\kappa}(x) =& \frac{2x}{\cosh(x/2)}, \\
\mathcal{G}(x) =& \frac{x}{2\sinh(x/2)},
\end{align}
and the effect of the magnetic field appears only in $\zeta_{l}^2$ which is given by
\begin{align}
\zeta_{l}^2 = 1 + \frac{2 \tau_{\text{e-e}}^{(l)}}{\pi^2}\left(\frac{1}{ \tau_{\text{imp}}^{(l)}} - i \eta_{l} \omega^{(l)}_{\text{c}}\right), \label{eq:def_zeta_l}
\end{align}
with $\omega^{(l)}_{\text{c}} = |e|B/m_{l}$. We take $\zeta_{l}$ so as to make $\text{Re}~\zeta_{l} > 0$. Equations (\ref{eq:int_eq_electrons}) and (\ref{eq:int_eq_holes}) are extensions of the previously used equations \cite{Oliva1982, Anderson1987, Li2018, Lee2021} for the case of a magnetic field.

The dimensionless real parameters $\lambda^{(l)}_{X}$ and $\beta^{(l)}_{X}$, which characterize the angular integration of potentials of the electron-electron scatterings \cite{Oliva1982, Anderson1987, Li2018, Lee2021}, are defined by 
\begin{align}
\lambda^{(l)}_{\sigma} =& \tau^{(l)}_{\text{e-e}} \left( \frac{ - \Lambda^{(ll)}_{2} + \Lambda^{(ll)}_{3} + \Lambda^{(ll)}_{4}}{\tau^{(ll)}_{\text{e-e}}} + \frac{\Lambda^{(l\overline{l})}_{3}}{\tau^{(l\overline{l})}_{\text{e-e}}} \right), \label{eq:def_lambda_sigma} \\
\beta^{(l)}_{\sigma} =& \tau^{(l)}_{\text{e-e}} \cdot \frac{v_{\text{F},\overline{l}}}{v_{\text{F},l}} \cdot \frac{- \Lambda^{(l\overline{l})}_{2} + \Lambda^{(l\overline{l})}_{4}}{\tau^{(l\overline{l})}_{\text{e-e}}}, \label{eq:def_beta_sigma} \\
\lambda^{(l)}_{\kappa} =& \tau^{(l)}_{\text{e-e}} \left( \frac{\Lambda^{(ll)}_{2} + \Lambda^{(ll)}_{3} + \Lambda^{(ll)}_{4}}{\tau^{(ll)}_{\text{e-e}}} + \frac{\Lambda^{(l\overline{l})}_{3}}{\tau^{(l\overline{l})}_{\text{e-e}}}\right), \label{eq:def_lambda_kappa} \\
\beta^{(l)}_{\kappa} =& \tau^{(l)}_{\text{e-e}} \cdot \frac{v_{\text{F},\overline{l}}}{v_{\text{F},l}} \cdot \frac{\Lambda^{(l\overline{l})}_{2} + \Lambda^{(l\overline{l})}_{4}}{\tau^{(l\overline{l})}_{\text{e-e}}}. \label{eq:def_beta_kappa}
\end{align}
Here, $\Lambda^{(ll')}_{i}$ is defined by $\Lambda^{(ll')}_{i} = \Lambda^{(ll')}_{i; n =1}$, which represents the geometrical factor of the electron-electron scattering. For general $n$, $\Lambda_{i;n}^{(ll')}$ is given by
\begin{align}
\Lambda_{i;n}^{(ll')} 
=& \left(\int \frac{d\Omega}{4\pi} \frac{W^{(ll')}(\theta,\varphi)}{R^{(ll')}(\theta)} \right)^{-1} \nonumber \\
& \times \int \frac{d\Omega}{4\pi} \frac{W^{(ll')}(\theta,\varphi)}{R^{(ll')}(\theta)} P_{n}(\cos \theta_{1i}), \label{eq:def_Lambda_n}
\end{align}
where $P_{n}$ is the Legendre polynomial and $\theta_{1i}$ is the angle between $\hat{\bm{k}}$ and $\hat{\bm{k}}_{i}$. Derivations are given in Appendix \ref{App:derivation_integral_eq}. 

We note several properties of the parameters. 
First, the momentum conservation $\bm{k} + \bm{k}_{2} = \bm{k}_{3} + \bm{k}_{4}$ is reflected in various identities. The following identity is satisfied:
\begin{align}
k_{\text{F},l} + k_{\text{F},l'}\cos \theta_{12} - k_{\text{F},l} \cos \theta_{13} - k_{\text{F},l'}\cos \theta_{14} = 0.
\end{align}
Therefore, from the definition of $\Lambda^{(ll')}_{i}$ in Eq.~(\ref{eq:def_Lambda_n}), we obtain
\begin{align}
k_{\text{F},l} + k_{\text{F},l'}\Lambda^{(ll')}_{2} - k_{\text{F},l}\Lambda^{(ll')}_{3} - k_{\text{F},l'}\Lambda^{(ll')}_{4}  = 0.
\end{align} 
In particular, we have $1 + \Lambda^{(ll)}_{2} - \Lambda^{(ll)}_{3} -\Lambda^{(ll)}_{4}  = 0$ when $l= l'$. With this identity, we find
\begin{align}
\lambda^{(l)}_{\sigma} =& \tau^{(l)}_{\text{e-e}} \left( \frac{1}{\tau^{(ll)}_{\text{e-e}}} + \frac{\Lambda^{(l\overline{l})}_{3}}{\tau^{(l\overline{l})}_{\text{e-e}}} \right) \leq 1,
\end{align} 
and 
\begin{align}
\lambda^{(l)}_{\sigma} \geq& \tau^{(l)}_{\text{e-e}} \frac{\Lambda^{(l\overline{l})}_{3}}{\tau^{(l\overline{l})}_{\text{e-e}}} \geq - 1,
\end{align} 
where we have used $-1 \leq \Lambda^{(ll')}_{i} \leq 1$ from Eq.~(\ref{eq:def_Lambda_n}). Thus, $-1 \leq \lambda^{(l)}_{\sigma} \leq 1$ is satisfied. We see that $\lambda^{(l)}_{\sigma} = 1$ is satisfied in the absence of the interband scattering, or $1/\tau^{(l\overline{l})}_{\text{e-e}} = 0$. In such a case, the electrical conductivity is not affected by the intraband electron-electron scattering \cite{Smith1969, Ah-Sam1971, Lee2020}. We get the following identity as well \cite{Anderson1987, Lee2021}:
\begin{align}
&\lambda^{(l)}_{\sigma} + \frac{m_{\overline{l}}}{m_l} \beta^{(l)}_{\sigma} \nonumber \\
&= \tau^{(l)}_{\text{e-e}} \left( \frac{1}{\tau^{(ll)}_{\text{e-e}}} + \frac{\Lambda^{(l\overline{l})}_{3}}{\tau^{(l\overline{l})}_{\text{e-e}}} + \frac{k_{\text{F},\overline{l}}}{k_{\text{F},l}} \cdot \frac{- \Lambda^{(l\overline{l})}_{2} + \Lambda^{(l\overline{l})}_{4}}{\tau^{(l\overline{l})}_{\text{e-e}}} \right) = 1. \label{eq:beta_constrainst} 
\end{align} 

For $\lambda^{(l)}_{\kappa}$, we have 
\begin{align}
-1 \leq \lambda^{(l)}_{\kappa} =& \tau^{(l)}_{\text{e-e}} \left( \frac{1 + 2\Lambda^{(ll)}_{2}}{\tau^{(ll)}_{\text{e-e}}} + \frac{\Lambda^{(l\overline{l})}_{3}}{\tau^{(l\overline{l})}_{\text{e-e}}} \right) \leq 3.
\end{align} 
$\lambda_{\kappa} = 3$ is the maximum value of $\lambda_{\kappa}$ and is the special case where the carriers are scattered only forward ($\theta = 0$) by the intraband scattering. 

We introduce a matrix $\hat{\lambda}_{X}$ which summarizes the electron-electron scatterings as 
\begin{align}
\hat{\lambda}_{X} = \left( 
\begin{array}{cc}
\lambda^{(1)}_{X} & - \frac{ \tau_{\text{e-e}}^{(2)}}{ \tau_{\text{e-e}}^{(1)}}\beta^{(1)}_{X}\\
- \frac{ \tau_{\text{e-e}}^{(1)}}{ \tau_{\text{e-e}}^{(2)}}\beta^{(2)}_{X} & \lambda^{(2)}_{X}
\end{array}
\right). \label{eq:lambda_matrix}
\end{align}
$\hat{\lambda}_{\sigma}$ has 1 as an eigenvalue due to the momentum conservation \cite{Anderson1987}. We can directly confirm this using Eq.~(\ref{eq:beta_constrainst}).

After solving the integral equation, the transport coefficients [Eqs.~(\ref{eq:def_sigma}) and (\ref{eq:def_kappa})] are given by
\begin{align}
\sigma_{xx} + i\sigma_{yx} =& \sum_{l = 1,2} \frac{e^2 n_{l} \tau_{\text{e-e}}^{(l)}}{m_{l}} \int_{-\infty}^{\infty}  \frac{1}{4\cosh(u/2)} \varphi^{(l)}_{\sigma}(u) du, \label{eq:el_cond_integral} \\
\kappa_{xx} + i\kappa_{yx} =& \sum_{l = 1,2} \frac{k_B^2 T n_{l} \tau_{\text{e-e}}^{(l)}}{m_{l}} \int_{-\infty}^{\infty} \frac{u}{4\cosh(u/2)} \varphi^{(l)}_{\kappa}(u) du, \label{eq:th_cond_integral} 
\end{align}
where $- \hat{L}_{21}\hat{L}_{11}^{-1}\hat{L}_{12}/T$ in $\hat{\kappa}$ is neglected as stated.

We would like to note some remarks on neglecting $- \hat{L}_{21}\hat{L}_{11}^{-1}\hat{L}_{12}/T$ in $\hat{\kappa}$. In the present formalism, $\widetilde{\Phi}_{n,m}^{(l)}(x)$ is an odd function when we consider thermal transport \cite{Abrikosov1957, Abrikosov1959}. This is because we fixed wavevectors on the Fermi surfaces as in Eq.~(\ref{eq:el_el_scattering}) to focus on the leading order of energy. In this formalism, $k_{x}$ and $k_{y}$ in both sides of Eq.~(\ref{eq:Boltzmann_eq}) are treated as $k_{\text{F},l}Y_{n = 1, m = \pm 1}(\theta_1,\phi_1)$. Then, even and odd parts of the functions are decoupled in Eqs.~(\ref{eq:int_eq_electrons}), (\ref{eq:int_eq_holes}), and (\ref{eq:th_cond_integral}). This leads to $\hat{L}_{12} = 0$ and $\hat{L}_{21} = 0$ and vanishing contribution in $\hat{L}_{22}$ from even functions. Note that, in this case, we cannot discuss the Seebeck effect.
However, when we include higher orders with respect to temperature, $\widetilde{\Phi}_{n,m}^{(l)}(x)$ should be a mixture of even and odd functions, and $\hat{L}_{12}$ and $\hat{L}_{21}$ become non-zero. To roughly discuss the order of $\hat{L}_{ij}$, we introduce two typical relaxation times: $\tau_{\text{tr},\sigma}$ for electrical transport and $\tau_{\text{tr},\kappa}$ for thermal transport.
We can estimate $\hat{L}_{21} = \mathcal{O}(\tau_{\text{tr},\sigma} T^2)$, $-\hat{L}_{21}\hat{L}_{11}^{-1}\hat{L}_{12}/T = \mathcal{O}(\tau_{\text{tr},\sigma} T^3)$, on the other hand, $\hat{L}_{22}/T = \mathcal{O}(\tau_{\text{tr},\kappa} T) + \mathcal{O}(\tau_{\text{tr},\sigma} T^3)$. In the usual case, we can neglect the correction terms of the order of $\mathcal{O}(\tau_{\text{tr},\sigma} T^3)$ unless $\tau_{\text{tr}, \sigma}$ is much larger than $\tau_{\text{tr}, \kappa}$.
Therefore, we have to take care of the case where $\tau_{\text{tr}, \sigma}$ diverges and $\tau_{\text{tr}, \sigma} T^3$ cannot be neglected in the setting of this paper. Actually, $k_{x}$ and $k_{y}$ are closely related to the divergence of $\tau_{\text{tr}, \sigma}$ since the momentum conservation makes $k_{x}$ and $k_{y}$ eigenfunctions with zero eigenvalues for the original scattering term Eq.~(\ref{eq:el_el_scattering_original}). Therefore, the relaxation of the modes $k_{x}$ and $k_{y}$ vanishes if the impurity scattering and Umklapp scattering are negligible. This leads to the divergence of $\tau_{\text{tr}, \sigma}$, and then, $\hat{L}_{ij}~(i,j \in \{1,2 \})$. However, it has been shown that the divergent term in $\hat{L}_{22}/T$, which is the order of $\mathcal{O}(\tau_{\text{tr},\sigma} T^3)$, is canceled by the divergence of $- \hat{L}_{21}\hat{L}_{11}^{-1}\hat{L}_{12}/T$ in the single-carrier case \cite{Lucas2018, Lee2020}. This cancellation is a consequence of the condition of no electric current $\bm{j} = \hat{L}_{11} \bm{E} + \hat{L}_{12} (-\nabla T/T) = 0$. We only have to consider the odd part $\widetilde{\Phi}_{n,m}^{(l)}(x)$ for thermal transport, and we can neglect $- \hat{L}_{21}\hat{L}_{11}^{-1}\hat{L}_{12}/T$ in the single-carrier case. This cancellation is not exact in multi-band systems. The remaining contribution, the ambipolar contribution, could diverge in the absence of the impurity scattering in a compensated case \cite{Zarenia2020, Lee2021, Takahashi2023}. However, considering that the impurity scattering will suppress the divergence of $\tau_{\text{tr}, \sigma}$, we neglect the contribution of $- \hat{L}_{21}\hat{L}_{11}^{-1}\hat{L}_{12}/T$ as it is the order of $\tau_{\text{tr},\sigma} T^3$ at low temperatures which we are interested in (see also Appendix~\ref{App:RTA} where we provide formulae in the RTA including this contribution).

\section{Baber scattering} \label{Sec:Baber}

In this section, we discuss Baber scattering, in which one of the carriers (assumed to be holes in this paper) is in equilibrium \cite{Baber1937, Bennett1969, Schriempf1969, Ah-Sam1971, Lin2015}. We give a set of eigenfunctions of the integral equation for the single-band case, obtain exact transport coefficients, and discuss their behaviors.

In Baber scattering, the distribution function of the strongly relaxed carriers (here holes) is neglected, i.e., $\varphi^{(2)}_{X}(x) = 0$. This will be realized, for example, if we take the limit of $1/\tau_{\text{imp}}^{(2)} \to \infty$. Then, we can find analytic formulae for $\varphi^{(1)}_{X}(x)$ and the electrical and thermal conductivities because the system is effectively a single-carrier system.

It is sufficient to consider the integral equation
\begin{align}
F_{X}(x) =& (\zeta^{2}\pi^2 + x^2) \varphi_{X}(x) \nonumber \\
 &- 2\lambda_{X} \int_{-\infty}^{\infty} \mathcal{G}(x - u) \varphi_{X}(u) du, \label{eq:single_int_eq}
\end{align}
where the parameters are abbreviated as $\tau_{\text{e-e}}^{(1)} \to \tau_{\text{e-e}}$, $\zeta_1 \to \zeta$, and $\lambda^{(1)}_{X} \to \lambda_{X}$. In the following, all the band indices $(l,l')$ are unnecessary.
The electrical and thermal conductivities are given by
\begin{align}
\sigma_{xx} + i\sigma_{yx} =& \frac{e^2 n \tau_{\text{e-e}}}{m} \int_{-\infty}^{\infty}  \frac{1}{4\cosh(u/2)} \varphi_{\sigma}(u) du, \label{eq:el_cond_integral_single} \\
\kappa_{xx} + i\kappa_{yx} =& \frac{k_B^2 T n \tau_{\text{e-e}}}{m} \int_{-\infty}^{\infty}  \frac{u}{4\cosh(u/2)} \varphi_{\kappa}(u) du. \label{eq:th_cond_integral_single} 
\end{align}

Although holes do not carry currents directly, they scatter off the electrons. The information is encoded in parameters. Especially, the interband scattering $I^{(12)}_{\text{e-e}}$ makes $\lambda_{\sigma}$ lower than 1 and the finite $T^2$ resistivity arises \cite{Ah-Sam1971}. This means that the momentum is lost through the holes in equilibrium.

\subsection{Eigen functions}
First, we solve the following eigenvalue problem appearing on the left-hand side of Eq.~(\ref{eq:single_int_eq}) \cite{Bennett1969, Brooker1972, Golosov1995, Golosov1998, Pethick2009, Lee2020}:
\begin{align}
(\zeta^{2} \pi^2 + x^2) \varphi(x) = 2\lambda \int_{-\infty}^{\infty} \mathcal{G}(x - u) \varphi(u) du. \label{eq:eigen_prob}
\end{align}
The eigenfunction immediately gives the solution of Baber scattering and is also important for the analysis of semimetals in the following sections. 

Fourier transformation, $\psi(k) = \int_{-\infty}^{\infty} \varphi(x) e^{-ikx} dx$, brings Eq.~(\ref{eq:eigen_prob}) to a differential equation,
\begin{align}
- \frac{d^2}{dk^2} \psi(k) + \zeta^2\pi^2 \psi(k) = 2\pi^2 \lambda \text{sech}^2(\pi k) \psi(k). \label{eq:sech_schrodinger}
\end{align}
The eigenfunction of Eq.~(\ref{eq:sech_schrodinger}) is given by \cite{Bennett1969, Brooker1972, Golosov1995, Golosov1998, Pethick2009, Lee2020}
\begin{align}
&\psi_{n;\zeta}(k) \nonumber \\
=& [\text{sech}(\pi k)]^{\zeta} {}_{2}F_{1}\left[-n, n + 2\zeta + 1, 1 + \zeta , \frac{1 - \tanh(\pi k)}{2} \right],
\end{align}
where ${}_{2} F_{1}$ is the hypergeometric function. This eigenfunction is equivalent to 
\begin{align}
\psi_{n;\zeta}(k) 
=& \frac{\Gamma(n + 1)\Gamma(2\zeta + 1)}{\Gamma(n + 2\zeta + 1 )} [\text{sech}(\pi k)]^{\zeta} C_{n}^{\zeta + 1/2}\left(\tanh(\pi k)\right) \\
=& \frac{\Gamma(n + 1)\Gamma(\zeta + 1)}{\Gamma(n + \zeta + 1)} [\text{sech}(\pi k)]^{\zeta} P_{n}^{(\zeta,\zeta)}\left( \tanh(\pi k) \right),   
\end{align}
where $C_{n}^{\alpha}$ is the Gegenbauer polynomial, and $P_{n}^{(\alpha, \beta)}$ is the Jacobi polynomial, respectively. The eigenvalue is given by $\lambda = \lambda_{n}(\zeta) = (n + \zeta)(n + \zeta + 1)/2$ for a complex parameter $\zeta$.

$\psi_{n;\zeta}(k)$ satisfies the orthogonal relation,
\begin{align}
&\int_{-\infty}^{\infty}\text{sech}^2(\pi k) \psi_{n;\zeta}(k) \psi_{m;\zeta}(k) dk \nonumber \\
&= \frac{ 2^{2\zeta + 1}\Gamma(n + 1)[\Gamma(\zeta + 1)]^2}{\pi(2n + 2\zeta + 1) \Gamma(n + 2\zeta + 1)} \delta_{n,m}, \label{eq:eigen_orthogonal_k}
\end{align}
which is inherited from the orthogonal relations of the Jacobi polynomials \cite{Table_of_Integrals}. Then, $\varphi_{n;\zeta}(x)$ is given by
\begin{align}
\varphi_{n;\zeta}(x) = \frac{1}{2\pi} \int_{-\infty}^{\infty} \psi_{n;\zeta}(k) e^{ikx} dk.
\end{align}
Note that $\psi_{n;\zeta}(k) = (-1)^{n}\psi_{n;\zeta}(-k)$ and $\varphi_{n;\zeta}(x) = (-1)^{n}\varphi_{n;\zeta}(-x)$ since $P_{n}^{(\alpha,\alpha)}(-x) = (-1)^{n} P_{n}^{(\alpha,\alpha)}(x)$.

As shown in Eq.~(\ref{eq:def_zeta_l}), when the impurity scattering and a magnetic field are absent, $\zeta = 1$. In this case, the eigenfunction, $\psi_{n;\zeta = 1}$, is given by
\begin{align}
\psi_{n;\zeta = 1}(k) = - \frac{2}{(n+1)(n+2)} P_{n+1}^{1}(\tanh(\pi k)),
\end{align}
where $P_{n}^{l}$ is the associated Legendre polynomial. 

\subsection{Exact formulae of the electrical and thermal conductivities}
For the electrical conductivity $(X = \sigma)$, the energy dependence of the distribution function can be expanded as
\begin{align}
\varphi_{\sigma}(x) = \frac{F_{\sigma}(x)}{x^2 + \zeta^2 \pi^2} + \sum_{n = 0}^{\infty} d_{2n} \varphi_{2n;\zeta}(x), \label{eq:single_el_expansion_ser}
\end{align}
where $\varphi_{2n;\zeta}(x)$ is an even function. This expansion is a generalization of that studied without the magnetic field \cite{Jensen1969, Smith_Jensen1989}.
Using the orthogonal relation Eq.~(\ref{eq:eigen_orthogonal_k}) and some integrals, we can find $d_{2n}$ as 
\begin{align}
d_{2n} =& \frac{2 (2n + \zeta + 1/2) \lambda_{\sigma}}{\pi \lambda_{2n}(\zeta)(\lambda_{2n}(\zeta) - \lambda_{\sigma})} \nonumber \\ 
& \times \frac{\Gamma(n + \zeta + 1/2)\Gamma(n + (\zeta + 1)/2)}{\Gamma(\zeta + 1)\Gamma(n + 1)\Gamma(n + \zeta/2 + 1)}, \label{eq:d_2n}
\end{align}
where $\Gamma$ is the gamma function.
By evaluating Eq.~(\ref{eq:el_cond_integral}), we obtain a formula for the electrical conductivity in a rapidly converging series:

\begin{widetext}
\begin{align}
\sigma_{xx} + i\sigma_{yx} = \frac{e^2n\tau_{\text{e-e}}}{m}\left[ \frac{1}{\pi^2 \zeta} \psi^{(1)}\left(\frac{1 + \zeta}{2}\right) + \frac{1}{4\pi} \sum_{n = 0}^{\infty} \frac{ \Gamma(\zeta + 1)\Gamma(n+1/2)\Gamma(n + (\zeta + 1)/2)}{\Gamma(n + \zeta/2 + 1)}d_{2n} \right], \label{eq:single_el_cond_ser}
\end{align}
\end{widetext}
where $\psi^{(1)}$ is the trigamma function.

In the limit of $1/\tau_{\text{e-e}} \ll 1/\tau_{\text{imp}} - i\omega_{\text{c}}$ or $1/\zeta \to 0$, we find
\begin{align}
\lim_{1/\zeta \to 0} (\sigma_{xx} + i\sigma_{yx})=& \lim_{1/\zeta \to 0} \frac{e^2n\tau_{\text{e-e}}}{m} \cdot \frac{1}{\pi^2 \zeta} \psi^{(1)}\left(\frac{1 + \zeta}{2}\right) \nonumber \\
=& \frac{e^2n}{m} \left( \frac{1}{\tau_{\text{imp}}} - i\omega_{\text{c}}\right)^{-1}, \label{eq:single_el_Drude_imp}
\end{align}
while other terms vanish since the first term in Eq.~(\ref{eq:single_el_expansion_ser}) gives the leading contribution in $1/\zeta \to 0$. In this case, we recover the usual Drude result by the impurity scattering and a magnetic field.

Let us turn to the thermal conductivity $(X = \kappa)$. As in the case of electrical conductivity, we expand the energy dependence of the distribution function as follows:
\begin{align}
\varphi_{\kappa}(x) = \frac{F_{\kappa}(x)}{x^2 + \zeta^2 \pi^2} + \sum_{n = 0}^{\infty} d_{2n+1} \varphi_{2n + 1;\zeta}(x), \label{eq:single_th_expansion_ser}
\end{align}
where $\varphi_{2n + 1;\zeta}(x)$ is an odd function. Then, we find $d_{2n + 1}$ as 

\begin{align}
d_{2n + 1} =& -\frac{2i(2n + \zeta+ 3/2) \lambda_{\kappa}}{\lambda_{2n + 1}(\zeta)(\lambda_{2n + 1}(\zeta) - \lambda_{\kappa}) } \nonumber \\ 
& \times \frac{\Gamma(n + \zeta+ 3/2)\Gamma(n + (\zeta+ 1)/2)}{\Gamma(n + \zeta/2 + 2)}. \label{eq:d_2n+1}
\end{align}
Finally, the thermal conductivity is given by 
\begin{widetext}
\begin{align}
\kappa_{xx} + i\kappa_{yx} = \frac{k_B^2Tn\tau_{\text{e-e}}}{m}\left[ 2 - \zeta \psi^{(1)}\left( \frac{1 + \zeta}{2} \right) - \frac{1}{4i} \sum_{n = 0}^{\infty} \frac{ \Gamma(\zeta + 1) \Gamma(n+3/2)\Gamma(n + (\zeta + 1)/2)}{\Gamma(n + \zeta + 1) \Gamma(n + \zeta/2 + 2)} d_{2n + 1}  \right]. \label{eq:single_th_cond_ser}
\end{align}
\end{widetext}

Similarly to the electrical conductivity, in the limit of $1/\tau_{\text{e-e}} \ll 1/\tau_{\text{imp}} - i\omega_{\text{c}}$, we recover the usual Drude conductivity from the first and second terms in Eq.~(\ref{eq:single_th_cond_ser}) as
\begin{align}
\lim_{1/\zeta \to 0} (\kappa_{xx} + i\kappa_{yx}) =& \lim_{1/\zeta \to 0} \frac{k_B^2Tn\tau_{\text{e-e}}}{m}\left[ 2 - \zeta \psi^{(1)}\left( \frac{1 + \zeta}{2} \right) \right] \nonumber \\
=& \frac{\pi^2k_B^2Tn}{3m} \left( \frac{1}{\tau_{\text{imp}}} - i\omega_{\text{c}}\right)^{-1}, \label{eq:single_th_Drude_imp} 
\end{align}
while other terms vanish again.

\subsection{Cases of $\lambda_{\sigma} = 1$ and $\lambda_{\kappa} = 3$}
If we consider a truly single carrier system rather than the effective single band system of Baber scattering, $\lambda_{\sigma} = 1$ is satisfied due to the momentum conservation. In this case, we can directly confirm that the solution of Eq.~(\ref{eq:single_int_eq}) is given by 
\begin{align}
\varphi_{\sigma}(x) = \frac{1}{\tau_{\text{e-e}}} \left(\frac{1}{\tau_{\text{imp}}} - i\omega_{\text{c}} \right)^{-1} \frac{1}{\cosh(x/2)} \propto \varphi_{n=0,\zeta = 1}(x)\label{eq:imp_el_dist_function}.
\end{align}
This leads to the usual Drude formula without the electron-electron scatterings,
\begin{align}
\sigma_{xx} + i\sigma_{yx} = \frac{e^2n}{m} \left(\frac{1}{\tau_{\text{imp}}} - i\omega_{\text{c}} \right)^{-1}, \label{eq:single_el_inter_absent}
\end{align}
which means that the electron-electron scatterings do not affect the electrical conductivity when $\lambda_{\sigma} = 1$ \cite{Smith1969, Ah-Sam1971, Lee2020}.

Let us consider the thermal transport for the case of $\lambda_{\kappa} = 3$, which is a fairly special case realized in a system without the interband scattering and with the intraband scattering only having value at $\theta = 0$. In this case, we find that the solution is given by
\begin{align}
\varphi_{\kappa}(x) = \frac{1}{\tau_{\text{e-e}}} \left(\frac{1}{\tau_{\text{imp}}} - i\omega_{\text{c}} \right)^{-1} \frac{x}{\cosh(x/2)} \propto \varphi_{n=1,\zeta = 1}(x),
\end{align} 
and the thermal conductivity is given by
\begin{align}
\kappa_{xx} + i\kappa_{yx} = \frac{\pi^2 k_B^2Tn}{3m} \left(\frac{1}{\tau_{\text{imp}}} - i\omega_{\text{c}} \right)^{-1}. \label{eq:single_th_inter_absent}
\end{align}
This means that the electron-electron scatterings do not affect the thermal current when $\lambda_{\kappa} = 3$ as in the case of the electrical conductivity with $\lambda_{\sigma} = 1$ \cite{Sykes1970}.

\subsection{Conductivities and thermal conductivities under the magnetic field}
\subsubsection{Magnetic field dependence of conductivities and resistivities}
\begin{figure}[tbp]
\begin{center}
\rotatebox{0}{\includegraphics[angle=0,width=1\linewidth]{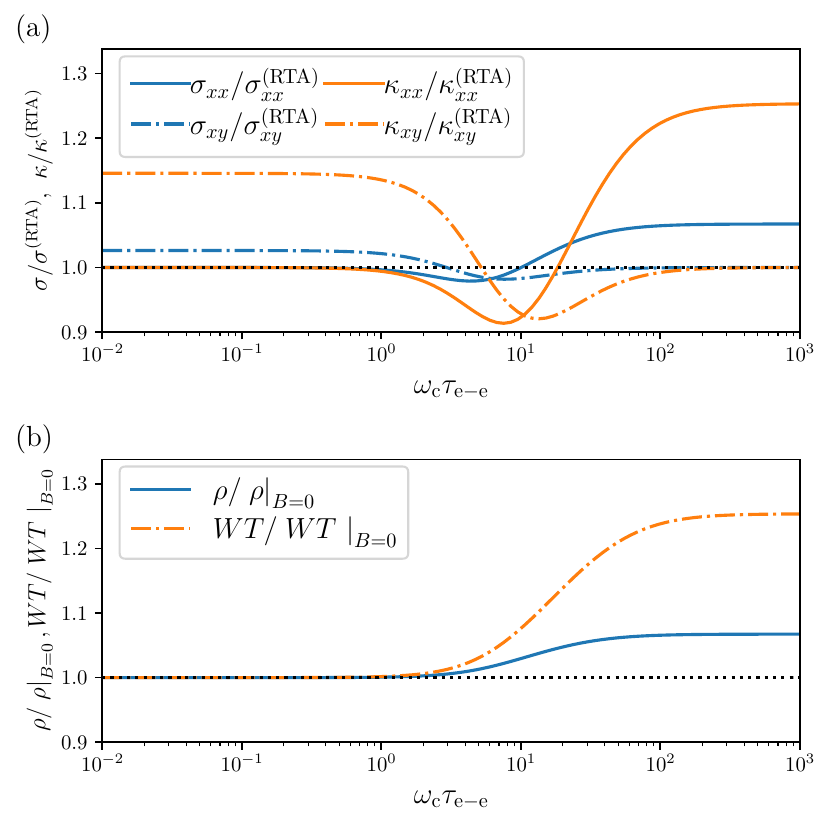}}
\caption{(a) A comparison of the magnetic field dependence of $\sigma_{xx}$, $\sigma_{xy}$, $\kappa_{xx}$, and $\kappa_{xy}$ for Baber scattering divided by the RTA results without the impurity scattering 
($\zeta^2 = 1-i(2/\pi^2)\omega_{\text{c}}\tau_{\text{e-e}}$) for $\lambda_{\sigma} = \lambda_{\kappa} = 1/3$. (b) The magnetic field dependence of the electrical resistivity $\rho$ and thermal resistivity $WT$ for the same parameters as in (a).}
\label{Fig:Baber_mag_dep}
\end{center}
\end{figure} 

As anticipated, the magnetic field dependence is close to the usual RTA result. Figure~\ref{Fig:Baber_mag_dep}(a) shows the magnetic field dependence of $\sigma_{ij}$ and $\kappa_{ij}$ without impurity scattering ($\zeta^2 = 1-i(2/\pi^2)\omega_{\text{c}}\tau_{\text{e-e}}$) divided by the results of the RTA as functions of $\omega_{\text{c}}\tau_{\text{e-e}}$. We do not specify the temperature since the conductivities divided by the result of the RTA become universal curves of $\omega_{\text{c}}\tau_{\text{e-e}}$ since the parameter $\zeta$ is a function of $\omega_{\text{c}}\tau_{\text{e-e}}$. We note that we show the result for a wider range of $\omega_{\text{c}}\tau_{\text{e-e}}$ but the validity of the model is in general limited to the range where Landau quantization is not significant. 
The conductivities in the RTA take the Drude formula given by 
\begin{align} 
\sigma^{(\text{RTA})}_{xx} =& \frac{e^2 n\tau_{\text{tr},\sigma}}{m} \cdot \frac{1}{1 + (\omega_{\text{c}}\tau_{\text{tr},\sigma})^2}, \label{eq:single_el_RTA_xx}\\
\sigma^{(\text{RTA})}_{yx} =& \frac{e^2 n\tau_{\text{tr},\sigma}}{m} \cdot \frac{\omega_{\text{c}}\tau_{\text{tr},\sigma}}{1 + (\omega_{\text{c}}\tau_{\text{tr},\sigma})^2}, \label{eq:single_el_RTA_xy}\\
\kappa^{(\text{RTA})}_{xx} =& \frac{\pi^2 k_B^2 Tn\tau_{\text{tr},\kappa}}{3m} \cdot \frac{1}{1 + (\omega_{\text{c}}\tau_{\text{tr},\kappa})^2}, \label{eq:single_th_RTA_xx}\\
\kappa^{(\text{RTA})}_{yx} =& \frac{\pi^2 k_B^2 Tn \tau_{\text{tr},\kappa}}{3m} \cdot \frac{\omega_{\text{c}}\tau_{\text{tr},\kappa}}{1 + (\omega_{\text{c}}\tau_{\text{tr},\kappa})^2}, \label{eq:single_th_RTA_xy}
\end{align}
where $\tau_{\text{tr},\sigma}$ and $\tau_{\text{tr},\kappa}$ are typical relaxation times for electrical and thermal transport \cite{Lucas2018, Lee2020} (see also Appendix \ref{App:RTA}). We choose these transport relaxation times and those for the RTA $\tau_{\text{tr},\sigma}$ and $\tau_{\text{tr},\kappa}$ at $\zeta = 1$ so as to reproduce the electrical and thermal conductivities in the absence of a magnetic field, i.e., the relaxation times are given by
\begin{align}
\tau_{\text{tr},\sigma} =& \left( \frac{e^2n}{m}\right)^{-1} \left. \sigma_{xx} \right|_{B = 0,~\zeta = 1},\\
\tau_{\text{tr},\kappa} =& \left( \frac{\pi^2 k_B^2Tn}{3m}\right)^{-1} \left. \kappa_{xx} \right|_{B = 0,~\zeta = 1},
\end{align}
where the right-hand sides are calculated by Eqs.~(\ref{eq:single_el_cond_ser}) and (\ref{eq:single_th_cond_ser}). Note that the impurity scattering is absent $(\zeta = 1)$. This condition corresponds to the case of $\tau_{\text{e-e}} \ll \tau_{\text{imp}}$. On the evaluation, we numerically evaluate Eqs.~(\ref{eq:single_el_cond_ser}) and (\ref{eq:single_th_cond_ser}), which give rapid convergences (see Appendix~\ref{App:solution_properties}). We set $\lambda_{\sigma} = \lambda_{\kappa} = 1/3$ which is realized when the short-ranged (Hubbard) interband scattering dominates \cite{Bennett1969, Ah-Sam1971} (see also Appendix~\ref{App:derivation_integral_eq}). 

Figure~\ref{Fig:Baber_mag_dep}(a) demonstrates that $\sigma_{ij}/\sigma^{(\text{RTA})}_{ij}$ and $\kappa_{ij}/\kappa^{(\text{RTA})}_{ij}$ are of order unity, but there are deviations from 1 as a function of $B$. 

In Fig.~\ref{Fig:Baber_mag_dep}(b), we plot the electrical resistivity $\rho$ and the thermal resistivity $WT$. Resistivities are normalized by the value at the zero-magnetic field. Other parameters are the same as in Fig.~\ref{Fig:Baber_mag_dep}(a). We find that both $\rho$ and $WT$ show small but non-zero magnetoresistance. This is in contrast to the RTA result independent of the magnetic field, where the resistivities are calculated as
\begin{align}
\rho^{(\text{RTA})} = \frac{m}{e^2 n\tau_{\text{tr},\sigma}},~WT^{(\text{RTA})} = \frac{m}{e^2 n\tau_{\text{tr},\kappa}}.
\end{align}
We can show that the resistivities for a large magnetic field are given by (see Appendix~\ref{App:solution_properties})
\begin{align}
\left. \rho \right|_{B \to \infty} =& \frac{m}{e^2n\tau_{\text{e-e}}} \cdot \frac{2\pi^2(1 - \lambda_{\sigma})}{3}, \label{eq:Baber_magnetoresistance_rho} \\
\left. WT \right|_{B \to \infty} =& \frac{m}{e^2n\tau_{\text{e-e}}} \cdot \frac{2\pi^2(3 - \lambda_{\kappa})}{5}. \label{eq:Baber_magnetoresistance_WT} 
\end{align}
We can also confirm that these resistivities satisfy
\begin{align}
\left. \rho \right|_{B \to \infty} \geq \left. \rho \right|_{B = 0},~\left. WT \right|_{B \to \infty} \geq \left. WT \right|_{B = 0}.
\end{align}

These departures from the RTA demonstrated in Fig.~\ref{Fig:Baber_mag_dep}(a) and (b) are attributed to the energy dependence of distribution functions [Eqs.~(\ref{eq:single_el_expansion_ser}) and (\ref{eq:single_th_expansion_ser})] originating from the inelastic scatterings. The RTA cannot describe longitudinal and transverse transport exactly at the same time because inelastic scatterings do not allow us to determine one unique characteristic relaxation time, with which all transport coefficients in an isotropic system can be correctly expressed \cite{Smith_Jensen1989}. In fact, in a weak magnetic field $\omega_{\text{c}} \tau_{\text{e-e}} \ll 1$, the transverse electrical and thermal conductivities do not match the results by the RTA, even though the relaxation times are chosen to be exact at the zero-magnetic field.

\subsubsection{Hall and thermal Hall effects}
\begin{figure}[tbp]
\begin{center}
\rotatebox{0}{\includegraphics[angle=0,width=1\linewidth]{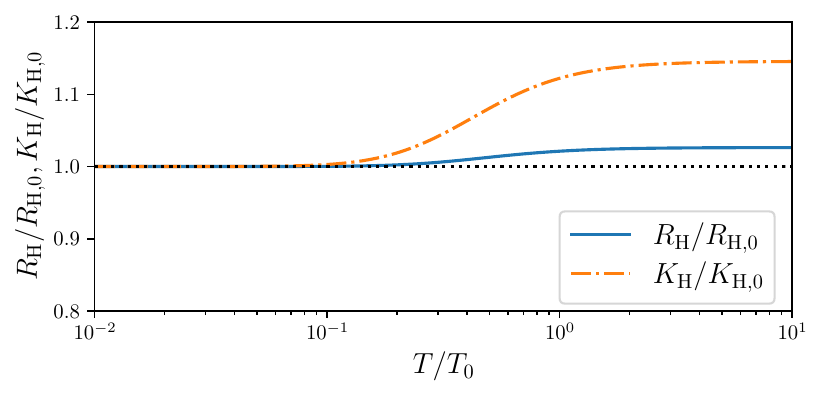}}
\caption{The temperature dependence of normalized $R_{\text{H}}$ and $K_{\text{H}}$ for Baber scattering. We set $\lambda_{\sigma} = \lambda_{\kappa} = 1/3$ and $\omega_{\text{c}}\tau_{\text{imp}} = 10^{-3}$.}
\label{Fig:Baber_Hall}
\end{center}
\end{figure}

Before we study the Lorenz ratios, we focus on the Hall and thermal Hall effects considering $R_{\text{H}}$ and $K_{\text{H}}$. Figure \ref{Fig:Baber_Hall} shows the temperature dependence of $R_{\text{H}}$ and $K_{\text{H}}$ normalized by constants $R_{\text{H},0}$ and $K_{\text{H},0}$:
\begin{align}
R_{\text{H},0} = \frac{1}{en},~K_{\text{H},0} = \frac{1}{enL_{0}}.
\end{align}
$R_{\text{H},0}$ and $K_{\text{H},0}$ are results expected by the RTA [Eqs.~(\ref{eq:single_el_RTA_xx})-(\ref{eq:single_th_RTA_xy})].
We set $\lambda_{\sigma} = \lambda_{\kappa} = 1/3$ as in Fig.~\ref{Fig:Baber_mag_dep}. We now set a finite impurity scattering rate and a weak magnetic field $\omega_{\text{c}}\tau_{\text{imp}} = 10^{-3}$. The temperature is normalized by $T_0$ where $\tau_{\text{imp}} = \tau_{\text{e-e}}$. Note that $1/\tau_{\text{e-e}} \propto T^{2}$ and $1/\tau_{\text{imp}} \propto T^{0}$.

\begin{figure}[tbp]
\begin{center}
\rotatebox{0}{\includegraphics[angle=0,width=1\linewidth]{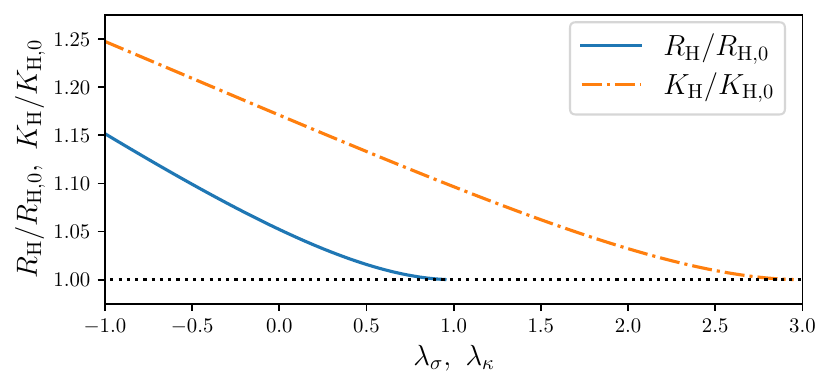}}
\caption{$\lambda_{X}$ dependence of $R_{\text{H}}$ and $K_{\text{H}}$ in the absence of the impurity scattering and a weak field limit.}
\label{Fig:Baber_Hall_lambda}
\end{center}
\end{figure}

The deviations from $R_{\text{H},0}$ and $K_{\text{H},0}$ are caused by the inelastic scattering, which prohibits us from finding unique relaxation times, as discussed in the previous section. The temperature dependence of $R_{\text{H}}$ is weak. This is consistent with a result obtained by a matrix formulation of the Boltzmann equation \cite{Schulz1995}. The temperature dependence of $K_{\text{H}}$ is not as small as that of $R_{\text{H}}$ but not so large. These departures lead to $\widetilde{L}_{\text{H}} = R_{\text{H}}/K_{\text{H}} \neq L_{0}$. The temperature dependence is a shift from $R_{\text{H},0}$ and $K_{\text{H},0}$ in the impurity scattering dominating regime ($T \ll T_{0}$, or $\tau_{\text{e-e}} \gg \tau_{\text{imp}}$) to the limiting values in the electron-electron scatterings dominating regime ($T \gg T_{0}$, or $\tau_{\text{e-e}} \ll \tau_{\text{imp}}$). The limiting values of $R_{\text{H}}$ and $K_{\text{H}}$ in $T/T_{0} = \sqrt{\tau_{\text{imp}}/\tau_{\text{e-e}}} \to \infty$ are parametrized by $\lambda_{\sigma}$ and $\lambda_{\kappa}$, respectively. Figure~\ref{Fig:Baber_Hall_lambda} shows $\lambda_{X}$ dependence of normalized $R_{\text{H}}$ and $K_{\text{H}}$ in the absence of the impurity scattering (or in the $T/T_{0} \to \infty$ limit) and weak magnetic field limit. As discussed before, the values are limited to $-1 \leq \lambda_{\sigma} < 1$ and $-1 \leq \lambda_{\kappa} < 3$, respectively. We see that both $R_{\text{H}}/R_{\text{H},0}$ and $K_{\text{H}}/K_{\text{H},0}$ monotonically decrease and approach 1 for $\lambda_{\sigma} \to 1$ and $\lambda_{\kappa} \to 3$ where the electron-electron scatterings do not affect currents and conductivities are given by Eqs.~(\ref{eq:single_el_inter_absent}) and (\ref{eq:single_th_inter_absent}).

\subsubsection{Lorenz ratio and Hall Lorenz ratio}

\begin{figure}[tbp]
\begin{center}
\rotatebox{0}{\includegraphics[angle=0,width=1\linewidth]{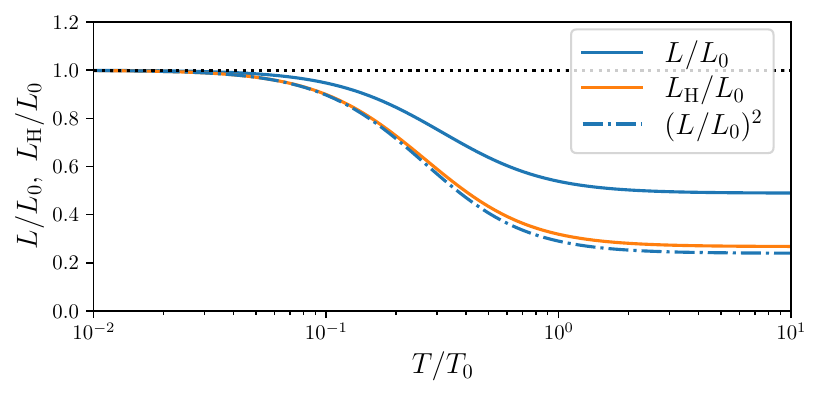}}
\caption{Temperature dependence of normalized Lorenz ratio and Hall Lorenz ratio in the case of Baber scattering for $\lambda_{\sigma} = \lambda_{\kappa} = 1/3$. We also plot $(L_{}/L_0)^2$ (blue dash-dotted lines). The black dotted lines indicate the WF law.}
\label{Fig:Baber_Lorenz}
\end{center}
\end{figure} 

The temperature dependences of the normalized Lorenz ratio $L_{}/L_0$ and Hall Lorenz ratio $L_{\text{H}}/L_0$ are shown in Fig.~\ref{Fig:Baber_Lorenz} with $\lambda_{\sigma} = \lambda_{\kappa} = 1/3$ the same as in Fig.~\ref{Fig:Baber_Hall}. We set $\omega_{\text{c}}\tau_{\text{imp}} = 10^{-3}$. In Fig.~\ref{Fig:Baber_Lorenz}, the Lorenz ratio and the Hall Lorenz ratio reach some non-zero value for $T/T_{0} \gg 1$ with $\rho \propto T^2$ and $WT \propto T^2$. We also plot $(L_{}/L_0)^2$ (blue dash-dotted lines).

$L_{\text{H}}/L_{0}$ is approximately close to  $(L_{}/L_0)^2$ as expected in Eq.~(\ref{eq:Hall_Lorenz_weak_mag}) whereas a small deviation is found for $T/T_{0} \gtrsim 0.1$, a relatively high-temperature regime because $\widetilde{L}_{\text{H}} = R_{\text{H}}/K_{\text{H}} = L_{0}$ does not hold in general when the electron-electron scatterings exist.

We also find that, in the regime $T/T_{0} \ll 1$, $L_{\text{H}}/L_0 \simeq (L_{}/L_0)^2$ is asymptotically satisfied. Actually, by treating the electron-electron scatterings as a perturbation, we can show (see Appendix~\ref{App:solution_properties}),
\begin{align}
\frac{L_{}}{L_0} &\simeq 1 - a\frac{\tau_{\text{imp}}}{\tau_{\text{e-e}}}, \label{eq:single_Lorenz_asym} \\
\frac{L_{\text{H}}}{L_0} &\simeq 1 - 2a \frac{\tau_{\text{imp}}}{\tau_{\text{e-e}}} \simeq \left( \frac{L_{}}{L_0} \right)^2, \label{eq:single_Hall_Lorenz_asym}
\end{align}
where a constant $a$ is given by 
\begin{align}
a = - \frac{2\pi^2}{3} (1 - \lambda_{\sigma}) + \frac{2\pi^2}{5} (3 - \lambda_{\kappa} ). \label{eq:single_Lorenz_asym_const}
\end{align}

\section{Semimetals} \label{Sec:Semimetal}
Next, we consider the transport properties in the two-band semimetals with intra- and interband scatterings. In the two-band system, we cannot expect general analytical solutions in the presence of the impurity scattering or the magnetic field except when $\text{diag}(\zeta_1^2, \zeta_2^2)$ and $\hat{\lambda}_{X}$ are simultaneously diagonalizable \cite{Maldague1979}. Therefore, we rely on the numerical calculation by the variational method \cite{Ziman2001}. We expand the energy-dependent parts of the distribution functions using a finite set of trial functions. We use the eigenfunctions of Eq.~(\ref{eq:single_int_eq}) at $\zeta = 1$, which diagonalize the part of electron-electron scatterings, as trial functions. The energy-dependent parts of the distribution functions are expanded as
\begin{align}
\varphi^{(l)}_{\sigma}(x) =& \sum_{n = 0}^{N - 1} c^{(l)}_{2n} \varphi_{2n;\zeta = 1}(x), \label{eq:semimetal_el_expansion} \\
\varphi^{(l)}_{\kappa}(x) =& \sum_{n = 0}^{N - 1} c^{(l)}_{2n + 1} \varphi_{2n + 1;\zeta = 1}(x). \label{eq:semimetal_th_expansion}
\end{align}
To obtain $c^{(l)}_{2n}$ and $c^{(l)}_{2n + 1}$, we numerically solve linear equations mapped from Eqs.~(\ref{eq:int_eq_electrons}) and (\ref{eq:int_eq_holes}) (see Appendix~\ref{App:solution_properties}). Then, we calculate the electrical and thermal conductivities using Eqs.~(\ref{eq:el_cond_integral}) and (\ref{eq:th_cond_integral}). In this paper, we set $N = 150$, which gives sufficient convergence and numerically exact solutions (see Appendix~\ref{App:solution_properties}). 

For interpretations of results, we use the expressions by the RTA. Although the RTA cannot describe transport coefficients correctly, as we have discussed for Baber scattering, the RTA, taking into account the momentum conservation, gives qualitatively good interpretations. 

For the intra- and interband electron-electron scatterings, we use the screened Coulomb interaction \cite{Li2018, Lee2021, Takahashi2023},
\begin{align}
W^{(ll')} = \frac{2\pi}{\hbar} \left( \frac{e^2}{\varepsilon_{0}} \cdot \frac{1}{q^2 + \alpha^{2}} \right)^2, \label{eq:screend_Coulomb}
\end{align}
where $q = |\bm{k} - \bm{k}_3| = k_{\text{F},l} \sin \theta \sin (\varphi/2)/R^{(ll')}(\theta) $ is the momentum transfer, $\varepsilon_{0}$ is the dielectric constant, and $\alpha$ is the inverse of the screening length. The calculations of $\lambda^{(l)}_{X}$ and $\beta^{(l)}_{X}$ are found in Appendix A. 
We neglect the exchange process for simplicity. $\alpha$ and $\varepsilon_{0}$ are related as $\alpha^2 = e^2(m_1k_{\text{F},1} + m_
2k_{\text{F},2})/\pi^2\hbar^2\varepsilon_{0}$ if the screeing is determined by the Thomas-Fermi screeing. Note that $\beta^{(l)}_{\kappa} = 0$ for the potential Eq.~(\ref{eq:screend_Coulomb}).

\subsection{Momentum conservation in the electron-electron scattering}
Before presenting the results, we consider the effect of momentum conservation 
in the electron-electron scattering. 
This effect is particularly pronounced for electrical transport at relatively high temperatures where the single-particle damping rate due to impurity scattering is much smaller than that due to electron-electron scattering, i.e., $1/\tau_{\text{imp}}^{(l)} \ll 1/\tau_{\text{e-e}}^{(l)}$. 
(Note that $1/\tau_{\text{e-e}}^{(l)} \propto T^2$ and that an even higher temperature range, where $1/\tau_{\text{e-e}}^{(l)}$ deviates from the square law of temperature, is beyond the scope of this paper).
Here, we consider the more general case of semimetals with any number of nonspherical Fermi surfaces and momentum-dependent impurity scattering, and derive the electrical resistivity and the Hall coefficient 
at such temperatures.

The linearized Boltzmann equation, Eq.~(\ref{eq:Boltzmann_eq}), can be written including the weak time dependence of the distribution function as
\begin{align}
\frac{d \delta f^{(l)} (\bm{k})}{dt} 
& = 
e \bm{v}^{(l)}_{\bm{k}} \cdot \bm{E} 
\left(- \frac{\partial f_0(\varepsilon_{l,\bm{k}})}{\partial \varepsilon_{l,\bm{k}}} \right) 
-  M^{(l)}[\Phi] 
\nonumber\\
&\quad 
- I^{(l)}_{\text{imp}}[\Phi] 
- \sum_{l'}I^{(ll')}_{\text{e-e}}[\Phi]. 
\label{eq:Boltzmann_eq2} 
\end{align}
The second line in this equation corresponds to the so-called collision integral.
Since $\bm{v}^{(l)}_{\bm{k}} = \hbar^{-1}\nabla_{\bm{k}} \varepsilon_{l,\bm{k}}$ for any dispersion
$\varepsilon_{l,\bm{k}}$, the total charge $Q$, and the electric current $\bm{j}$ of the entire system, which includes both the electron and hole Fermi surfaces, satisfy
\begin{align}
Q \bm{E}&= \frac{2}{V} \sum_{\bm{k},l} \hbar \bm{k}
(e \bm{v}^{(l)}_{\bm{k}} \cdot \bm{E}) \left(- \frac{\partial f_0(\varepsilon_{l,\bm{k}})}{\partial \varepsilon_{l,\bm{k}}} \right) ,
\label{eq:overlap}
\\
\bm{j} \times \bm{B} &= - \frac{2}{V} \sum_{\bm{k},l} \hbar \bm{k} M^{(l)}[\Phi].
\end{align}
By multiplying Eq.~(\ref{eq:Boltzmann_eq2}) by $\hbar \bm{k}$ and summing over $\bm{k}$ and $l$, 
we obtain the equation of motion for the total momentum $\bm{P}$ induced by the electric field $\bm{E}$,
\begin{align}
\frac{d\bm{P}}{dt} = Q\bm{E} + \bm{j} \times \bm{B} + \bm{F}_{\text{imp}} +\bm{F}_{\text{e-e}},
\label{eq:forcerelation}
\end{align}
where $\bm{F}_{\text{imp}}$ and $\bm{F}_{\text{e-e}}$ represent the damping forces caused by the collision integrals of the impurity scattering and  of the electron-electron scattering, respectively, and they 
are defined by
\begin{align}
\bm{F}_{\text{imp}} &= - \frac{2}{V}\sum_{\bm{k},l} 
\hbar\bm{k} I^{(l)}_{\text{imp}}[\Phi] ,
\label{eq:forceimp}
\\
\bm{F}_{\text{e-e}} &= - \frac{2}{V} \sum_{\bm{k},l} \hbar \bm{k} \sum_{l'}  I^{(ll')}_{\text{e-e}}[\Phi] .
\end{align}
When $Q \neq 0$, as shown below, 
the steady state condition $d\bm{P}/dt = \bm{0}$ in Eq.~(\ref{eq:forcerelation}) relates the electric field $\bm{E}$ to the electric current $\bm{j}$ in the form
\begin{align}
\bm{E} = \rho \bm{j} + R_{\text{H}} \bm{B} \times \bm{j},
\label{eq:rhoRH}
\end{align}
where $\rho$ and $R_{\text{H}}$ are the electrical resistivity and the Hall coefficient, respectively.
Here we consider the case where the external electric field $\bm{E}_\text{ext}$, 
the Hall electric field $\bm{E}_{\text{H}}$, and the magnetic field $\bm{B}$ are all perpendicular to each other. We can write the electric field as $\bm{E} = \bm{E}_\text{ext} + \bm{E}_{\text{H}}$, and require that $\bm{E}_\text{ext}$ should be parallel to $\bm{j}$ and 
that $\bm{E}_{\text{H}}$ should be perpendicular to $\bm{j}$. 
The Hall electric field is then given by $\bm{E}_{\text{H}} = (R_{\text{H}}/\rho) \bm{B} \times \bm{E}_\text{ext}$, 
which is equivalent to that obtained from the boundary condition that $\bm{j}$ in the direction perpendicular to $\bm{E}_\text{ext}$ should be zero.
Equation~(\ref{eq:forcerelation}) shows that if the damping force $\bm{F}_{\text{imp}} +\bm{F}_{\text{e-e}}$ is parallel to $\bm{j}$, $R_{\text{H}}$ is given simply by the inverse of $Q$.

The equation of motion, Eq.~(\ref{eq:forcerelation}), is a general consequence of the Boltzmann equation, 
which can also be applied to the presence of Umklapp scattering in the electron-electron scattering. 
In the case of semimetals, however, the Umklapp scattering is ineffective because of the small Fermi surfaces, 
and the momentum conservation in electron-electron scattering yields
\begin{align}
\bm{F}_{\text{e-e}}= \bm{0}.
\label{eq:damingforcee-e}
\end{align}
In Sec.~\ref{Sec:integral_eq}, we have seen that for the two-band system with spherical Fermi surfaces, the linearized Boltzmann equation can be understood in the matrix form, where $\hat{\lambda}_\sigma$ 
defined by Eq.~(\ref{eq:lambda_matrix}) has an eigenvalue of $1$ due to the momentum conservation. 
This can be generalized to the case of any number of nonspherical Fermi surfaces. 
If we consider the collision integral in Eq.~(\ref{eq:Boltzmann_eq2}) 
as a matrix operation acting on $\Phi$, the matrix, 
which is often called the memory matrix, has a zero eigenvalue due to the momentum conservation 
in the absence of impurity scattering 
\cite{Maebashi1997}, 
and its eigenfunction is given by
\begin{align}
\Phi^{(l)}(\hat{\bm{k}},\xi_{l,\bm{k}}) 
= \frac{\beta\hbar}{Q} \bm{k} \cdot \bm{j}.
\label{eq:eigenfunction}
\end{align}
Note that this eigenfunction satisfies $\sum_{l'}I^{(ll')}_{\text{e-e}}[\Phi] =0$ and Eq.~(\ref{eq:def_el_current}). 
Then, Eq.~(\ref{eq:damingforcee-e}) follows from the zero eigenvalue of the memory matrix, 
and Eq.~(\ref{eq:eigenfunction}) is the solution of the Boltzmann equation 
in the limit of weak impurity scattering.
By substituting this solution into Eq.~(\ref{eq:forceimp}), we find that the damping force due to impurity scattering is parallel to the electric current and is written in terms of the mean mobility $\bar{\mu}_{\text{imp}}$ due to impurity scattering, 
\begin{align}
\bm{F}_{\text{imp}} 
= - \frac{|Q|}{Q}
\frac{\bm{j}}{\bar{\mu}_{\text{imp}}}.
\label{eq:forceimp2}
\end{align}
For a cubic system, the inverse of $\bar{\mu}_{\text{imp}}$ is given by
\begin{align}
\frac{1}{\bar{\mu}_{\text{imp}}} = \frac{2}{3|Q|V}\sum_{\bm{k},l} 
\frac{\hbar^2\bm{k}^2}{\tau_{\text{imp}}^{(l)} (\bm{k})}
\left(- \frac{\partial f_0(\varepsilon_{l,\bm{k}})}{\partial \varepsilon_{l,\bm{k}}} \right).
\label{}
\end{align}
With Eqs.~(\ref{eq:damingforcee-e}) and (\ref{eq:forceimp2}),  the steady state condition of $d\bm{P}/dt= 0$ in Eq.~(\ref{eq:forcerelation}) yields
\begin{align}
\bm{E} = \frac{\bm{j}}{
|Q| \bar{\mu}_{\text{imp}}}
+ \frac{\bm{B} \times \bm{j}}{Q}.
\label{eq:Eforuncompensated}
\end{align}
Comparing this equation with Eq.~(\ref{eq:rhoRH}), we obtain the electrical resistivity and the Hall coefficient of uncompensated semimetals at high temperatures where the electron-electron scattering is dominant, as
\begin{align}
\rho &= 1/|Q|\bar{\mu}_{\text{imp}},
\label{eq:rhoQnonzero}
\\
R_{\text{H}}&= 1/Q.
\label{eq:RHQnonzero}
\end{align}

As clearly seen in Eq.~(\ref{eq:overlap}), the total charge $Q$ represents the overlap of momentum and electric current. 
Therefore, when $Q=0$, i.e., for compensated semimetals with equal numbers of electrons and holes, the total momentum does not contribute to the electric current, so that $\rho$ and $R_{\text{H}}$ cannot be obtained from Eq.~(\ref{eq:forcerelation}). 
In this case, we classify the Fermi surfaces into electron-like  ($l \in e$) and hole-like ($l \in h$), and derive two equations of motion: one for the total momentum $\bm{P}^{(e)}$ of the electrons and one for the total momentum $\bm{P}^{(h)}$ of the holes, 
\begin{align}
\frac{d\bm{P}^{(e)}}{dt} &= Q^{(e)}\bm{E} + \bm{j}^{(e)}\times \bm{B} + \bm{F}_{\text{imp}}^{(e)}+\bm{F}_{\text{e-e}}^{(e)},
\label{eq:EOMelectron}
\\
\frac{d\bm{P}^{(h)}}{dt} &= Q^{(h)}\bm{E} + \bm{j}^{(h)}\times \bm{B} + \bm{F}_{\text{imp}}^{(h)}+\bm{F}_{\text{e-e}}^{(h)},
\label{eq:EOMhole}
\end{align}
where $\bm{j}^{(e)}$ and $\bm{j}^{(h)}$ are the electric currents carried by electrons and holes, respectively, 
In a similar way to Eq.~(\ref{eq:eigenfunction}), we take the solution of the Boltzmann equation as
\begin{align}
\Phi^{(l)}(\hat{\bm{k}},\xi_{l,\bm{k}}) 
= \left\{
\begin{array}{rl}
\displaystyle{\frac{\beta\hbar}{Q^{(e)}} \bm{k} \cdot \bm{j}^{(e)}} 
& \quad\mbox{for $l \in e$}
\\
&
\\
\displaystyle{\frac{\beta\hbar}{Q^{(h)}} \bm{k} \cdot \bm{j}^{(h)} }
& \quad\mbox{for $l \in h$}
\end{array}
\right. .
\label{eq:eigenfunction2}
\end{align}
This solution gives the damping forces due to impurity scattering in the same form 
as Eq.~(\ref{eq:forceimp2}),
\begin{align}
\bm{F}_{\text{imp}}^{(e)} &
=- \frac{|Q^{(e)}|}{Q^{(e)}}
\frac{\bm{j}^{(e)} }{\bar{\mu}^{(e)}_{\text{imp}}},
\\
\bm{F}_{\text{imp}}^{(h)} &=- \frac{|Q^{(h)}|}{Q^{(h)}}
\frac{\bm{j}^{(h)} }{\bar{\mu}^{(h)}_{\text{imp}}}.
\end{align}
Note that when $\bm{j}^{(e)}$ and $\bm{j}^{(h)}$ are not parallel to the total electric current, $\bm{j} = \bm{j}^{(e)} + \bm{j}^{(h)}$, $\bm{F}_{\text{imp}}^{(e)}$ and $\bm{F}_{\text{imp}}^{(h)}$ have the components perpendicular to $\bm{j}$.
By substituting Eq.~(\ref{eq:eigenfunction2}) into Eq.~(\ref{eq:el_el_scattering_original}), on the other hand, we find that the damping forces due to electron-electron scattering 
are always parallel to $\bm{j}$ and have no components perpendicular to it,
\begin{align}
\bm{F}_{\text{e-e}}^{(e)} &= - \frac{2}{V} \sum_{\bm{k}}\sum_{l \in e}   \hbar \bm{k}  
\sum_{l'} I^{(ll')}_{\text{e-e}}[\Phi] = \frac{1}{\bar{\mu}_{\text{e-h}}} \bm{j},
\label{eq:damp^e}
\\
\bm{F}_{\text{e-e}}^{(h)} &= - \frac{2}{V} \sum_{\bm{k}}\sum_{l \in h}   \hbar \bm{k}  
\sum_{l'} I^{(ll')}_{\text{e-e}}[\Phi] = - \frac{1}{\bar{\mu}_{\text{e-h}}} \bm{j},
\label{eq:damp^h}
\end{align}
where $1/\bar{\mu}_{\text{e-h}}$ is proportional to $T^2$ and is produced by electron-hole scattering. These equations show that $\bm{F}_{\text{e-e}} = \bm{F}_{\text{e-e}}^{(e)} + \bm{F}_{\text{e-e}}^{(h)} =\bm{0}$ 
in coincidence with Eq.~(\ref{eq:damingforcee-e}).

Here, we introduce the drift velocity of electrons $\bm{v}_{\text{d}}^{(e)}$ and 
the drift velocity of holes $\bm{v}_{\text{d}}^{(h)}$ by 
\begin{align}
\bm{v}_{\text{d}}^{(e)} &= \bm{j}^{(e)}/Q^{(e)},
\\
\bm{v}_{\text{d}}^{(h)} &= \bm{j}^{(h)}/Q^{(h)}.
\end{align}
From Eqs.~(\ref{eq:EOMelectron}) and (\ref{eq:EOMhole}), 
the equation of motion for the total momentum, $\bm{P} = \bm{P}^{(e)} + \bm{P}^{(h)}$,
is given by
\begin{align}
\frac{d\bm{P}}{dt} = Q \bm{E} + \bm{j} \times \bm{B} 
- \frac{|Q^{(e)}|}{\bar{\mu}^{(e)}_{\text{imp}}} \bm{v}_{\text{d}}^{(e)}
- \frac{|Q^{(h)}|}{\bar{\mu}^{(h)}_{\text{imp}}} \bm{v}_{\text{d}}^{(h)},
\label{eq:forceQ=0}
\end{align}
where $Q=Q^{(e)} + Q^{(h)}$ and $\bm{j} = Q^{(e)} \bm{v}_{\text{d}}^{(e)} + Q^{(h)} \bm{v}_{\text{d}}^{(h)}$. 
This equation is also applied to the case of $Q \neq 0$, 
comparing Eqs.~(\ref{eq:eigenfunction}) and~(\ref{eq:eigenfunction2}), 
it follows that the drift velocities of electrons and holes are equal to each other and 
are parallel to the total electric current, 
\begin{align}
\bm{v}_{\text{d}}^{(e)} = \bm{v}_{\text{d}}^{(h)} = \bm{j}/Q.
\label{eq:driftQnonzero}
\end{align}
Hence, the momentum conservation generates the unusual behavior of 
minority carriers: for $Q<0$, the hole flows in the opposite direction to the external electric field, 
and for $Q>0$, the electron flows in the same direction as the external electric field.
Substituting Eq.~(\ref{eq:driftQnonzero}) for Eq.~(\ref{eq:forceQ=0}) again yields Eq.~(\ref{eq:Eforuncompensated}) for the uncompensated semimetals.

For the compensated case of $Q=0$, on the other hand, 
we take $Q^{(e)} = - Q^{(h)} = en$, 
where $n$ is the electron or hole number density.
The steady state condition of $d\bm{P}/dt = \bm{0}$
leads to the fact that the electron and hole drift velocities are given by
\begin{align}
\bm{v}_{\text{d}}^{(e)}  
&= \frac{1}{|e|n} \cdot \frac{\bar{\mu}^{(e)}_{\text{imp}}}{\bar{\mu}^{(e)}_{\text{imp}}+\bar{\mu}^{(h)}_{\text{imp}}} 
\left( - \bm{j} + \bar{\mu}^{(h)}_{\text{imp}} \, \bm{j} \times \bm{B} \right),
\label{eq:vde}
\\
\bm{v}_{\text{d}}^{(h)} 
&= \frac{1}{|e|n} \cdot 
\frac{\bar{\mu}^{(h)}_{\text{imp}}}{\bar{\mu}^{(e)}_{\text{imp}}+\bar{\mu}^{(h)}_{\text{imp}}} 
\left( \bm{j} + \bar{\mu}^{(e)}_{\text{imp}}\, \bm{j} \times \bm{B} \right).
\label{eq:vdh}
\end{align}
Thus, the electrons and holes move in opposite directions, parallel to $\bm{j}$, with equal velocities perpendicular to it.
The electrical resistivity $\rho$ and the Hall coefficient $R_{\text{H}}$ for $Q = 0$ are then derived from the equation of motion for the relative momentum, 
$\Delta\bm{P} = \bm{P}^{(e)} - \bm{P}^{(h)}$, 
\begin{align}
\frac{d \Delta \bm{P}}{dt} &= 2 en  \bm{E} 
+ en \left(\bm{v}_{\text{d}}^{(e)} + \bm{v}_{\text{d}}^{(h)}\right) \times \bm{B} 
\nonumber\\
&\quad +
en \left(\frac{\bm{v}_{\text{d}}^{(e)}}{\bar{\mu}^{(e)}_{\text{imp}}} 
- \frac{\bm{v}_{\text{d}}^{(h)}}{\bar{\mu}^{(h)}_{\text{imp}}}\right)
+\frac{2}{\bar{\mu}_{\text{e-h}}} \bm{j} .
\label{eq:relativeQ=0}
\end{align}
By the steady state condition of $d\Delta\bm{P}/dt = \bm{0}$ 
with Eqs.~(\ref{eq:vde}) and (\ref{eq:vdh}), we obtain 
\begin{align}
\bm{E} &= \frac{1}{|e|n}
\left(\frac{1}{\bar{\mu}^{(e)}_{\text{imp}}+\bar{\mu}^{(h)}_{\text{imp}}}
+\frac{1}{\bar{\mu}_{\text{e-h}}}
\right)
\bm{j}
\nonumber \\
&\quad
+ \frac{1}{en}\cdot \frac{\bar{\mu}^{(e)}_{\text{imp}}-\bar{\mu}^{(h)}_{\text{imp}}}{\bar{\mu}^{(e)}_{\text{imp}}+\bar{\mu}^{(h)}_{\text{imp}}} \,\bm{B} \times \bm{j} 
\nonumber
\\
&\quad + \frac{1}{|e|n}\cdot
\frac{\bar{\mu}^{(e)}_{\text{imp}} \bar{\mu}^{(h)}_{\text{imp}}}{\bar{\mu}^{(e)}_{\text{imp}}+\bar{\mu}^{(h)}_{\text{imp}}}\,\bm{B} \times (\bm{j} \times \bm{B}) . 
\label{eq:Eforcompensated}
\end{align}
The last term, which is proportional to $\bm{B} \times (\bm{j} \times \bm{B})$, 
is of a type not found in Eq.~(\ref{eq:Eforuncompensated}). 
It represents the magnetoresistance effect, 
which occurs because the electron and hole drift velocities have terms proportional to $\bm{j} \times \bm{B}$. 
Comparing Eq.~(\ref{eq:Eforcompensated}) with Eq.~(\ref{eq:rhoRH}), 
we finally obtain
\begin{align}
\rho &= \frac{1}{|e|n} 
\left(\frac{1 + \bar{\mu}^{(e)}_{\text{imp}} \bar{\mu}^{(h)}_{\text{imp}} B^2}{\bar{\mu}^{(e)}_{\text{imp}}+\bar{\mu}^{(h)}_{\text{imp}}}
+\frac{1}{\bar{\mu}_{\text{e-h}}}
\right),
\label{eq:monotonic rho}
\\
R_{\text{H}}&= \frac{1}{en} \cdot \frac{\bar{\mu}^{(e)}_{\text{imp}}-\bar{\mu}^{(h)}_{\text{imp}}}{\bar{\mu}^{(e)}_{\text{imp}}+\bar{\mu}^{(h)}_{\text{imp}}} .
\label{eq:TindepRH}
\end{align}

\subsection{Resistivity and magnetoresistance}
First, we discuss the electric resistivity $\rho$ and thermal resistivity $WT$ of semimetals. In Fig.~\ref{Fig:semimetal_magnetoresistance}, we plot temperature dependences of the electrical resistivity $\rho$ and thermal resistivity $WT$ (a) of the compensated semimetal and (b) of the uncompensated semimetal ($\chi = 0.9$ and $n_2 = (0.9)^3 n_1$) for different magnetic field strengths $\omega_{\text{c}}^{(1)}\tau_{\text{imp}}^{(1)} = 0, 1.0, 3.0,$ and $10.0$. In Fig.~\ref{Fig:semimetal_magnetoresistance} (c) and (d), we show $d \ln \rho /d \ln T = T /\rho \cdot d\rho/dT$ (solid lines) and $d \ln WT /d \ln T$ (dash-dotted lines), which are helpful to see local minimum and maximum values, for the two cases. We show the temperature dependence of their ratios $\widetilde{L} = \rho/WT$ (e) of the compensated semimetal and (f) of the uncompensated semimetal. We normalize $\rho$ and $WT$ by $\rho_0 = [e^2(n_{1}\tau^{(1)}_{\text{imp}}/m_1 + n_{2}\tau^{(2)}_{\text{imp}}/m_2)]^{-1}$, which is the resistivity in the absence of electron-electron scatterings and the magnetic field. We set $\tau^{(2)}_{\text{imp}}/\tau^{(1)}_{\text{imp}} = 1$, $m_2/m_1 = 2$, and $\alpha = k_{\text{F},1}$. Temperature is normalized by $T^{(1)}_0$ where $\tau^{(1)}_{\text{imp}} = \tau^{(1)}_{\text{e-e}}$. Note that $1/\tau^{(l)}_{\text{e-e}} \propto T^{2}$ and $1/\tau^{(l)}_{\text{imp}} \propto T^{0}$.

\subsubsection{Temperature dependence for the compented case ($n_2 = n_1$)}
\begin{figure*}[tbp]
\begin{center}
\rotatebox{0}{\includegraphics[angle=0,width=1\linewidth]{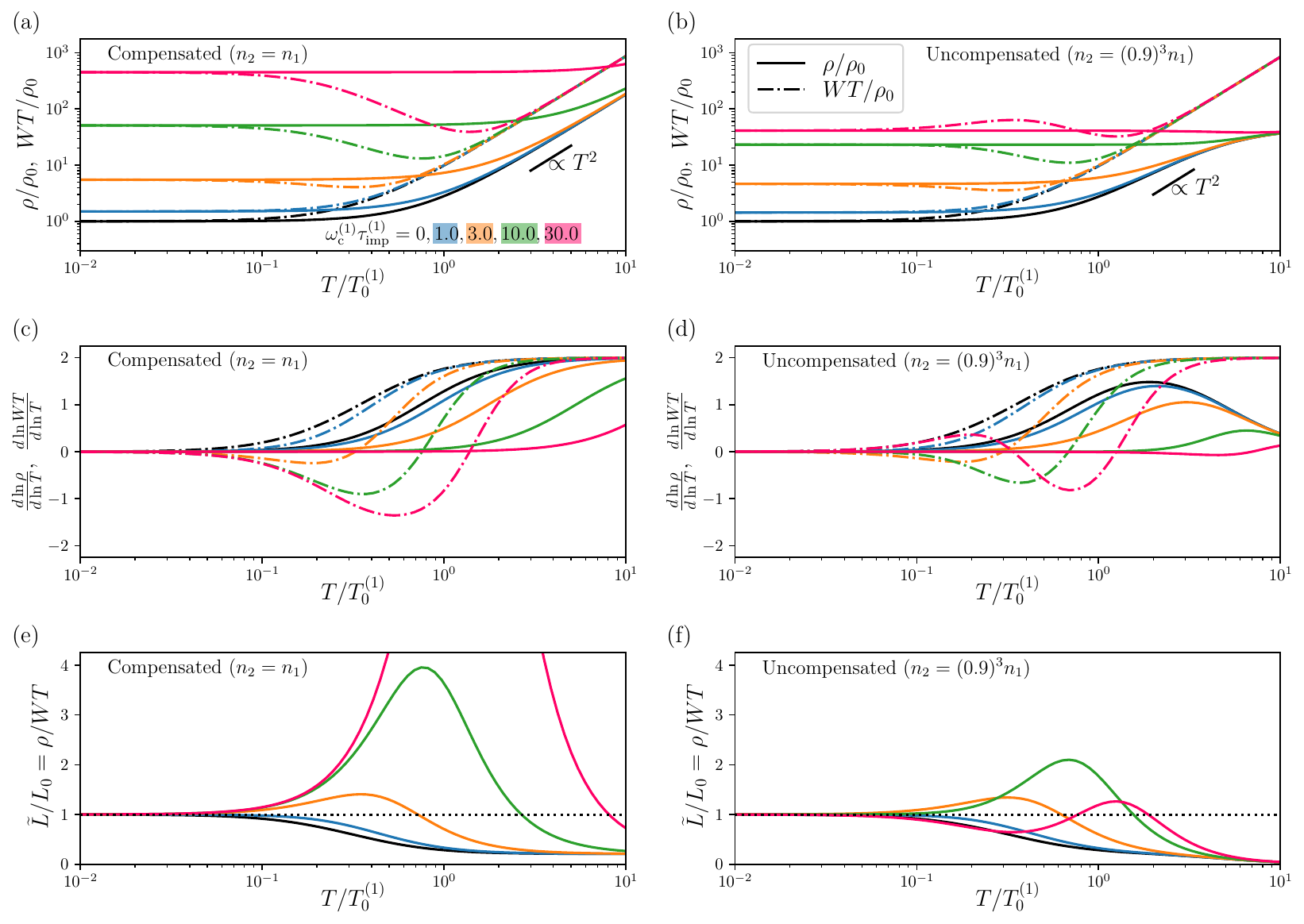}}
\caption{Temperature dependence of the electrical resistivity $\rho$ (solid lines) and thermal resistivity $WT$ (dash-dotted lines) of (a) the compensated semimetal and (b) the uncompensated semimetal ($n_2 = (0.9)^3$), $d \ln \rho/d \ln T$ (solid lines) and $d \ln WT/d \ln T$ (dash-dotted lines) of (c) the compensated semimetal and (d) the uncompensated semimetal, and $\widetilde{L}/L_{0} = \rho/WT$ of (e) the compensated semimetal and (f) the uncompensated semimetal for different strengths of the magnetic field $\omega_{\text{c}}^{(1)}\tau_{\text{imp}}^{(1)} = 0, 1.0, 3.0, 10.0,$ and $30.0$. }
\label{Fig:semimetal_magnetoresistance}
\end{center}
\end{figure*}

First, let us discuss the compensated case, where only the relative motion contributes to the longitudinal transport. As a function of the magnetic field, $\omega^{(1)}_{\text{c}}$, we see that both the electrical and thermal resistivities show magnetoresistance, which is larger at lower temperatures. As for the temperature dependences, $\rho$ shows a monotonic temperature dependence. On the contrary, $WT$ decreases with temperature for the lowest temperature region and the magnitude becomes larger in large magnetic fields. In Fig.~\ref{Fig:semimetal_magnetoresistance}~(e), we see that an upward violation of the WF law occurs in the intermediate temperature, $\lim_{T \to 0} (\widetilde{L} - L_{0})/T^2 > 0$, for large magnetic fields. The momentum conservation enhances the upward violation, as we discuss later.

We can interpret the monotonic temperature dependence of $\rho$ 
as the temperature dependence of $1/\bar{\mu}_{\text{e-h}}$ in Eq.~(\ref{eq:monotonic rho}). 
The origin of the $B$-squared term in Eq.~(\ref{eq:monotonic rho}) is most easily understood by considering the limit of a strong magnetic field.
Since this limit is equivalent to that of weak impurity scattering, 
both the electron and hole drift velocities are equal to the velocity 
of a free charged particle moving in perpendicular electric and magnetic fields.
Furthermore,  for $Q=0$ the Hall electric field is of the order of $1/B$ and therefore $|\bm{E}_{\text{H}}| \ll |\bm{E}_\text{ext}|$, so that 
\begin{align}
\bm{v}_\text{d}^{(e)} = \bm{v}_\text{d}^{(h)}= (\bm{E}_\text{ext} \times \bm{B}) /B^2.
\label{eq:driftfreeext}
\end{align}
Substituting this equation into Eq.~(\ref{eq:forceQ=0}), 
we can easily derive the $B$-squared term in Eq.~(\ref{eq:monotonic rho}).  
Hence, the motion of electrons and holes perpendicular to the external electric field generates the resistivity proportional to $B^2$.
This is an interpretation of how magnetoresistance arises in semimetals. 
In the compensated system, the relative momentum can contribute to the electric current, 
which can be relaxed by electron-hole scattering, as described by the last term in Eq.~(\ref{eq:relativeQ=0}). 
As discussed in Ref.~\cite{Kukkonen1979}, 
the electron-hole scattering does not affect the transverse transport 
and only increases the longitudinal resistivity.
This is because the damping force due to the electron-hole scattering 
is parallel to the electric current, as shown in Eqs.~(\ref{eq:damp^e}) and~(\ref{eq:damp^h}).

Let us check this behavior with the RTA, which gives qualitatively reasonable formulae. Due to the momentum-conserving interband scattering, the electrical conductivity cannot be described by the sum of a single-carrier model. In the RTA, the electrical resistivity $\rho$ is given by
\begin{align}
\left. \rho^{(\text{RTA})}\right|_{n_1 = n_2} = \rho_{0} \left(1 + \frac{\tau_{\text{imp}}^{(1)}}{\widetilde{\tau}^{(12)}_{\text{e-e}}} + \frac{\tau_{\text{imp}}^{(2)}}{\widetilde{\tau}^{(21)}_{\text{e-e}}} + \tau_{\text{imp}}^{(1)}\tau_{\text{imp}}^{(2)}\omega_{\text{c}}^{(1)}\omega_{\text{c}}^{(2)}\right), \label{eq:semimetal_rho_RTA}
\end{align}
where $\widetilde{\tau}^{(l\overline{l})}_{\text{e-e}} \propto T^{-2}$ is the relaxation time of the interband scattering in the RTA. See Refs.~\cite{Gantmakher1978, Kukkonen1979} or Appendix~\ref{App:RTA} for the full expression of $\rho^{(\text{RTA})}$ including the uncompensated case. It is found that the electrical resistivity shows the monotonic temperature dependence for the combination of the impurity scattering and the electron-electron scatterings.

In thermal transport, in contrast, the momentum conservation of the electron-electron scatterings does not play an important role. The thermal conductivity in the RTA is given by a sum of that of the single-carrier system (see Appendix~\ref{App:RTA}). Therefore, we can roughly grasp the behavior using a thermal transport relaxation time $1/\tau^{(l)}_{\kappa,\text{tr}} \sim 1/\tau^{(l)}_{\text{imp}} + A^{(l)}/\tau^{(l)}_{\text{e-e}}$ for each carrier system, where $A^{(l)}$ is a transport factor and can be extracted from the exact value if needed. Then, the thermal resistivity takes the usual form of the resistivity in the two-band model with the mobility replaced by the thermal one \cite{AshcroftMermin} (see also Appendix~\ref{App:RTA}),
\begin{align}
&WT^{(\text{RTA})} \nonumber \\
&\simeq \frac{1}{|e|} \cdot \frac{ n_1\mu^{(1)}_{\kappa} + n_2\mu^{(2)}_{\kappa} + \mu^{(1)}_{\kappa}\mu^{(2)}_{\kappa}(n_1\mu^{(2)}_{\kappa} + n_2\mu^{(1)}_{\kappa})B^2}{[n_1\mu^{(1)}_{\kappa} + n_2\mu^{(2)}_{\kappa}]^2 + [\mu^{(1)}_{\kappa}\mu^{(2)}_{\kappa}(n_1 - n_2)B]^2}, 
\label{eq:semimetal_WT_RTA} 
\end{align}
where $\mu^{(l)}_{\kappa} = |e|\tau^{(l)}_{\text{tr},\kappa}/m_{l}$ is the mobility including contributions from the impurity scattering, intra- and interband electron-electron scatterings. The behavior of resistivity is determined by the relative strength between $1/\mu^{(l)}_{\kappa}$ and $B$. 

In the compensated case, $n = n_1 = n_2$, $WT^{(\text{RTA})} = B^2/|e|n \cdot (1/\mu^{(1)}_{\kappa} + 1/\mu^{(2)}_{\kappa})^{-1}$ when the magnetic field is large enough while $WT^{(\text{RTA})} = 1/|e|n \cdot (\mu^{(1)}_{\kappa} + \mu^{(2)}_{\kappa})^{-1}$ when the magnetic field is small. For a large magnetic field such as $\omega_{\text{c}}^{(1)}\tau_{\text{imp}}^{(1)} = 3.0, 10.0$, and $30.0$ in Fig.~\ref{Fig:semimetal_magnetoresistance}(a), $WT$ decreases with tempearture at intermediate region since $1/\mu^{(l)}_{\kappa} \sim 1/\tau^{(l)}_{\text{imp}} + A^{(l)}/\tau^{(l)}_{\text{e-e}}$ increases with tempearture. As the tempearture increases, $1/\mu^{(l)}_{\kappa}$ in the numerator outweighs $B$ and $WT$ starts increasing. As a result, the thermal resistivity exhibits non-monotonic temperature dependence for a large magnetic field.

Next, we focus on $\widetilde{L}_{}$ in Fig.~\ref{Fig:semimetal_magnetoresistance}(e). We note that $\widetilde{L} = L_{}$ for zero magnetic field shown in $\omega_{\text{c}}^{(1)}\tau_{\text{imp}}^{(1)} = 0$ (black line). $\widetilde{L}|_{B = 0}$ decreases monotonically with increasing temperature and reaches some constant value since both $\rho$ and $WT$ are proportional to $T^2$, originating from the electron-electron scattering. We see that the WF law for $\widetilde{L}_{}$ in the compensated system for a large magnetic field is upwardly violated in intermediate temperatures due to the different behavior of $\rho$ and $WT$. This violation is estimated using the RTA as 
\begin{align}
\frac{\widetilde{L}^{}_{}|_{B \to \infty}}{L_{0}}  \sim \frac{1/\mu^{(1)}_{\kappa} + 1/\mu^{(2)}_{\kappa}}{1/\mu^{(1)}_{\text{imp}} + 1/\mu^{(2)}_{\text{imp}}} \gg \frac{L_{0}}{L_{}|_{B = 0}} . \label{eq:semimetal_compensated_Lorenz_res_estimate}
\end{align}
This upward violation in large magnetic fields is a feature of the compensated system, and this is enhanced by the momentum conservation, which leads to the monotonic temperature dependence of $\rho$.

\subsubsection{Temperature dependence for the uncompensated case ($n_2 \neq n_1$)}
We show the results for the uncompensated case in Figs.~\ref{Fig:semimetal_magnetoresistance} (b) and (d). In the uncompensated semimetal, the electrical resistivity saturates when the electron-electron scattering dominates over the impurity scattering, as shown in Fig.~\ref{Fig:semimetal_magnetoresistance}(b). This is because the total momentum, which cannot be relaxed by the electron-electron scatterings, contributes to the electrical transport, unlike the compensated system \cite{Kukkonen1976, Maldague1979}.
In the uncompensated system, the magnetoresistance also saturates in a large magnetic field \cite{Smith_Jensen1989}. This is found in the case of $\omega^{(1)}_{\text{c}}\tau^{(1)}_{\text{imp}} = 30.0$ in the low-temperature regime. We see that the temperature dependence is quite weak in this situation, and the two limiting values, strong electron-electron scattering and strong magnetic field, become the same. 

Such saturated electrical resistivity is obtained when electrons and holes move in unison in the direction of the current at the same drift velocity as in Eq.~(\ref{eq:driftQnonzero}). Therefore, the value of saturated electrical resistivity is that of the electrical resistivity of an effective single carrier system with charge density 
$Q=e(n_1-n_2)$ and the mean mobility $\bar{\mu}_\text{imp}=|n_1-n_2|(n_1/\mu_\text{imp}^{(1)}+n_2/\mu_\text{imp}^{(2)})^{-1}$ as in Eq.~(\ref{eq:rhoQnonzero}). 
In the weak magnetic field and high temperature regime, as explained previously, 
Eq.~(\ref{eq:driftQnonzero}) is the result of momentum conservation in the electron-electron scattering. In the low temperature and high magnetic field regime, on the other hand, both the electron and hole drift velocities are equal to 
the velocity of a free charged particle moving in perpendicular electric and magnetic fields, 
as in the compensated case of $Q=0$. For the present case of $Q \neq 0$, 
however, the Hall electric field is proportional to $B$ and therefore $|\bm{E}_{\text{H}}| \gg |\bm{E}_\text{ext}|$. 
In this case,  the electron and hole drift velocities are given by
\begin{align}
\bm{v}_\text{d}^{(e)} = \bm{v}_\text{d}^{(h)} = (\bm{E}_{\text{H}} \times \bm{B})/B^2, 
\end{align}
so that they are of the order of $1$ in the limit of $B \to \infty$ 
and become parallel to the current $\bm{j}$. 
From Eq.~(\ref{eq:forceQ=0}) in the direction perpendicular to $\bm{j}$, 
the Hall electric field is then given by $\bm{E}_{\text{H}} =  (\bm{B} \times \bm{j})/Q$, 
which leads to Eq.~(\ref{eq:driftQnonzero}). 
Hence, the two limiting values of the electrical resistivity 
in a strong magnetic field and in strong electron-electron scattering 
become the same.

Using the RTA, we can also estimate the limiting values of the resistivity and confirm these behaviors. In the strong electron-electron scattering limit, the saturated electrical resistivity is determined by the impurity scattering and given by 
\begin{align}
\lim_{T \to \infty} \rho^{(\text{RTA})} = \frac{n_1/\mu^{(1)}_{\text{imp}} + n_2/\mu^{(2)}_{\text{imp}}}{|e|(n_1 - n_2)^2}. \label{eq:semimetal_resistivity_limit_T}
\end{align}
For the strong magnetic field limit, the saturated electrical resistivity is estimated as
\begin{align}
\lim_{B \to \infty} \rho^{(\text{RTA})} = \frac{n_1/\mu^{(1)}_{\text{imp}} + n_2/\mu^{(2)}_{\text{imp}}}{|e|(n_1 - n_2)^2}, \label{eq:semimetal_resistivity_limit_B}
\end{align}
which is the same limiting value as $T \to \infty$. 

Next, we discuss the thermal resistivity. Since thermal transport is hardly affected by the momentum conservation, the behavior of the thermal resistivity is qualitatively similar to the compensated case except for an increase of $WT$ with temperatures in the intermediate temperature regime seen in the case of $\omega_{\text{c}}^{(1)} \tau_{\text{imp}}^{(1)} = 30.0$. This is related to the saturation of $WT$ with $B \to \infty$, which is estimated in the RTA as
\begin{align}
\lim_{B \to \infty}  WT^{(\text{RTA})}  \simeq \frac{ n_1/\mu^{(1)}_{\kappa} + n_2/\mu^{(2)}_{\kappa}}{|e|(n_1 - n_2)^2}. \label{eq:th_resistivity_semimetal_limit}
\end{align}
The increase of $WT$ with temperatures for $\omega_{\text{c}}^{(1)} \tau_{\text{imp}}^{(1)} = 30.0$ is understood from the tempearture dependences of $1/\mu^{(l)}_{\kappa}$ in Eq.~(\ref{eq:th_resistivity_semimetal_limit}).

In the uncompensated case, $\widetilde{L}|_{B = 0} = L|_{B = 0}$ approaches zero in high temperature since $\rho$ is saturated as in Eq.~(\ref{eq:semimetal_resistivity_limit_T}) whereas $WT \propto T^2$. For $\omega_{\text{c}}^{(1)} \tau_{\text{imp}}^{(1)} = 3.0$ and $10.0$, $\widetilde{L}$ shows the upward violation of the WF law in the intermediate temperatures. 
In further strong magnetic field as $\omega_{\text{c}}^{(1)} \tau_{\text{imp}}^{(1)} = 30.0$, the WF law for $\widetilde{L}$ is violated downwardly, where $\widetilde{L}$ is estimated in the RTA as
\begin{align}
\frac{\widetilde{L}_{}|_{B \to \infty}}{L_{0}}  \sim \frac{n_1/\mu^{(1)}_{\text{imp}} + n_2/\mu^{(2)}_{\text{imp}}}{n_1/\mu^{(1)}_{\kappa} + n_2/\mu^{(2)}_{\kappa}} < 1. \label{eq:semimetal_uncompensated_Lorenz_res_estimate}
\end{align}
This is in contrast to the compensated case and the upward violation in the intermediate temperature regime found in Fig.~\ref{Fig:semimetal_magnetoresistance}~(f) is attributed to the fact that the system has electrons and holes and is not far from the compensated case.

\begin{figure}[tbp]
\begin{center}
\rotatebox{0}{\includegraphics[angle=0,width=1\linewidth]{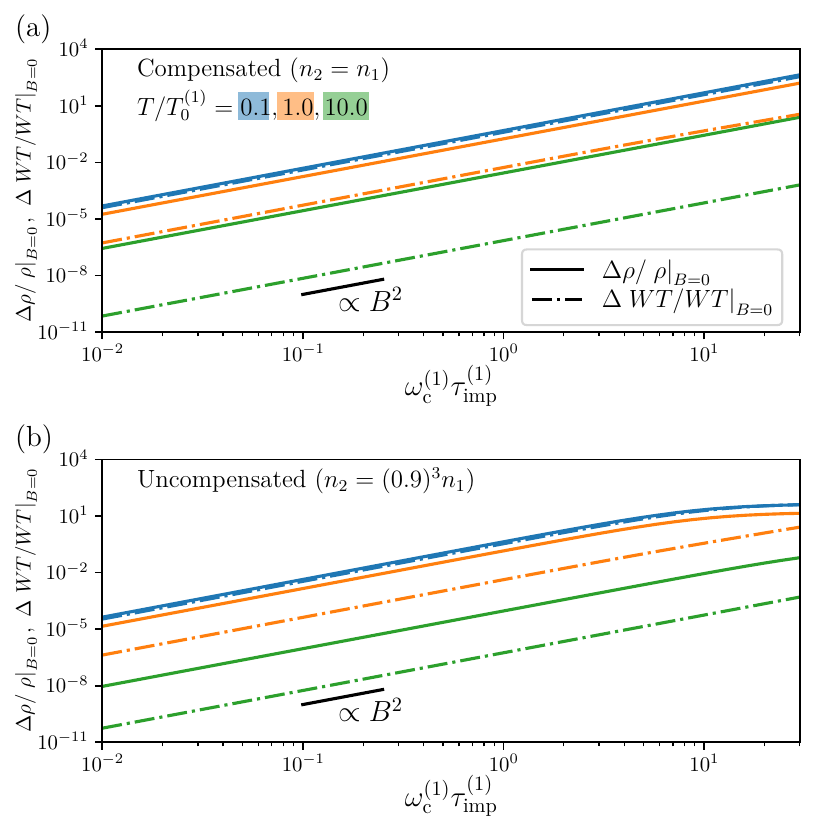}}
\caption{Magnetic field dependence of renormalized magnetoresistance $\Delta \rho$ and $\Delta WT$ of (a) the compensated semimetal and (b) the uncompensated semimetal ($n_2 = (0.9)^3 n_1$) for different temperatures $T/T^{(1)}_{0} = 0.1, 1.0,$ and $10.0$.}
\label{Fig:semimetal_magnetoresistance_mag_dep}
\end{center}
\end{figure} 

\subsubsection{Magnetic field dependence}
Figure~\ref{Fig:semimetal_magnetoresistance_mag_dep} shows the magnetic field dependences of magnetoresistance $\Delta \rho = \rho - \left. \rho \right|_{B = 0}$ and $\Delta WT = WT - \left. WT \right|_{B = 0}$ (a) in the compensated semimetal and (b) in the uncompensated semimetal ($n_2 = (0.9)^3 n_1$) for three different temperatures $T/T^{(1)}_{0} = 0.1, 1.0,$ and $10.0$. As we discussed above, we see the non-saturating behavior in the compensated case and saturation of the magnetoresistance in the uncompensated case for $T/T^{(1)}_{0} = 0.1$.

Note that whether the system has the saturation of the magnetoresistance or not is determined by the carrier number, regardless of scattering mechanisms.

\subsection{Hall and thermal Hall effects}
\label{SecVC}
\subsubsection{Temperature dependence}
\begin{figure}[tbp]
\begin{center}
\rotatebox{0}{\includegraphics[angle=0,width=1\linewidth]{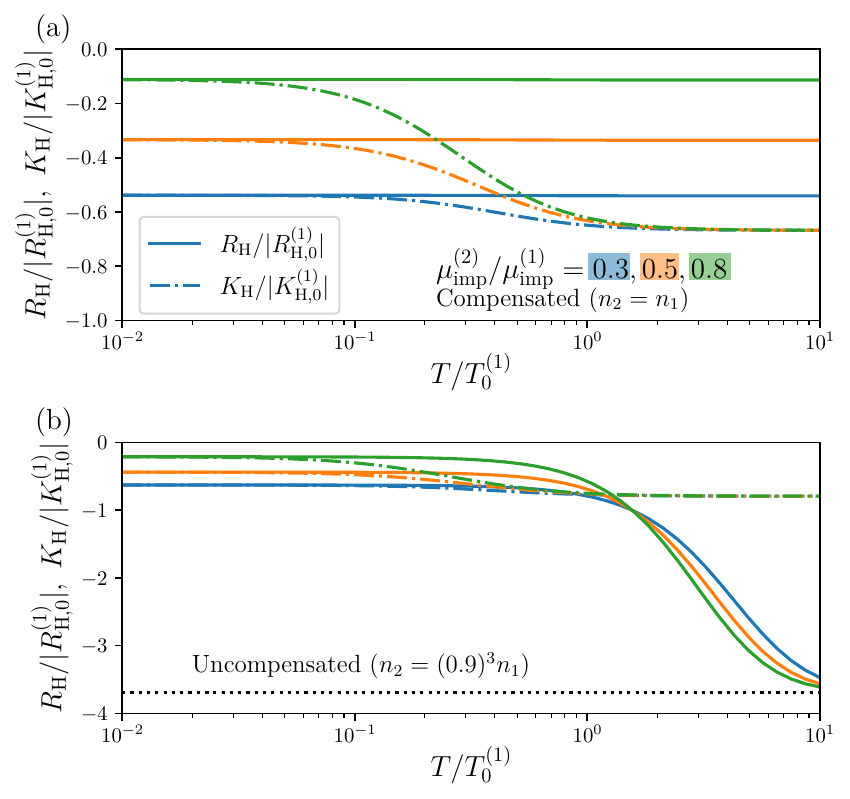}}
\caption{Temperature dependence of normalized $R_{\text{H}}$ and $K_{\text{H}}$ in a weak magnetic field of (a) the compensated semimetal and (b) the uncompensated semimetal ($n_2 = (0.9)^3 n_1$) for three different values of $\mu_{\text{imp}}^{(2)}/\mu_{\text{imp}}^{(1)} = 0.3, 0.5,$ and $0.8$. The black dotted line in (b) indicates $\lim_{T \to \infty} R^{(\text{RTA})}_{\text{H}}/R^{(1)}_{\text{H},0} = n_1/(n_1 - n_2)$ from Eq.~(\ref{eq:semimetal_Hall_uncompensated_strong_ee_RTA})}
\label{Fig:semimetal_Hall}
\end{center}
\end{figure}

In Fig.~\ref{Fig:semimetal_Hall}, we show the temperature dependence of $R_{\text{H}}/|R^{(1)}_{\text{H},0}|$ and $K_{\text{H}}/|K^{(1)}_{\text{H},0}|$ ($R^{(1)}_{\text{H},0} = 1/en_1$ and $K^{(1)}_{\text{H},0} = 1/en_1L_{0}$) of (a) the compensated semimetal and (b) the uncompensated semimetal ($n_2 = (0.9)^3 n_1$) for three different cases of $\mu_{\text{imp}}^{(2)}/\mu_{\text{imp}}^{(1)} = 0.3, 0.5,$ and $0.8$. We set $m_2/m_1 = 2$, $\omega_{\text{c}}^{(1)}\tau_{\text{imp}}^{(1)} = 10^{-3}$, and $\alpha = k_{\text{F},1}$.

As discussed in Ref.~\cite{Kukkonen1979} with the RTA, 
$R_{\text{H}}$ in the compensated case is almost temperature-independent, determined by $\mu^{(2)}_{\text{imp}}/\mu^{(1)}_{\text{imp}}$, and not affected by the electron-electron scatterings. 
This is because the damping force due to the electron-hole scattering 
and therefore the last term in Eq.~(\ref{eq:relativeQ=0}) are parallel to the electric current.
As a result, $R_{\text{H}}$ is given in the temperature-independent form as Eq.~(\ref{eq:TindepRH}).
Our calculation involving the energy-dependent parts of the distribution functions, which confirms that the effect of the electron-electron scattering does not appear, is consistent with Eq.~(\ref{eq:TindepRH}).

The Hall coefficient of a compensated semimetal in the RTA is given by \cite{Kukkonen1979}
\begin{align}
\left. R^{(\text{RTA})}_{\text{H}}\right|_{n = n_1 = n_2} = \frac{1}{en} \cdot \frac{\mu_{\text{imp}}^{(1)} - \mu_{\text{imp}}^{(2)}}{\mu_{\text{imp}}^{(1)} + \mu_{\text{imp}}^{(2)}}, \label{eq:semimetal_Hall_RTA} 
\end{align}
with $\mu^{(l)}_{\text{imp}} = |e|\tau^{(l)}_{\text{imp}}/m_{l}$ being the mobility of the band $l$ by the momentum dissipative scattering, which is impurity scattering in this paper (see also Appendix~\ref{App:RTA}). We can see that $R_{\text{H}}$ measures the difference of Hall signals of electrons and holes in terms of mobilities due to the impurity scatterings.

In the uncompensated case, we observe that the Hall coefficient has significant temperature dependence. We can understand that this is a shift from the low-temperature value determined by the impurity scattering to the high-temperature value where the electron-electron scattering dominates. The limiting value in the weak electron-electron scattering and in the weak magnetic field limit in the RTA is given by
\begin{align}
\lim_{T \to 0} R_{\text{H}}^{(\text{RTA})} 
= \frac{1}{e} \cdot \frac{n_1(\mu^{(1)}_{\text{imp}})^2 - n_2(\mu^{(2)}_{\text{imp}})^2}{[n_1\mu^{(1)}_{\text{imp}} + n_2\mu^{(2)}_{\text{imp}}]^2} ~(\tau^{(l)}_{\text{imp}} \ll \tau^{(l)}_{\text{e-e}}), \label{eq:semimetal_Hall_uncompensated_weak_ee_RTA}
\end{align}
In the strong electron-electron scattering limit, we find
\begin{align}
\lim_{T \to \infty} R_{\text{H}}^{(\text{RTA})} = \frac{1}{e(n_1 - n_2)}~(\tau^{(l)}_{\text{imp}} \gg \tau^{(l)}_{\text{e-e}}), 
\label{eq:semimetal_Hall_uncompensated_strong_ee_RTA} 
\end{align}
which is equivalent to Eq.~(\ref{eq:RHQnonzero}).
See Refs.~\cite{Gantmakher1978, Kukkonen1979} or Appendix~\ref{App:RTA} for the full expression of $R^{(\text{RTA})}_{\text{H}}$. Then, these RTA results are in accord with the results shown in Fig.~\ref{Fig:semimetal_Hall}. We can understand that the latter limiting value is the consequence of the strong electron-electron scattering, which locks the movement of electrons and holes together and makes the system an effective single carrier system with net charge $e(n_1 - n_2)$ for the Hall response \cite{Kukkonen1979}. 

$K_{\text{H}}$ is temperature dependent for both compensated and uncompensated cases, and there is no qualitative difference between the two cases in a weak magnetic field since the momentum conservation hardly affects the thermal current. $K_{\text{H}}$ in a weak magnetic field is evaluated in the RTA as
\begin{align}
K^{(\text{RTA})}_{\text{H}} 
\simeq \frac{1}{eL_{0}} \cdot \frac{n_1(\mu^{(1)}_{\kappa})^2 - n_2(\mu^{(2)}_{\kappa})^2}{[n_1\mu^{(1)}_{\kappa} + n_2\mu^{(2)}_{\kappa}]^2}.  \label{eq:semimetal_th_Hall_RTA} 
\end{align}
The expression corresponds to Eq.~(\ref{eq:semimetal_Hall_uncompensated_weak_ee_RTA}) however the mobilities of the impurity scattering are replaced with the thermal mobilities and the temperature dependence appears through the temperature dependence of $\mu^{(l)}_{\kappa}$.

\subsubsection{Mass ratio dependence of $K_{\text{H}}$ in the strong electron-electron scattering limit}
\begin{figure}[tbp]
\begin{center}
\rotatebox{0}{\includegraphics[angle=0,width=1\linewidth]{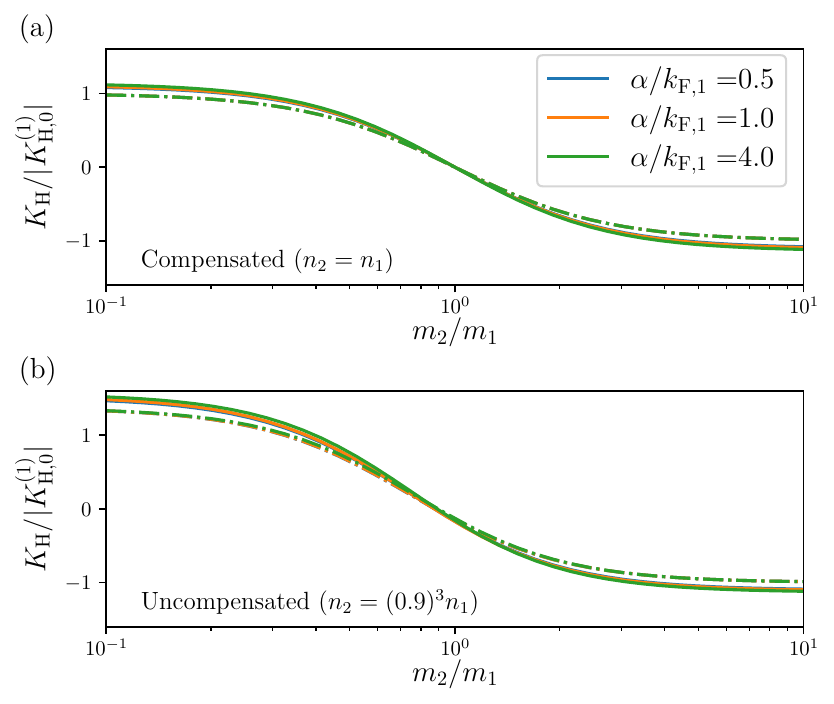}}
\caption{Mass ratio $m_2/m_1$ dependences of normalized $K_{\text{H}}$ of (a) the compensated semimetal and (b) the uncompensated semimetal ($n_2 = (0.9)^3 n_1$) for three different screening lengths. The dash-dotted lines are results obtained by the RTA.}
\label{Fig:semimetal_KH_mass_TF}
\end{center}
\end{figure}

The limiting values of $K_{\text{H}}$ for $T/T_{0}^{(1)} \gg 1$ are determined by the offset of transverse thermal transport between electrons and holes by electron-electron scatterings, as we can see in Eq.~(\ref{eq:semimetal_th_Hall_RTA}). To understand the limiting value, we show in Fig.~\ref{Fig:semimetal_KH_mass_TF} the mass ratio $m_2/m_1$ dependence of $K_{\text{H}}$ of (a) the compensated semimetal and (b) the uncompensated semimetal ($n_2 = (0.9)^3 n_1$) in the absence of the impurity scattering and weak magnetic field limits. We plot the result of three cases of different inverse screening lengths $\alpha/k_{\text{F},1} = 0.5, 1,$ and $4$. The dash-dotted line shows the result obtained by Eq.~(\ref{eq:semimetal_th_Hall_RTA}) where we use $\tau_{\text{e-e}}^{(l)}$ given by Eq.~(\ref{eq:sum_rel_time_ee}) as a relaxation time \footnote{To be precise, Eq.~(\ref{eq:sum_rel_time_ee}) should acquire a factor originating from geometrical factors and the discrepancy between Eq.~(\ref{eq:sum_rel_time_ee}) and the quasiparticle relaxation time around $2/\pi^2$. However, these factors are canceled out in Eq.~(\ref{eq:semimetal_th_Hall_RTA}).}. We note that, in the compensated case, the RTA result is simply given by $K^{(\text{RTA})}_{\text{H}} = \left(m_1^2 - m_2^2\right)/\left(m_1^2 + m_2^2\right)$. 

$K_{\text{H}}$ for both compensated and uncompensated cases in the strong electron-electron scattering limit is determined by the carrier number and the mass ratio, which is influential on the relaxation times $\tau_{\text{e-e}}^{(l)}$ in Eqs.~(\ref{eq:sum_rel_time_ee}) and (\ref{eq:def_rel_time_ee}). We see that the screening length dependence is weak.
These behaviors are qualitatively described by the RTA, although the exact value is slightly different from $K^{(\text{RTA})}_{\text{H}}$. This is due to the inelastic feature of the electron-electron scatterings, which cannot be described by the RTA, as we have discussed in Baber scattering case. Since $\beta_{\kappa}^{(l)} = 0$ as noted, the integral equations Eqs.~(\ref{eq:int_eq_electrons}) and (\ref{eq:int_eq_holes}) are decoupled. Therefore, we can straightforwardly compare semimetals to Baber scattering focusing on $\lambda_{\kappa}^{(l)}$, whose value varies from $\lambda_{\kappa}^{(l)} \sim 0.5$ for $\alpha/k_{\text{F},1} = 0.5$ to $\lambda_{\kappa}^{(l)} \sim 1$ for $\alpha/k_{\text{F},1} = 4$ depending on the screening length and the carrier numbers. By comparing the value of $\lambda_{\kappa}^{(l)}$ to Fig.~\ref{Fig:Baber_Hall} discussed for Baber scattering, we can estimate $K_{\text{H}}/K^{(\text{RTA})}_{\text{H}} \sim 1.1$, which is close to the actual values in Fig.~\ref{Fig:semimetal_Hall}. From this estimation, we can also understand the weak screening length dependence.

It should be noted that $K_{\text{H}}$ is sensitive to the relative strength of the intra- and interband electron-electron scatterings. Thus, $K_{\text{H}}$ can be a useful tool to estimate the relative strength of scatterings. 
For example, in the compensated case, $K_{\text{H}} \propto \left(m_1^4 - m_2^4\right)/\left(m_1^4 + m_2^4\right)$ in the absence of the interband scattering and $K_{\text{H}} = 0$ in the absence of the intraband scattering. For the uncompensated case, we do not have simple expressions of the RTA as in the compensated case, although we can expect a similar mass ratio dependence as in the compensated case.

\subsection{Lorenz ratio and Hall Lorenz ratio}
\begin{figure*}[tbp]
\begin{center}
\rotatebox{0}{\includegraphics[angle=0,width=1\linewidth]{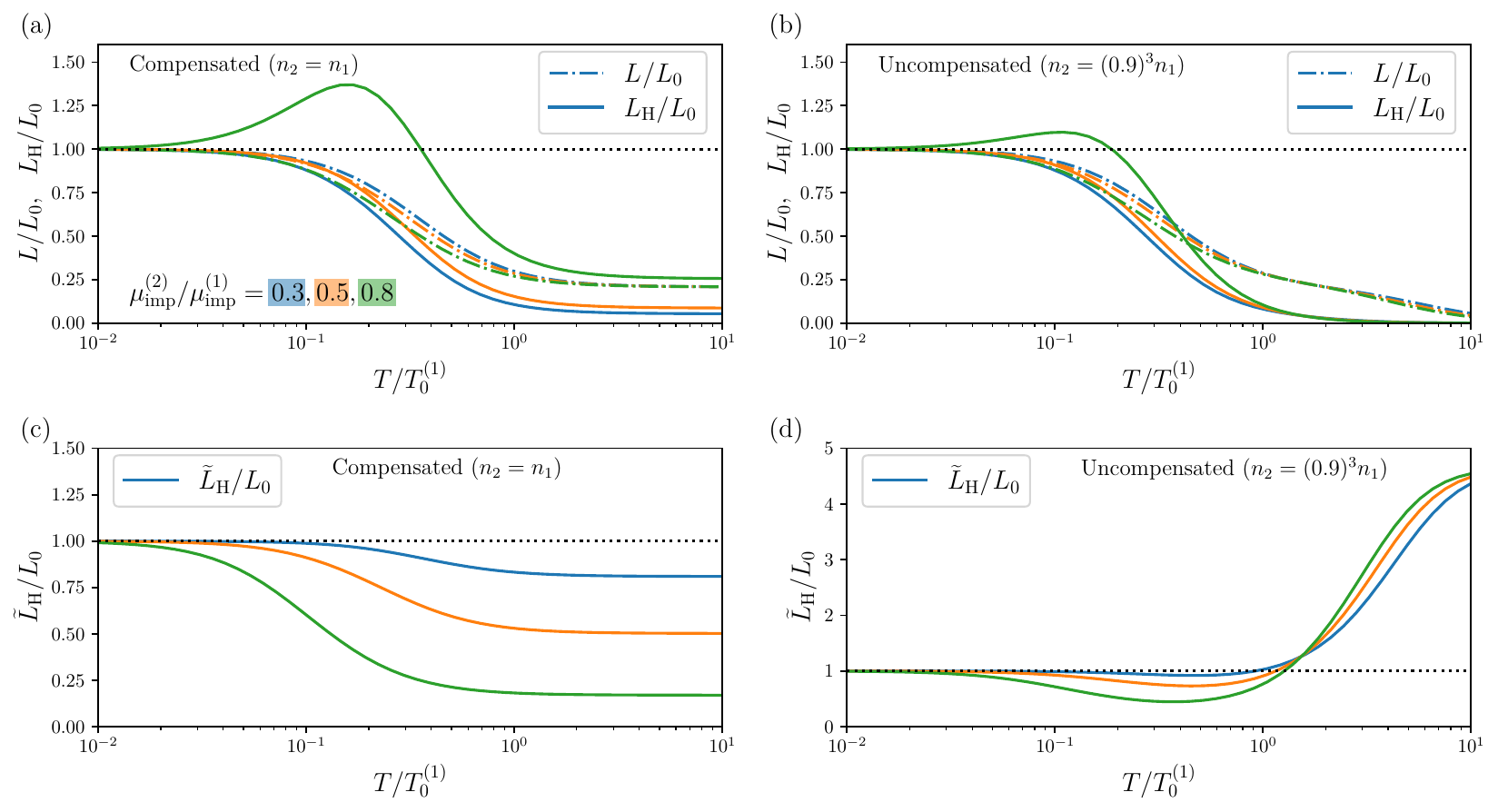}}
\caption{Temperature dependence of the Lorenz ratio (dash-dotted lines) and Hall Lorenz ratio (solid lines) of (a) the compensated semimetal and (b) the uncompensated semimetal ($n_2 = (0.9)^3 n_1$) and $\widetilde{L}_{\text{H}}$ normalized by $L_{0}$ of (c) the compensated semimetal and (d) the uncompensated semimetal ($n_2 = (0.9)^3 n_1$) for three different values of $\mu_{\text{imp}}^{(2)}/\mu_{\text{imp}}^{(1)} = 0.3, 0.5,$ and $0.8$. The parameters are the same as in Fig.~\ref{Fig:semimetal_Hall}. The black dotted lines indicate the WF law.}
\label{Fig:semimetal_Lorenz}
\end{center}
\end{figure*} 

Having the results of $R_{\text{H}}$ and $K_{\text{H}}$ in mind, we discuss the behavior of the Lorenz ratio and the Hall Lorenz ratio. Figure~\ref{Fig:semimetal_Lorenz} shows the temperature dependence of the Lorenz ratio (dash-dotted lines) and the Hall Lorenz ratio (solid lines) in (a) the compensated semimetal and (b) the uncompensated semimetal ($n_2 = (0.9)^3 n_1$) for three different cases of $\mu_{\text{imp}}^{(2)}/\mu_{\text{imp}}^{(1)} = 0.3, 0.5,$ and $0.8$ as in Fig.~\ref{Fig:semimetal_Hall}. Other parameters are also the same as in Fig.~\ref{Fig:semimetal_Hall}. We also plot temperature dependence of normalized $\widetilde{L}_{\text{H}} = R_{\text{H}}/K_{\text{H}}$ of (c) the compensated semimetal and (d) the uncompensated semimetal for reference since $L_{\text{H}} \simeq L^2/\widetilde{L}_{\text{H}}$ in a weak magnetic field. The WF law holds for both the Lorenz ratio and the Hall Lorenz ratio for $T \ll T^{(1)}_0$.  The Lorenz ratio monotonically decreases and reaches the value as we have discussed for the zero-magnetic field in Fig.~\ref{Fig:semimetal_magnetoresistance}(e) and the limiting value is independent of $\mu_{\text{imp}}^{(2)}/\mu_{\text{imp}}^{(1)}$. In the uncompensated case, $L$ approaches zero as in Fig.~\ref{Fig:semimetal_magnetoresistance}(f). On the contrary, the behavior and limiting value of the high-temperature side of the Hall Lorenz ratio depend on $\mu_{\text{imp}}^{(2)}/\mu_{\text{imp}}^{(1)}$ and a non-monotonic temperature dependence is found for $\mu_{\text{imp}}^{(2)}/\mu_{\text{imp}}^{(1)} = 0.8$.  This is caused by the different behavior between $R_{\text{H}}$ and $K_{\text{H}}$, found in Fig.~\ref{Fig:semimetal_Hall}, due to the momentum conservation of the electron-electron scatterings. In particular, the upward violation of $L_{\text{H}}$ in the intermediate temperature is attributed to the reduction of $\widetilde{L}_{\text{H}}$ overcoming the decrease of the Lorenz ratio. Although we expect that the small Hall Lorenz ratio is found in the system with the small Lorenz ratio, we still need considerations on $\widetilde{L}_{\text{H}}$, and this is one example of such a case.

Note that, in the hole-excess system $(n_2 > n_1)$, the Hall coefficient behaves as $R_{\text{H}} \to 1/e(n_1 - n_2) > 0$ when the electron-electron scatterings are strong but the sign of $K_{\text{H}}$ still depends on the mobilities of electron-electron scatterings. This can lead to $\widetilde{L}_{\text{H}} < 0$ and $L_{\text{H}} < 0$.

\section{Discussion}

\subsection{A possible application to the violation of the Wiedemann-Franz law in semimetals}

For the violation of the WF law in compensated semimetals, two scenarios have been proposed within the framework of the theory of the interband scattering \cite{Li2018}. One possibility is that the potential of the interband scattering is a weakly screened Coulomb interaction. In this case, the different relaxations between electrical and thermal transport are caused by the weakness of the screening, and the relative strength of the intra- and interband scatterings does not matter. The other scenario is that the intraband electron-electron scattering is relatively strong compared to the interband scattering. The intraband electron-electron scattering hardly affects the electric current but relaxes the thermal current well. Therefore, relatively strong intraband scattering causes a strong violation of the WF law. In this case, the screening of interaction does not need to be weak.

A three-dimensional Dirac electron system was predicted to have a large dielectric constant, which, in the limit of zero mass gap and zero chemical potential, exhibits a logarithmic divergence, which corresponds to the ultraviolet divergence of quantum electrodynamics~\cite{Maebashi2017lorentz}. Since the large dielectric constant results in a very long screening length of the Coulomb interaction between electrons,  an interesting situation is expected in such a system, where transport phenomena are dominated by electron-electron scattering due to weakly screened Coulomb interaction. 
This is also the case for a Weyl semimetal, which can be considered as the massless limit of the three-dimensional Dirac electron system~\cite{Maebashi2019nuclear}. 
Thus, the first scenario may explain the experimentally observed violation of the WF law in the Weyl semimetal $\text{WP}_{2}$~\cite{Gooth2018, Jaoui2018}, as discussed in Ref.~\cite{Li2018}. 
As seen in Sec.~\ref{SecVC}, when compensated, the electron-electron scattering has little effect on $R_{\text{H}}$, affecting only $K_{\text{H}}$. Therefore, we expect that in the case of $\text{WP}_{2}$, $R_{\text{H}}$ has only a very weak temperature dependence compared to $K_{\text{H}}$ as in Fig.~\ref{Fig:semimetal_Hall} (a), resulting in the violation of the transverse WF law as well as the longitudinal case as shown in Fig.~\ref{Fig:semimetal_Lorenz} (a).

The second scenario is important because it allows ordinary semimetals with not so large dielectric constants to violate the WF law. As mentioned previously, 
$K_{\text{H}}$ is sensitive to the relative strength of intraband and interband electron-electron scatterings, whereas $R_{\text{H}}$ is not. 
On the assumption that the only momentum dissipative process at low temperatures is the impurity scattering, 
we can estimate $\mu_{\text{imp}}^{(l)}$ and $\mu^{(l)}_{\kappa}$ from $R_{\text{H}}$, $K_{\text{H}}$, and other transport properties. By combining the information on effective masses, we can estimate the relative strength of scatterings, and we get a precise understanding of the violation of the WF law. 

\subsection{Inclusion of other scattering processes}
In this paper, we only consider the impurity scattering as a momentum-relaxing process. Other momentum-relaxing processes (e.g., the electron-phonon scattering) can be taken into account by adding the scattering terms in the Boltzmann equation. For the RTA, the inverse relaxation times can be added to $1/\tau^{(l)}_{\text{imp}}$. In that case, the monotonic temperature dependence of $\rho$ in the compensated case found in Fig.~\ref{Fig:semimetal_magnetoresistance}(a) can be altered due to the temperature dependence of other momentum-relaxing processes.

\section{Conclusion}\label{Sec:Conclusion}
In this paper, we have studied the electrical and thermal magnetotransport properties and violations of the WF law in the effective single-carrier system with Baber scattering and the two-band semimetals, solving the Boltzmann equation in the presence of the impurity, electron-electron scatterings, and the magnetic field. The effect of the magnetic field is taken into account by introducing complex-valued distribution functions for the energy dependence. We have used the analytic solutions in Baber scattering and numerically exact solutions in semimetals. The exact solutions enable us to take the inelastic nature into account correctly. We have sorted out the transport properties using $R_{\text{H}}$, $K_{\text{H}}$, and another set of Lorenz ratios defined in terms of the resistivity and the Hall coefficient, $\widetilde{L}_{}$ and $\widetilde{L}_{\text{H}}$. In particular, the Hall Lorenz ratio is expressed as $L_{\text{H}} \simeq K_{\text{H}}/R_{\text{H}} \cdot L_{}^2 = L_{}^2/\widetilde{L}_{\text{H}}$ in a weak magnetic field.

For Baber scattering, we have demonstrated features of the electron-electron scatterings originating from the inelastic nature encoded in the energy dependence of distribution functions. We quantify the non-zero magnetoresistance, $R_{\text{H}} \neq 1/en$, $K_{\text{H}} \neq 1/enL_{0}$, and $\widetilde{L}_{\text{H}} \neq L_{0}$. This leads to a small modification from $L_{\text{H}}/L_{0} = (L_{}/L_{0})^2$ in a weak magnetic field. These effects can be described only by methods beyond the RTA, even though the RTA formulae, which are given in the Drude forms, offer relatively good approximations upon choosing proper relaxation times.

In semimetals, we first have shown that the temperature dependence is different between $\rho$ and $WT$ reflecting the fact that momentum conservation is critical for electrical transport but not for thermal transport. $\rho$ shows the monotonic temperature dependence in our model, which has the impurity scattering and electron-electron scatterings. In the uncompensated case, $\rho$ saturates with increasing temperature and increasing the strength of a magnetic field. Then, the two saturated values become the same [Eqs.~(\ref{eq:semimetal_resistivity_limit_T}) and (\ref{eq:semimetal_resistivity_limit_B})]. This leads to quite weak temperature dependence when a magnetic field is large. In contrast, $WT$ shows the non-monotonic temperature dependence when the magnetic field is large. The sign of $\lim_{T \to 0} [\widetilde{L} - L_{0}]/T^2$ depends on $B$ in both compensated and uncompensated systems. The momentum conservation boosts the violation of the WF law for $\widetilde{L}$.
$R_{\text{H}}$ and $K_{\text{H}}$ behave quite differently as well due to the momentum conservation. 
Our numerically exact calculations have validated the previously known results of the Hall coefficient by the RTA, which is significantly affected by the momentum conservation and sensitive to the carrier numbers.
In contrast, $K_{\text{H}}$ in a weak magnetic field does not show a qualitative difference between the compensated and the uncompensated cases because the thermal current is hardly affected by the momentum conservation. These bring an impurity scattering dependent complex behavior to the Hall Lorenz ratio through $L_{\text{H}} \simeq L_{}^2/\widetilde{L}_{\text{H}}$ in a weak magnetic field. Also, the thermal Hall effect is sensitive to the relative strength between intra- and interband electron-electron scatterings. We have shown that analysis of the thermal Hall effect and the Hall Lorenz ratio can further elucidate the nature of the electron-electron scatterings in addition to the investigation of the longitudinal transport. We hope that these analyses help clarify the effects of the electron-electron scatterings in materials from various perspectives.

In this paper, calculations are limited to isotropic band structures and the low-temperature region $k_BT \ll \varepsilon_{\text{F}}$. In particular, analyses of the Hall Lorenz ratio and other transport properties in anisotropic systems are subjects of future work.
 
\section*{Acknowledgments}
This work is supported by Grants-in-Aid for Scientific Research from the Japan Society for the Promotion of Science (Grants No. JP20K03802, No. JP21K03426, No. JP22K18954, and No. JP25KJ0924), and JST-Mirai Program Grant (Grant No. JPMJMI19A1). K. T. is supported by Forefront Physics and Mathematics Program to Drive Transformation (FoPM), University of Tokyo.

\appendix 
\section{Derivation of the integral equations} \label{App:derivation_integral_eq}
In this Appendix, we derive several equations used in Sec.~\ref{Sec:integral_eq} by examining acts of the scattering terms and the magnetic field operator on the expanded distribution function Eq.~(\ref{eq:distr_expansion_spherical}). We also define and calculate the dimensionless parameters $\lambda^{(l)}_{X}$ and $\beta^{(l)}_{X}$. 

\subsection{Electron-electron scatterings}
We follow the treatment for the electron-electron scatterings in Refs. \cite{Abrikosov1957, Abrikosov1959, Brooker1968, Jensen1968, Smith1969, Jensen1969, Bennett1969, Sykes1970, Ah-Sam1971, Brooker1972, Egilsson1977, Oliva1982, Anderson1987, Golosov1995, Golosov1998, Pethick2009, Li2018, Lee2020, Lee2021}. Note that we fix the momenta on the Fermi surfaces throughout the calculations. Later, it will be shown that only $Y_{1,1}(\theta_{1}, \phi_{1})$ and $Y_{1,-1}(\theta_{1}, \phi_{1})$ are necessary in the present case. However, in the following, we show that the formalism applies to general $Y_{n,m}(\theta_{1}, \phi_{1})$.

By substituting Eq.~(\ref{eq:distr_expansion_spherical}) into the scattering term of the electron-electron scattering [Eq.~(\ref{eq:el_el_scattering})] we obtain

\begin{widetext}
\begin{align}
I^{(ll')}_{\text{e-e}}[\Phi] = 
& \sum_{n,m} \frac{m_{l} m_{l'}^2}{8\pi^4 \hbar^{6}} \int_{-\infty}^{\infty} d\xi_{2} d\xi_{3} d\xi_{4} \frac{1}{e^{\beta \xi_{l,\bm{k}}} + 1} \frac{1}{e^{\beta \xi_{2}} + 1} \frac{1}{1 + e^{- \beta \xi_{3}}} \frac{1}{1 + e^{- \beta \xi_{4}}} \delta(\xi_{l,\bm{k}} + \xi_2 - \xi_3 - \xi_4) Y_{n,m}(\theta_1, \phi_1) \nonumber \\ 
& \times\int \frac{d\Omega}{4\pi} \frac{W^{(ll')}(\theta,\varphi)}{R^{(ll')}(\theta)} [\widetilde{\Phi}^{(l)}_{n,m}(\beta \xi_{l,\bm{k}}) + P_{n}(\cos \theta_{12})\widetilde{\Phi}^{(l')}_{n,m}(\beta\xi_{2}) - P_{n}(\cos \theta_{13})\widetilde{\Phi}^{(l)}_{n,m}(\beta\xi_{3}) - P_{n}(\cos \theta_{14})\widetilde{\Phi}^{(l')}_{n,m}(\beta\xi_{4})].
\end{align}
Here, we have used the following identity \cite{Sykes1970}:
\begin{align}
\int \frac{d\varphi_{2}}{2\pi} Y_{n,m}(\theta_{i}, \phi_{i}) = Y_{n,m}(\theta_1,\phi_1) P_{n}(\cos \theta_{1i}),~(i = 2,3,~\text{and}~4),
\end{align}
where $\theta_{i}$ ($\phi_{i}$) is the polar (azimuth) angle of $\bm{k}_i$ and $\theta_{1i}$ is the angle between $\hat{\bm{k}}$ and $\hat{\bm{k}}_{i}$. Then, using the definition of $\Lambda_{i;n}^{(ll')}$ in Eq.~(\ref{eq:def_Lambda_n}), we can rewrite the electron-electron scattering as
\begin{align}
I^{(ll')}_{\text{e-e}}[\Phi] = 
&\frac{1}{\tau^{(ll')}_{\text{e-e}}} \sum_{n,m} \int_{-\infty}^{\infty} dx_{2} dx_{3} dx_{4} \frac{1}{e^{x_1} + 1} \frac{1}{e^{x_2} + 1} \frac{1}{1 + e^{-x_3}} \frac{1}{1 + e^{-x_4}} \delta(x_1 + x_2 - x_3 - x_4)\nonumber \\
& \times Y_{n,m}(\theta_{1}, \phi_{1}) [\widetilde{\Phi}^{(l)}_{n,m}(x_1) + \Lambda_{2;n}^{(ll')}\widetilde{\Phi}^{(l')}_{n,m}(x_{2}) - \Lambda_{3;n}^{(ll')}\widetilde{\Phi}^{(l)}_{n,m}(x_{3}) - \Lambda_{4;n}^{(ll')}\widetilde{\Phi}^{(l')}_{n,m}(x_{4})],
\end{align}
where $x_1 = \beta\xi_{l,\bm{k}}$,~$x_i = \beta\xi_{i}~(i = 2$-$4$), and $\tau^{(ll')}_{\text{e-e}}$ is defined in Eq.~(\ref{eq:def_rel_time_ee}). The energy integrals of $x_2,x_3$, and $x_4$ lead to \cite{Abrikosov1957, Abrikosov1959, Brooker1968, Jensen1968, Smith1969, Jensen1969, Bennett1969, Sykes1970, Ah-Sam1971, Brooker1972, Egilsson1977, Oliva1982, Anderson1987, Golosov1995, Golosov1998, Pethick2009, Li2018, Lee2020, Lee2021}
\begin{align}
I^{(ll')}_{\text{e-e}}[\Phi] = 
& \frac{1}{\tau^{(ll')}_{\text{e-e}}} \sum_{n,m} Y_{n,m}(\theta_1, \phi_1) \frac{1}{(e^{x_1} + 1)(1 + e^{-x_1})} \nonumber \\
& \times \left[ \frac{\pi^2 + x_1^2}{2}\widetilde{\Phi}^{(l)}_{n,m}(x_1) - \int_{-\infty}^{\infty} \mathcal{F}(x_1,u)[- \Lambda_{2;n}^{(ll')}\widetilde{\Phi}^{(l')}_{n,m}(-u) + \Lambda_{3;n}^{(ll')} \widetilde{\Phi}^{(l)}_{n,m}(u) + \Lambda_{4;n}^{(ll')} \widetilde{\Phi}^{(l')}_{n,m}(u)] du \right], \label{eq:el_el_scattering_simplified}
\end{align}
where $\mathcal{F}(x,u)$ is given by
\begin{align}
\mathcal{F}(x,u) = \frac{\cosh(x/2)}{\cosh(u/2)} \mathcal{G}(x - u) = \frac{\cosh(x/2)}{\cosh(u/2)} \cdot \frac{x-u}{2\sinh[(x-u)/2]}.
\end{align}
From the coefficient of $\widetilde{\Phi}^{(l)}_{n,m}(x_1)$ in Eq.~(\ref{eq:el_el_scattering_simplified}), we see that the lifetime of carriers by the electron-electron scattering on the Fermi surfaces is given by $2/\pi^2 \cdot \tau^{(ll')}_{\text{e-e}}$.

\subsection{Magnetic field}
By substituting Eq.~(\ref{eq:distr_expansion_spherical}) into the effect of a magnetic field, Eq.~(\ref{eq:magnetic_field_operator}), we obtain
\begin{align}
M^{(l)}[\Phi] 
=& \frac{e}{\hbar} \left( - \frac{1}{\beta} \frac{\partial f_{0}(\varepsilon_{l,\bm{k}})}{\partial \varepsilon_{l,\bm{k}}} \right) \sum_{n,m}( \bm{v}^{(l)}_{\bm{k}} \times \bm{B}) \cdot \nabla_{\bm{k}} Y_{n,m}(\theta_1,\phi_1) \widetilde{\Phi}^{(l)}_{n,m}(\beta \xi_{l,\bm{k}}), \nonumber \\
=& - \eta_{l} \frac{eB}{m_{l}} \left( - \frac{1}{\beta} \frac{\partial f_{0}(\varepsilon_{l,\bm{k}})}{\partial \varepsilon_{l,\bm{k}}} \right) \sum_{n,m} \frac{\partial}{\partial \phi_1} Y_{n,m}(\theta_1,\phi_1) \widetilde{\Phi}^{(l)}_{n,m}(\beta \xi_{l,\bm{k}}).
\end{align}
where we have used $\bm{v}_{\bm{k}}^{(l)} = \eta_{l}\hbar\bm{k}/m_{l}$ and $(k_{y} \partial/\partial k_{x} - k_{x} \partial/\partial k_{y}) Y_{n,m}(\theta_1,\phi_1) = - \partial/\partial \phi_1 \cdot  Y_{n,m}(\theta_1,\phi_1)$.

\subsection{Integral equations}
The left-hand side of the Boltzmann equation Eq.~(\ref{eq:Boltzmann_eq}) is proportional to $v_{\bm{k};x}^{(l)}  \propto k_x \propto Y_{1,1}(\theta_1,\phi_1)$. The impurity and electron-electron scatterings are diagonal with respect to $Y_{n,m}(\theta_1,\phi_1)$ while the magnetic field operator connects $Y_{1,1}(\theta_1,\phi_1)$ and $Y_{1,-1}(\theta_1,\phi_1)$. Therefore, we only need to consider the terms with $Y_{1,1}(\theta_1,\phi_1) \propto k_x$ and $Y_{1,-1}(\theta_1,\phi_1) \propto k_y$.
Then, by multiplying the left- and right-hand sides of Eq.~(\ref{eq:Boltzmann_eq}) with $Y_{1,\pm 1}(\theta_1,\phi_1)$ and integrating with respect to $(\theta_1,\phi_1)$, we obtain the following integral equations for the energy freedom as 
\begin{align}
&\eta_{l} \sqrt{\frac{4\pi}{3}} \beta v_{\text{F},l} F^{(l)}_{\text{ext}} \nonumber \\
&= \frac{1}{\tau^{(l)}_{\text{imp}}} \widetilde{\Phi}^{(l)}_{1,1}(x_1) - \eta_{l} \frac{eB}{m_{l}} \widetilde{\Phi}^{(l)}_{1,-1}(x_1) \nonumber \\
&~~~+ \sum_{l' = 1,2}\frac{1}{\tau^{(ll')}_{\text{e-e}}} \left[ \frac{\pi^2 + x_1^2}{2}\widetilde{\Phi}^{(l)}_{1,1}(x_1) - \int_{-\infty}^{\infty} \mathcal{F}(x_1,u) [- \Lambda_{2}^{(ll')} \widetilde{\Phi}^{(l')}_{1,1}(-u) + \Lambda_{3}^{(ll')} \widetilde{\Phi}^{(l)}_{1,1}(u) + \Lambda_{4}^{(ll')} \widetilde{\Phi}^{(l')}_{1,1}(u)] du \right], \label{eq:int_eq_x_axis}\\
0 
&= \frac{1}{\tau^{(l)}_{\text{imp}}} \widetilde{\Phi}^{(l)}_{1,-1}(x_1) + \eta_{l} \frac{eB}{m_{l}} \widetilde{\Phi}^{(l)}_{1,1}(x_1) \nonumber \\
&~~~+ \sum_{l' = 1,2}\frac{1}{\tau^{(ll')}_{\text{e-e}}} \left[ \frac{\pi^2 + x_1^2}{2}\widetilde{\Phi}^{(l)}_{1,-1}(x_1) - \int_{-\infty}^{\infty}  \mathcal{F}(x_1,u)[- \Lambda_{2}^{(ll')}\widetilde{\Phi}^{(l')}_{1,-1}(-u) + \Lambda_{3}^{(ll')} \widetilde{\Phi}^{(l)}_{1,-1}(u) + \Lambda_{4}^{(ll')} \widetilde{\Phi}^{(l')}_{1,-1}(u)] du \right], \label{eq:int_eq_y_axis} 
\end{align}
where we have used the abbreviation $\Lambda_{i;n = 1}^{(ll')} \to \Lambda_{i}^{(ll')}$ since we only consider $n = 1$. As we noted in Eqs.~(\ref{eq:distr_x_parametrize}) and (\ref{eq:distr_y_parametrize}), we can parametrize $\widetilde{\Phi}^{(l)}_{1,\pm 1}(u)$ in terms of even and odd functions. In both cases, we find that Eqs.~(\ref{eq:int_eq_x_axis}) and (\ref{eq:int_eq_y_axis}) are combined into a single integral equation if we add Eqs.~(\ref{eq:int_eq_x_axis}) and (\ref{eq:int_eq_y_axis}) with the latter being multiplied by $i$. By substituting Eqs.~(\ref{eq:distr_x_parametrize}) and (\ref{eq:distr_y_parametrize}) into the above-obtained single integral equation, we obtain

\begin{align}
1 
=& \frac{\tau^{(l)}_{\text{e-e}}}{\tau^{(l)}_{\text{imp}}} \cosh \left( \frac{x_1}{2} \right)\varphi_{\sigma}^{(l)}(x_1) + i \eta_{l} \frac{eB}{m_{l}} \tau^{(l)}_{\text{e-e}} \cosh \left( \frac{x_1}{2} \right)\varphi_{\sigma}^{(l)}(x_1)\nonumber \\
&+ \sum_{l' = 1,2} \frac{1}{\tau^{(ll')}_{\text{e-e}}} \Bigg[ \frac{\pi^2 + x_1^2}{2} \cosh \left( \frac{x_1}{2} \right) \tau^{(l)}_{\text{e-e}} \varphi_{\sigma}^{(l)}(x_1) \nonumber \\
&- \int_{-\infty}^{\infty}\mathcal{F}(x_1,u)\cosh \left( \frac{u}{2} \right)  \Bigg( - \frac{\eta_{l'}v_{\text{F},l'}}{\eta_{l}v_{\text{F},l}} \Lambda_{2}^{(ll')} \tau^{(l')}_{\text{e-e}} \varphi_{\sigma}^{(l')}(u) + \Lambda_{3}^{(ll')} \tau^{(l)}_{\text{e-e}} \varphi_{\sigma}^{(l)}(u)  + \frac{\eta_{l'}v_{\text{F},l'}}{\eta_{l} v_{\text{F},l}} \Lambda_{4}^{(ll')} \tau^{(l')}_{\text{e-e}} \varphi_{\sigma}^{(l')}(u) \Bigg) du \Bigg], \label{eq:el_int_eq_unified} \\
x_1
=& \frac{\tau^{(l)}_{\text{e-e}}}{\tau^{(l)}_{\text{imp}}} \cosh \left( \frac{x_1}{2} \right)\varphi_{\kappa}^{(l)}(x_1) + i \eta_{l} \frac{eB}{m_{l}} \tau^{(l)}_{\text{e-e}} \cosh \left( \frac{x_1}{2} \right)\varphi_{\kappa}^{(l)}(x_1)\nonumber \\
&+ \sum_{l' = 1,2} \frac{1}{\tau^{(ll')}_{\text{e-e}}} \Bigg[ \frac{\pi^2 + x_1^2}{2} \cosh \left( \frac{x_1}{2} \right) \tau^{(l)}_{\text{e-e}} \varphi_{\kappa}^{(l)}(x_1) \nonumber \\
&- \int_{-\infty}^{\infty}\mathcal{F}(x_1,u)\cosh \left( \frac{u}{2} \right)  \Bigg( \frac{\eta_{l'}v_{\text{F},l'}}{\eta_{l}v_{\text{F},l}} \Lambda_{2}^{(ll')} \tau^{(l')}_{\text{e-e}} \varphi_{\kappa}^{(l')}(u) + \Lambda_{3}^{(ll')} \tau^{(l)}_{\text{e-e}} \varphi_{\kappa}^{(l)}(u)  + \frac{\eta_{l'}v_{\text{F},l'}}{\eta_{l}v_{\text{F},l}} \Lambda_{4}^{(ll')} \tau^{(l')}_{\text{e-e}} \varphi_{\kappa}^{(l')}(u) \Bigg) du \Bigg] \label{eq:th_int_eq_unified}. 
\end{align}
\end{widetext}
By arranging terms, we obtain Eqs.~(\ref{eq:int_eq_electrons}) and (\ref{eq:int_eq_holes}).

\subsection{Calculation of angular integrals} 
To explicitly calculate the transport coefficients, we have to relate the scattering potential to parameters $\lambda^{(l)}_{X}$ and $\beta^{(l)}_{X}$ through $\Lambda_{i}^{(ll')}$. The calculations involve the angular integral,
\begin{align}
\int \frac{d\Omega}{4\pi} \frac{A(\theta,\varphi)}{R^{(ll')}(\theta)},
\end{align}
where $A$ is some function. If $A(\theta,\varphi)$ is a function of $q = |\bm{k} - \bm{k}_3| = k_{\text{F},l} \sin \theta \sin(\varphi/2)/ R^{(ll')}(\theta)$, we can simplify the angular integral as 
\begin{align}
\int \frac{d\Omega}{4\pi} \frac{A(q)}{R^{(ll')}(\theta)} =& \frac{1}{k_{\text{F},l}}\int_{0}^{2q^{(ll')}_{\text{F}}} dq A(q), \\
\int \frac{d\Omega}{4\pi} \frac{A(q)}{R^{(ll')}(\theta)} \cos \theta =& - \frac{1}{k_{\text{F},l}}\int_{0}^{2q^{(ll')}_{\text{F}}} dq A(q) \frac{q^2}{4k_{\text{F},l}k_{\text{F},l'}}, 
\end{align}
where $q^{(ll')}_{\text{F}} = \min\{k_{\text{F},l},k_{\text{F},l'}\}$.
Since $\cos \theta_{13} = 1 - q^2/2k_{\text{F},l}^2$, we obtain
\begin{align}
\int \frac{d\Omega}{4\pi} \frac{A(q)}{R^{(ll')}(\theta)} \cos \theta_{13} = \frac{1}{k_{\text{F},l}}\int_{0}^{2q^{(ll')}_{\text{F}}} dq A(q) \left( 1 - \frac{q^2}{2k_{\text{F},l}^2}\right).
\end{align}
The momentum conservation leads to $\cos \theta_{14} = 1 + \cos \theta_{12} - \cos \theta_{13}$ and
\begin{align}
\int \frac{d\Omega}{4\pi} \frac{A(q)}{R^{(ll')}(\theta)} \cos \theta_{14} = \frac{1}{k_{\text{F},l}}\int_{0}^{2q^{(ll')}_{\text{F}}} dq A(q) \frac{q^2}{4k_{\text{F},l}k_{\text{F},l'}}.
\end{align}
Using these integrals, we can calculate the relaxation time $\tau^{(ll')}_{\text{e-e}}$ and the parameter $\Lambda_{i}^{(ll')}$. For the screened Coulomb potential Eq.~(\ref{eq:screend_Coulomb}), $1/\tau^{(ll')}_{\text{e-e}}$ and $\Lambda_{i}^{(ll')}$ are given by
\begin{align}
\frac{1}{\tau^{(ll')}_{\text{e-e}}} =& \frac{m_{l}m_{l'}^2(k_BT)^2}{8\pi^4\hbar^6} \cdot \frac{2\pi}{\hbar} \left(\frac{e}{\varepsilon_{0}}\right)^2 \frac{1}{k_{\text{F},l}\alpha^3} \mathcal{I}_0(2q^{(ll')}_{\text{F}}/\alpha), \\
\Lambda_{2}^{(ll')} =& - \Lambda_{4}^{(ll')} = - \frac{\alpha^2}{4k_{\text{F},l}k_{\text{F},l'}} \cdot \frac{\mathcal{I}_2(2q^{(ll')}_{\text{F}}/\alpha)}{\mathcal{I}_0(2q^{(ll')}_{\text{F}}/\alpha)}, \\
\Lambda_{3}^{(ll')} =& 1 - \frac{\alpha^2}{2k_{\text{F},l}^2} \cdot \frac{\mathcal{I}_2(2q^{(ll')}_{\text{F}}/\alpha)}{\mathcal{I}_0(2q^{(ll')}_{\text{F}}/\alpha)},
\end{align}
where $\mathcal{I}_0(x)$ and $\mathcal{I}_2(x)$ are given by
\begin{align}
\mathcal{I}_0(x) =& \int_{0}^{x} \frac{1}{(z^2 + 1)^2} dz = \frac{1}{2}\left[\frac{x}{x^2 + 1} + \arctan(x) \right],\\
\mathcal{I}_2(x) =& \int_{0}^{x} \frac{z^2}{(z^2 + 1)^2} dz = \frac{1}{2}\left[-\frac{x}{x^2 + 1} + \arctan(x) \right]. 
\end{align}
In the limit of strong screening $2q^{(ll')}_{\text{F}}/\alpha \to 0$, $\Lambda_{i}^{(ll')}$ reaches the value of the case of the short-ranged potential, $W^{(ll')} = \text{const}$. In particular, $\Lambda_{2}^{(ll')} \to - (q^{(ll')}_{\text{F}})^2/3k_{\text{F},l}k_{\text{F},l'}$ and $\Lambda_{3}^{(ll')} \to 1/3$. Note that $\beta^{(l)}_{\kappa} = 0$ for the potential being a function of $q$ as in this paper.

\section{Detailed analysis of solutions} \label{App:solution_properties}
In this Appendix, we focus on analyzing the integral equations Eqs.~(\ref{eq:int_eq_electrons}),~(\ref{eq:int_eq_holes}), and (\ref{eq:single_int_eq}). We derive the conductivities of Baber scattering Eqs.~(\ref{eq:single_el_cond_ser}) and (\ref{eq:single_th_cond_ser}). 

\subsection{Derivation of formulae of Baber scattering}
First, we consider the electrical conductivity. Substituting Eq.~(\ref{eq:single_el_expansion_ser}) into Eq.~(\ref{eq:single_int_eq}), we get
\begin{align}
0 =& (\zeta^2\pi^2 + x^2) \sum_{n = 0}^{\infty} d_{2n} \varphi_{2n;\zeta}(x) - 2\lambda_{\sigma} \int_{-\infty}^{\infty}  \mathcal{G}(x-u)\nonumber \\
& \times \left[ \frac{F_{\sigma}(u)}{\zeta^2\pi^2 + u^2} + \sum_{n = 0}^{\infty} d_{2n} \varphi_{2n;\zeta}(u) \right] du. 
\end{align}
By multiplying $\varphi_{2m;\zeta}(-x)$ and integrating with respect to $x$, we obtain
\begin{align}
&2(\lambda_{2m}(\zeta) - \lambda_{\sigma}) d_{2m} \int_{-\infty}^{\infty}  \varphi_{2m;\zeta}(-x)\mathcal{G}(x-u)\varphi_{2m;\zeta}(u) dxdu \nonumber \\
&=\frac{\lambda_{\sigma}}{\lambda_{2m}(\zeta)} \int_{-\infty}^{\infty}  \varphi_{2m;\zeta}(u) F_{\sigma}(u) du.
\end{align}
Using the Fourier transformation of Eq.~(\ref{eq:eigen_orthogonal_k}) and
\begin{align}
&\int_{-\infty}^{\infty} \psi_{2n;\zeta}(k) \frac{1}{\cosh(\pi k)} dk \nonumber \\
&= \frac{1}{\pi}\cdot \frac{\Gamma(\zeta + 1)\Gamma(n+1/2)\Gamma(n + (\zeta + 1)/2)}{\Gamma(n + \zeta/2 + 1)\Gamma(n + \zeta + 1) }, \label{eq:int_psi_2n_sech} 
\end{align}
we find $d_{2n}$ [Eq.~(\ref{eq:d_2n})].

Then, substituting the energy-dependent parts of the distribution function into Eq.~(\ref{eq:el_cond_integral_single}) and using the integral,
\begin{align}
\int_{-\infty}^{\infty}  \frac{1}{\cosh^2(x/2)(x^2 + \zeta^2 \pi^2)} dx = \frac{2}{\zeta\pi^2} \psi^{(1)}\left(\frac{1 + \zeta}{2}\right), 
\end{align}
we obtain Eq.~(\ref{eq:single_el_cond_ser}).

In a similar way to the electrical conductivity, the calculation of the thermal conductivity can be carried out where we use the following integral:
\begin{align}
&\int_{-\infty}^{\infty} \psi_{2n + 1;\zeta}(k) \frac{\tanh(\pi k) }{\cosh(\pi k)} dk \nonumber \\
&= \frac{1}{\pi}\cdot \frac{\Gamma(\zeta+ 1)\Gamma(n+3/2)\Gamma(n + (\zeta+ 1)/2)}{\Gamma(n + \zeta/2 + 2)\Gamma(n + \zeta+ 1) }, \label{eq:int_psi_2n+1_sechtanh}
\end{align}
instead of Eq.~(\ref{eq:int_psi_2n_sech}). Then, we obtain Eqs.~(\ref{eq:d_2n+1}) and (\ref{eq:single_th_cond_ser}).
\subsection{Asymptotic behavior}
We analyze the integral equation for Baber scattering by treating the electron-electron scatterings perturbatively. We will show Eqs.~(\ref{eq:Baber_magnetoresistance_rho}) and (\ref{eq:Baber_magnetoresistance_WT}) when the electron-electron scatterings are weak compared with the magnetic field and Eqs.~(\ref{eq:single_Lorenz_asym})-(\ref{eq:single_Lorenz_asym_const}) when the electron-electron scatterings and the magnetic field are weak compared with the impurity scattering.

We start from another form of Eq.~(\ref{eq:single_int_eq}),
\begin{align}
F_{X}(x) =& 2\tau_{\text{e-e}}\left(\frac{1}{\tau_{\text{imp}}} - i\omega_{\text{c}} \right)\varphi_{X}(x) + (\pi^2 + x^2) \varphi_{X}(x) \nonumber \\
&
- 2\lambda_{X} \int_{-\infty}^{\infty}  \mathcal{G}(x - u) \varphi_{X}(u) du,
\end{align}
where the last two terms on the right-hand side, coming from the electron-electron scatterings, are treated perturbatively.
We now expand the energy-dependent part of the distribution function as
\begin{align}
\varphi_{X}(x) = \varphi_{X,0}(x) + \varphi_{X,1}(x) + \cdots.
\end{align}
Then, we can determine $\varphi_{X,0}(x)$ and $\varphi_{X,1}(x)$ as 
\begin{align}
\varphi_{X,0}(x)  =& \frac{1}{2\tau_{\text{e-e}}} \left(\frac{1}{\tau_{\text{imp}}} - i\omega_{\text{c}} \right)^{-1} F_{X}(x), \label{eq:single_dist_perturb_el_0th} \\
\varphi_{\sigma,1}(x) =& - \frac{1}{4\tau_{\text{e-e}}^2} \left(\frac{1}{\tau_{\text{imp}}} - i\omega_{\text{c}} \right)^{-2} \nonumber \\
& \times (1 - \lambda_{\sigma}) (\pi^2 + x^2) F_{\sigma}(x), \label{eq:single_dist_perturb_el_1st} \\
\varphi_{\kappa,1}(x) =& - \frac{1}{12\tau_{\text{e-e}}^2} \left(\frac{1}{\tau_{\text{imp}}} - i\omega_{\text{c}} \right)^{-2} \nonumber \\
& \times (3 - \lambda_{\kappa}) (\pi^2 + x^2) F_{\kappa}(x). \label{eq:single_dist_perturb_th_1st}
\end{align}
From these, the energy-dependent parts of the distribution functions, we find 

\begin{widetext}
\begin{align}
\sigma_{xx} + i\sigma_{yx} =& \frac{e^2n}{m}\left[ \left(\frac{1}{\tau_{\text{imp}}} - i\omega_{\text{c}} \right)^{-1} - \frac{2\pi^2}{3}(1 - \lambda_{\sigma})\frac{1}{\tau_{\text{e-e}}}\left(\frac{1}{\tau_{\text{imp}}} - i\omega_{\text{c}} \right)^{-2} + \cdots \right],\\
\kappa_{xx} + i\kappa_{yx} =& \frac{k_B^2Tn}{m}\left[ \frac{\pi^2}{3}\left(\frac{1}{\tau_{\text{imp}}} - i\omega_{\text{c}} \right)^{-1} - \frac{2\pi^4}{15}(3 - \lambda_{\kappa})\frac{1}{\tau_{\text{e-e}}}\left(\frac{1}{\tau_{\text{imp}}} - i\omega_{\text{c}} \right)^{-2} + \cdots \right].
\end{align}
From these conductivities, we obtain Eqs.~(\ref{eq:Baber_magnetoresistance_rho}) and (\ref{eq:Baber_magnetoresistance_WT}) by setting $1/\tau_{\text{imp}} = 0$ and $\tau_{\text{e-e}} \ll \omega_{\text{c}}$. Higher order terms are $O(\omega_{c}^{-3})$ and can be neglected. We obtain Eqs.~(\ref{eq:single_Lorenz_asym})-(\ref{eq:single_Lorenz_asym_const}) with $1/\tau_{\text{e-e}}, \omega_{\text{c}} \ll 1/\tau_{\text{imp}}$. We see that, in a weak magnetic field, corrections by $\varphi_{X,1}(x)$ to the transverse responses are twice as large as those to the longitudinal responses. This is essential in Eq.~(\ref{eq:single_Hall_Lorenz_asym}). 

\subsection{Calculation of the variational method}
Here, we derive linear equations mapped from the integral equation for the two-band semimetals on the basis of the variational method.
\subsubsection{Electrical transport}
Substituting the expansion of the energy-dependent parts of the distribution function Eq.~(\ref{eq:semimetal_el_expansion}) into Eq.~(\ref{eq:int_eq_electrons}), we obtain
\begin{align}
F_{\sigma}(x) =& 2\tau^{(1)}_{\text{e-e}} \left( \frac{1}{\tau^{(1)}_{\text{imp}}} - i\omega^{(1)}_{\text{c}} \right) \sum_{n = 0}^{N - 1} c^{(1)}_{2n} \varphi_{2n;\zeta = 1}(x)+ \sum_{n = 0}^{N - 1} 2(\lambda_{2n}(1) - \lambda^{(1)}_{\sigma}) c^{(1)}_{2n} \int_{-\infty}^{\infty} \mathcal{G}(x - u) \varphi_{2n;\zeta = 1}(u) du \nonumber \\ 
& + \sum_{n = 0}^{N - 1} \frac{ 2\tau_{\text{e-e}}^{(2)}}{ \tau_{\text{e-e}}^{(1)}}\beta^{(1)}_{\sigma} c^{(2)}_{2n} \int_{-\infty}^{\infty} \mathcal{G}(x - u) \varphi_{2n;\zeta = 1}(u) du.
\end{align}
Then, we perform the Fourier transformation as
\begin{align}
\int_{-\infty}^{\infty}  F_{\sigma}(x) e^{-ikx} dx =& 2\tau^{(1)}_{\text{e-e}}\left( \frac{1}{\tau^{(1)}_{\text{imp}}} - i\omega^{(1)}_{\text{c}} \right) \sum_{n = 0}^{N - 1} c^{(1)}_{2n} \psi_{2n;\zeta = 1}(k) \nonumber \\ 
& + \sum_{n = 0}^{N - 1} 2\left[ (\lambda_{2n}(1) - \lambda^{(1)}_{\sigma}) c^{(1)}_{2n} + \frac{\tau_{\text{e-e}}^{(2)}}{ \tau_{\text{e-e}}^{(1)}}\beta^{(1)}_{\sigma} c^{(2)}_{2n} \right] \int_{-\infty}^{\infty} \mathcal{G}(x - u) e^{-ikx} \varphi_{2n;\zeta = 1}(u) du dx.
\end{align}
Using 
\begin{align}
\int_{-\infty}^{\infty}  F_{\sigma}(x) e^{-ikx} dx = \int_{-\infty}^{\infty}  \frac{2}{\cosh(x/2)} e^{-ikx} dx = 4\pi \text{sech} (\pi k),
\end{align}
and
\begin{align}
\int_{-\infty}^{\infty} \mathcal{G}(x - u) e^{-ikx} dx = e^{-iku} \int_{-\infty}^{\infty}  \frac{x - u}{2 \sinh[(x - u)/2]} e^{-ik(x - u)} dx = e^{-iku} \cdot \pi^2 \text{sech}^2 (\pi k),
\end{align}
we obtain
\begin{align}
4\pi \text{sech} (\pi k) =& 2\tau^{(1)}_{\text{e-e}}\left( \frac{1}{\tau^{(1)}_{\text{imp}}} - i\omega^{(1)}_{\text{c}} \right) \sum_{n = 0}^{N - 1} c^{(1)}_{2n} \psi_{2n;\zeta = 1}(k) \nonumber \\ 
&+ \sum_{n = 0}^{N - 1} 2\left[ (\lambda_{2n}(1) - \lambda^{(1)}_{\sigma}) c^{(1)}_{2n} + \frac{\tau_{\text{e-e}}^{(2)}}{ \tau_{\text{e-e}}^{(1)}}\beta^{(1)}_{\sigma} c^{(2)}_{2n} \right]\pi^2 \text{sech}^2 (\pi k) \psi_{2n;\zeta = 1}(k). \label{eq:semimetal_var_k}
\end{align}
As $\psi_{n;\zeta = 1}(k) = - 2/(n+1)(n+2) \cdot P_{n + 1}^{1}(\tanh(\pi k))$, we introduce $\widetilde{c}^{(l)}_{2n}$ instead of $c^{(l)}_{2n}$ as
\begin{align}
\widetilde{c}^{(l)}_{2n} = - \frac{1}{(2n+1)(n+1)} c^{(l)}_{2n},
\end{align}
where we have
\begin{align}
\sum_{n = 0}^{N - 1} c^{(l)}_{2n} \psi_{2n;\zeta = 1}(k) = \sum_{n = 0}^{N - 1} \widetilde{c}^{(l)}_{2n} P_{2n+1}^{1}(\tanh(\pi k)).
\end{align}
Then, multiplying Eq.~(\ref{eq:semimetal_var_k}) by $P_{2m+1}^{1}(\tanh(\pi k))$ and integrating with respect to $k$, we obtain the equation for electrons as 
\begin{align}
- 8 =& 2\tau^{(1)}_{\text{e-e}}\left( \frac{1}{\tau^{(1)}_{\text{imp}}} - i\omega^{(1)}_{\text{c}} \right) \sum_{n = 0}^{N - 1} \frac{1}{\pi} \min\{2m + 1,2n + 1\}( \min\{2m + 1,2n + 1\} + 1) \widetilde{c}^{(1)}_{2n} \nonumber \\
& + \frac{8\pi (2m + 1)(m + 1)}{4m + 3} \left[ (\lambda_{2m}(1) - \lambda^{(1)}_{\sigma}) \widetilde{c}^{(1)}_{2m} + \frac{\tau_{\text{e-e}}^{(2)}}{ \tau_{\text{e-e}}^{(1)}}\beta^{(1)}_{\sigma} \widetilde{c}^{(2)}_{2m} \right] .
\end{align}
The equation for holes is obtained by exchanging band indices and replacing $i\omega^{(1)}_{\text{c}}$ with $-i\omega^{(2)}_{\text{c}}$. From $\widetilde{c}^{(l)}_{2n}$, we can calculate the transport coefficient as 
\begin{align}
\sigma_{xx} + i\sigma_{yx} 
=& \sum_{l = 1,2}\frac{e^2 n_{l} \tau^{(l)}_{\text{e-e}}}{m_{l}} \int_{-\infty}^{\infty}  \frac{1}{4\cosh(u/2)} \varphi^{(l)}_{\sigma}(u) du \nonumber \\
=& \sum_{l = 1,2}\frac{e^2 n_{l} \tau^{(l)}_{\text{e-e}}}{m_{l}} \int_{-\infty}^{\infty}  \frac{1}{4\cosh(\pi k)} \sum_{n = 0}^{N - 1} c^{(l)}_{2n} \psi_{2n;\zeta = 1}(k) dk
\nonumber \\
=& \sum_{l = 1,2}\frac{e^2 n_{l} \tau^{(l)}_{\text{e-e}}}{m_{l}} \cdot \frac{1}{\pi}\int_{-1}^{1} \frac{1}{4\sqrt{1 - \xi^2}} \sum_{n = 0}^{N - 1} \widetilde{c}^{(l)}_{2n} P_{2n + 1}^{1}(\xi) d\xi = \sum_{l = 1,2} \frac{e^2 n_{l} \tau^{(l)}_{\text{e-e}}}{m_{l}} \cdot \left( - \frac{1}{2\pi} \sum_{n = 0}^{N - 1} \widetilde{c}^{(l)}_{2n}\right).
\end{align}
Following integrals related to the associated Legendre polynomials $P_{n}^{1}$ are used:
\begin{align}
\int_{-1}^{1} P_{n}^{1}(\xi) P_{m}^{1}(\xi) d\xi =& \frac{2n(n + 1)}{2n + 1}\delta_{n,m}, \\
\int_{-1}^{1} \frac{P_{n}^{1}(\xi) P_{m}^{1}(\xi)}{1 - \xi^2} d\xi =& \min \{n, m\} (\min \{n, m\} + 1)~(\text{for}~n - m = even), \\
\int_{-1}^{1} \frac{P_{2n + 1}^{1}(\xi)}{\sqrt{1 - \xi^2}} d\xi =& -2.
\end{align}

\subsubsection{Thermal transport}
Next, we consider the equation for thermal transport. The calculation is almost parallel with electrical transport. Substituting Eq.~(\ref{eq:semimetal_th_expansion}) into Eq.~(\ref{eq:int_eq_electrons}), using 
\begin{align}
\int_{-\infty}^{\infty}  F_{\kappa}(x) e^{-ikx} dx = \int_{-\infty}^{\infty}  \frac{2x}{\cosh(x/2)} e^{-ikx} dx = - 4i\pi^2 \tanh(\pi k) \text{sech} (\pi k),
\end{align}
and defining
\begin{align}
i\widetilde{c}_{2n + 1} = - \frac{1}{(n+1)(2n+3)} c_{2n + 1},
\end{align}
we obtain
\begin{align}
- 4i\pi^2 \tanh(\pi k) \text{sech} (\pi k) =& 2\tau^{(1)}_{\text{e-e}} \left( \frac{1}{\tau^{(1)}_{\text{imp}}} - i\omega^{(1)}_{\text{c}} \right) \sum_{n = 0}^{N - 1} i\widetilde{c}^{(1)}_{2n + 1} P_{2n + 2}^{1}(\tanh(\pi k)) \nonumber \\
& + \sum_{n = 0}^{N - 1} 2\pi^2 \left[ (\lambda_{2n + 1}(1) - \lambda^{(1)}_{\kappa}) i\widetilde{c}^{(1)}_{2n + 1} + \frac{ \tau_{\text{e-e}}^{(2)}}{ \tau_{\text{e-e}}^{(1)}}\beta^{(1)}_{\kappa} i\widetilde{c}^{(2)}_{2n + 1} \right]\text{sech}^2 (\pi k) P_{2n + 2}^{1}(\tanh(\pi k)).  
\end{align}
By multiplying by $P_{2m + 2}^{1}(\tanh(\pi k))$ and integrating with respect to $k$, we get the equation
\begin{align}
8 \pi =& 2\tau^{(1)}_{\text{e-e}} \left( \frac{1}{\tau^{(1)}_{\text{imp}}} - i\omega^{(1)}_{\text{c}} \right) \sum_{n = 0}^{N - 1} \frac{1}{\pi} \min\{2m + 2,2n + 2\}( \min\{2m + 2,2n + 2\} + 1) \widetilde{c}^{(1)}_{2n + 1} \nonumber \\
& + \frac{8\pi (2m + 3)(m + 1)}{4m + 5} \left[ (\lambda_{2m + 1}(1) - \lambda^{(1)}_{\kappa}) \widetilde{c}^{(1)}_{2m + 1} + \frac{ \tau_{\text{e-e}}^{(2)}}{ \tau_{\text{e-e}}^{(1)}}\beta^{(1)}_{\kappa} \widetilde{c}^{(2)}_{2m + 1} \right],
\end{align}
where we used
\begin{align}
\int_{-1}^{1} \frac{\xi P_{2n}^{1}(\xi)}{\sqrt{1 - \xi^2}} d\xi =& -2.
\end{align}
Another equation is obtained by exchanging band indices and replacing $i\omega^{(1)}_{\text{c}}$ with $-i\omega^{(2)}_{\text{c}}$. Finally, the thermal conductivity is given by
\begin{align}
\kappa_{xx} + i\kappa_{yx} 
=& \sum_{l = 1,2} \frac{k_B^2Tn_{l} \tau^{(l)}_{\text{e-e}}}{m_{l}} \int_{-\infty}^{\infty}  \frac{u}{4\cosh(u/2)} \varphi^{(l)}_{\kappa}(u) du \nonumber \\
=& \sum_{l = 1,2} \frac{k_B^2Tn_{l} \tau^{(l)}_{\text{e-e}}}{m_{l}} \cdot \frac{1}{\pi}\int_{-1}^{1} \frac{i\pi \xi}{4\sqrt{1 - \xi^2}} \sum_{n = 0}^{N - 1} i\widetilde{c}^{(l)}_{2n} P_{2n + 2}^{1}(\xi) d\xi = \sum_{l = 1,2} \frac{k_B^2Tn_{l} \tau^{(l)}_{\text{e-e}}}{m_{l}} \cdot \left( \frac{1}{2}\sum_{n = 0}^{N - 1} \widetilde{c}_{2n + 1}\right).
\end{align}
\end{widetext}

\subsection{Convergence}
We discuss the convergence of solutions to check the accuracy.
\subsubsection{Baber scattering}
First, we consider Baber scattering. We test not only the solutions of Eqs.~(\ref{eq:single_el_cond_ser}) and (\ref{eq:single_th_cond_ser}), but also solutions obtained by the variational method by the expansions,
\begin{align}
\varphi_{\sigma}(x) =& \sum_{n = 0}^{N - 1} c_{2n} \varphi_{2n;\zeta = 1}(x), \label{eq:single_el_expansion_var} \\
\varphi_{\kappa}(x) =& \sum_{n = 0}^{N - 1} c_{2n + 1} \varphi_{2n + 1;\zeta = 1}(x), \label{eq:single_th_expansion_var} 
\end{align}
which are the single-carrier versions of Eqs.~(\ref{eq:semimetal_el_expansion}) and (\ref{eq:semimetal_th_expansion}) for comparison.

\begin{figure}[tbp]
\begin{center}
\rotatebox{0}{\includegraphics[angle=0,width=1\linewidth]{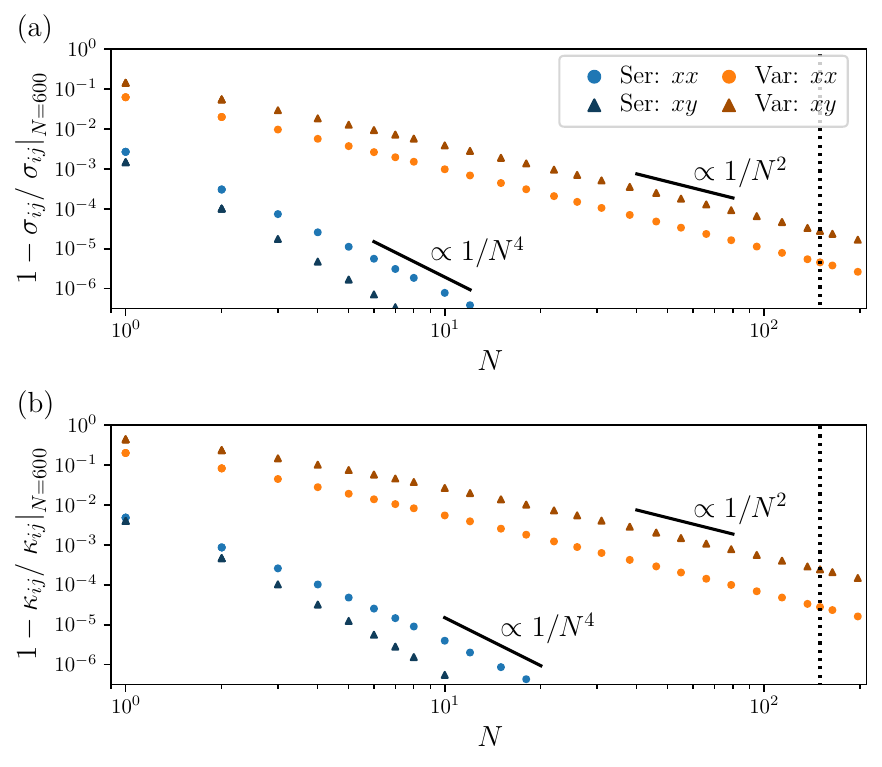}}
\caption{Relative errors of (a) the electrical conductivity and (b) the thermal conductivity in the case of Baber scattering to values evaluated with formulae Eqs.~(\ref{eq:single_el_cond_ser}) and (\ref{eq:single_th_cond_ser}) summing up to $N = 600$. Blue plots are calculated using the formulae in the form of series. Orange plots are calculated using the variational method. The black dotted lines indicate $N = 150$.}
\label{Fig:Baber_conv}
\end{center}
\end{figure}

In Fig.~\ref{Fig:Baber_conv}, we show the relative errors of (a) the electrical conductivity and (b) the thermal conductivity to those calculated by Eqs.~(\ref{eq:single_el_cond_ser}) and (\ref{eq:single_th_cond_ser}) with $N = 600$. Parameters are the same as in Fig.~\ref{Fig:Baber_Hall}. We set $T/T_{0} = \sqrt{\tau_{\text{imp}}/\tau_{\text{e-e}}} = 10$. Blue plots (labeled as "Ser") are calculated by summing the series Eqs.~(\ref{eq:single_el_cond_ser}) and (\ref{eq:single_th_cond_ser}) up to $N$. Orange plots (labeled as "Var") are calculated using Eqs.~(\ref{eq:single_el_expansion_var}) and (\ref{eq:single_th_expansion_var}) with $N$ trial functions.

Results by the series give rapid convergences almost proportional to $N^{-4}$. The solutions calculated by the variational method give convergence proportional to $N^{-2}$ toward the solutions by the series. This confirms that the two methods give the same result.

\subsubsection{Semimetals}
\begin{figure}[tbp]
\begin{center}
\rotatebox{0}{\includegraphics[angle=0,width=1\linewidth]{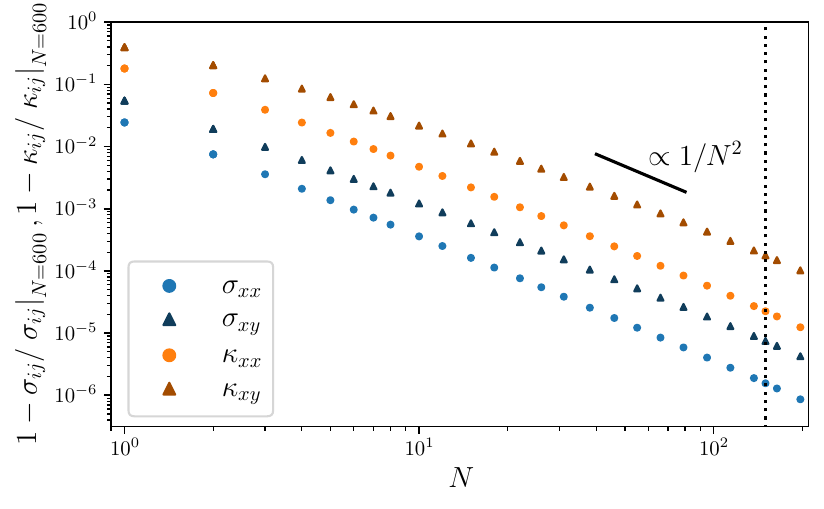}}
\caption{Relative errors of conductivities of the compensated semimetal to values evaluated with $N = 600$ trial functions. The black dotted line indicates $N = 150$.}
\label{Fig:semimetal_conv}
\end{center}
\end{figure}

We now consider the case of semimetals. In Fig.~\ref{Fig:semimetal_conv}, the relative errors of conductivities calculated with $N$ functions to those with $N = 600$. The parameters are the same as in Fig.~\ref{Fig:semimetal_Hall}(a) (the compensated semimetal) with $T/T^{(1)}_{0} = \sqrt{\tau^{(1)}_{\text{imp}}/\tau^{(1)}_{\text{e-e}}} = 10$ and $\mu_{\text{imp}}^{(2)}/\mu_{\text{imp}}^{(1)} = 0.5$. We can see that $N = 150$, which we used in the main text, gives sufficient convergence.

\section{Relaxation time approximation} \label{App:RTA}
In this Appendix, we derive the transport coefficients of the two-band system using the RTA while considering the momentum conservation of the electron-electron scatterings \cite{Gantmakher1978, Kukkonen1979, Entin2013, Gusev2018, Lee2020magneto, Lee2021, Olshanetsky2021}, and we provide electric, thermoelectric, and thermal transport coefficients in a magnetic field. We can solve the Boltzmann equation by introducing the RTA only for the electron-electron scattering while keeping other terms as they are.

The electron-electron scatterings conserve the total momentum, although they relax the relative movements of the carriers. Therefore, we can consider that the electron-electron scatterings relax carriers towards the system moving at the drift velocity $\bm{v}_{\text{d}}^{(l)}$, which is given by
\begin{align}
Q^{(l)}\bm{v}_{\text{d}}^{(l)} = e\eta^{(l)} n_{l}\bm{v}_{\text{d}}^{(l)}  = \frac{2}{V} \sum_{\bm{k}} e\bm{v}_{\bm{k}}^{(l)} \delta f^{(l)}(\bm{k}).
\end{align}
Then, we approximate the electron-electron scatterings for small $\bm{v}_{\text{d}}^{(l)}$ as follows \cite{Lee2021}:

\begin{align}
I^{(ll')}_{\text{e-e}}
&\simeq \frac{f^{(l)}(\bm{k}) - f_{0}(\varepsilon_{l,\bm{k}} - \bm{v}_{\text{d}}^{(l')} \cdot \hbar \bm{k})}{\widetilde{\tau}^{(ll')}_{\text{e-e}}}, 
\end{align}
where $\widetilde{\tau}^{(ll')}_{\text{e-e}} \propto T^{-2}$ is a relaxation time of electron-electron scattering in the RTA independent of $B$.
Furthermore, by expanding the scattering terms with respect to drift velocities up to the linear order, we obtain 
\begin{align}
I^{(ll')}_{\text{e-e}}
&\simeq \frac{1}{\widetilde{\tau}^{(ll')}_{\text{e-e}}} \left[ \delta f^{(l)}(\bm{k}) - \bm{v}_{\text{d}}^{(l')} \cdot \hbar \bm{k} \left( - \frac{\partial f_{0}(\varepsilon_{l,\bm{k}})}{\partial \varepsilon_{l,\bm{k}}} \right) \right],  
\end{align}

The momentum conservation requires a constraint on the relaxation times of the interband scattering as
\begin{align}
\frac{m_{1}n_{1}}{\widetilde{\tau}^{(12)}_{\text{e-e}}} = \frac{m_{2}n_{2}}{\widetilde{\tau}^{(21)}_{\text{e-e}}}.
\end{align}
With the approximation of the electron-electron scattering, the Boltzmann equation can be cast into the following form \cite{Lee2021}:
\begin{align}
&\left(e\widetilde{\bm{E}}^{(l)} + \xi_{l,\bm{k}}\left(-\frac{\nabla T}{T}\right) \right)\cdot \bm{v}^{(l)}_{\bm{k}} \left( -\frac{\partial f_{0}(\varepsilon_{l,\bm{k}})}{\partial \varepsilon_{l,\bm{k}}}\right) \nonumber \\
&= \frac{\delta f^{(l)}(\bm{k})}{\widetilde{\tau}^{(l)}} + M^{(l)}[\Phi], \label{eq:boltzmann_eq_RTA}
\end{align}
with 
\begin{align}
e\widetilde{\bm{E}}^{(1)} =& e\bm{E} + \frac{m_1 \bm{v}_{\text{d}}^{(1)}}{\widetilde{\tau}^{(11)}_{\text{e-e}}} + \frac{m_1 \bm{v}_{\text{d}}^{(2)}}{\widetilde{\tau}^{(12)}_{\text{e-e}}}, \\
e\widetilde{\bm{E}}^{(2)} =& e\bm{E} - \frac{m_2 \bm{v}_{\text{d}}^{(2)}}{\widetilde{\tau}^{(22)}_{\text{e-e}}} - \frac{m_2 \bm{v}_{\text{d}}^{(1)}}{\widetilde{\tau}^{(21)}_{\text{e-e}}}, \\
\frac{1}{\widetilde{\tau}^{(l)}} =& \frac{1}{\tau^{(l)}_{\text{imp}}} + \frac{1}{\widetilde{\tau}^{(l1)}_{\text{e-e}}} + \frac{1}{\widetilde{\tau}^{(l2)}_{\text{e-e}}}.
\end{align}

The solution of the Boltzmann equation, still including undetermined $\bm{v}_{\text{d}}^{(l)}$, is given by \cite{Ziman2001, Lee2020magneto}
\begin{widetext}
\begin{align}
\delta f^{(l)}(\bm{k}) = \frac{\widetilde{\tau}^{(l)}}{1 + (\omega_{\text{c}}^{(l)}\widetilde{\tau}^{(l)})^2} \left( - \frac{\partial f_{0}(\varepsilon_{l,\bm{k}})}{\partial \varepsilon_{l,\bm{k}}}\right) (v_{\bm{k};x}^{(l)}, v_{\bm{k};y}^{(l)}) \cdot \left( 
\begin{array}{cc}
1 & - \eta_{l} \omega_{\text{c}}^{(l)} \widetilde{\tau}^{(l)}\\
\eta_{l} \omega_{\text{c}}^{(l)} \widetilde{\tau}^{(l)} & 1\\
\end{array}
\right)\left[
e\left( 
\begin{array}{c}
\widetilde{E}_{x}^{(l)} \\
\widetilde{E}_{y}^{(l)} \\
\end{array}
\right) + \xi_{l,\bm{k}}
\left( 
\begin{array}{c}
- \nabla T / T \\
0 \\
\end{array}
\right) \right]. \label{eq:distr_RTA}
\end{align}
From this distribution function, we obtain an equation to determine the drift velocities $\bm{v}_{\text{d}}^{(l)}$ as
\begin{align}
&\left( 
\begin{array}{cccc}
m_1 \left(1/\tau^{(1)}_{\text{imp}} + 1/\widetilde{\tau}^{(12)}_{\text{e-e}} \right) & m_1/\widetilde{\tau}^{(12)}_{\text{e-e}} &m_1\omega^{(1)}_{\text{c}}&0\\
m_2/\widetilde{\tau}^{(21)}_{\text{e-e}} & m_2 \left(1/\tau^{(2)}_{\text{imp}} + 1/\widetilde{\tau}^{(21)}_{\text{e-e}} \right) & 0&-m_2\omega^{(2)}_{\text{c}}\\
-m_1\omega^{(1)}_{\text{c}}& 0 &m_1 \left(1/\tau^{(1)}_{\text{imp}} + 1/\widetilde{\tau}^{(12)}_{\text{e-e}} \right) & m_1/\widetilde{\tau}^{(12)}_{\text{e-e}} \\
0 &m_2\omega^{(2)}_{\text{c}}& m_2/\widetilde{\tau}^{(21)}_{\text{e-e}} & m_2 \left(1/\tau^{(2)}_{\text{imp}} + 1/\widetilde{\tau}^{(21)}_{\text{e-e}} \right) \\
\end{array} 
\right)
\left( 
\begin{array}{c}
v^{(1)}_{\text{d};x} \\
-v^{(2)}_{\text{d};x} \\
v^{(1)}_{\text{d};y} \\
-v^{(2)}_{\text{d};y} \\
\end{array}
\right) \nonumber \\
&=\left( 
\begin{array}{c}
eE + \braket{\xi_{1,\bm{k}}}_{1} \left(- \nabla T/T \right) \\
eE + \braket{\xi_{2,\bm{k}}}_{2} \left(- \nabla T/T \right) \\
0 \\
0 \\
\end{array}
\right), \label{eq:two-carrier_kinetic}
\end{align}
where $\braket{A_{\bm{k}}}_{l}$ for some function $A_{\bm{k}}$ is defined by 
\begin{align}
\braket{A_{\bm{k}}}_{l} = \frac{2m_{l}}{3n_{l}} \cdot \frac{1}{V}\sum_{\bm{k}} (\bm{v}^{(l)}_{\bm{k}})^2 A_{\bm{k}} \left( - \frac{\partial f_{0}(\varepsilon_{l,\bm{k}})}{\partial \varepsilon_{l,\bm{k}}} \right),
\end{align}
which satisfies $\braket{1}_{l} = 1$.
Eq.~(\ref{eq:two-carrier_kinetic}) is a two-carrier kinetic equation \cite{Gantmakher1978, Kukkonen1979, Entin2013, Gusev2018, Nguyen2020, Olshanetsky2021}. By projecting the linear equation onto the subspace of $v^{(1)}_{\text{d};x/y}$, the equation for Baber scattering is obtained. Then, the electrical conductivities for Baber scattering [Eqs.~(\ref{eq:single_el_RTA_xx}) and (\ref{eq:single_el_RTA_xy})] are derived with $1/\tau_{\text{tr},\sigma} = 1/\tau^{(1)}_{\text{imp}} + 1/\widetilde{\tau}^{(12)}_{\text{e-e}}$ where the intraband electron-electron scattering does not enter because of the momentum conservation. 

Using the solution, the electric current $\bm{j} = e(n_1 \bm{v}_{\text{d}}^{(1)} - n_2 \bm{v}_{\text{d}}^{(2)})$ and the transport coefficients can be calculated. Since our model enjoys isotropy for each band, it is convenient to introduce complex variables \cite{Ziman2001} by exploiting the following identification:
\begin{align}
\hat{L}_{ij} = 
\left( 
\begin{array}{cc}
L_{ij;xx} & L_{ij;xy}\\
L_{ij;yx} & L_{ij;yy}\\
\end{array}
\right) =
\left( 
\begin{array}{cc}
L_{ij;xx} & - L_{ij;yx}\\
L_{ij;yx} & L_{ij;xx}\\
\end{array}
\right) \leftrightarrow L_{ij;xx} + iL_{ij;yx}.
\end{align}
Then, $\hat{L}^{(\text{RTA})}_{11} = \hat{\sigma}^{(\text{RTA})}$ and $\hat{L}^{(\text{RTA})}_{12}$ are given by \cite{Gantmakher1978, Kukkonen1979, Entin2013, Gusev2018, Olshanetsky2021}
\begin{align}
\sigma^{(\text{RTA})}_{xx} + i\sigma^{(\text{RTA})}_{yx} =& \frac{e^2}{C}\left[\frac{n_1}{m_1}\left(\frac{1}{\tau_{\text{imp}}^{(2)}} + \frac{1}{\widetilde{\tau}^{(21)}_{\text{e-e}}} + i \omega_{\text{c}}^{(2)}\right) - \frac{n_1}{m_2 \widetilde{\tau}^{(12)}_{\text{e-e}}} + \frac{n_2}{m_2}\left(\frac{1}{\tau^{(1)}_{\text{imp}}} + \frac{1}{\widetilde{\tau}^{(12)}_{\text{e-e}}} - i \omega_{\text{c}}^{(1)}\right) - \frac{n_2}{m_1 \widetilde{\tau}^{(21)}_{\text{e-e}}}\right], \nonumber \\
& \label{eq:L_11_RTA} \\
L^{(\text{RTA})}_{12;xx} + iL^{(\text{RTA})}_{12;yx} =& \frac{e}{C}\Bigg[ \frac{n_1}{m_1}\left(\frac{1}{\tau_{\text{imp}}^{(2)}} + \frac{1}{\widetilde{\tau}^{(21)}_{\text{e-e}}} + i \omega_{\text{c}}^{(2)}\right)\braket{\xi_{1,\bm{k}}}_{1} - \frac{n_1 }{m_2 \widetilde{\tau}^{(12)}_{\text{e-e}}}\braket{\xi_{2,\bm{k}}}_{2} \nonumber \\
& + \frac{n_2}{m_2}\left(\frac{1}{\tau^{(1)}_{\text{imp}}} + \frac{1}{\widetilde{\tau}^{(12)}_{\text{e-e}}} - i \omega_{\text{c}}^{(1)}\right)\braket{\xi_{2,\bm{k}}}_{2} - \frac{n_2}{m_1 \widetilde{\tau}^{(21)}_{\text{e-e}}} \braket{\xi_{1,\bm{k}}}_{1}\Bigg], \label{eq:L_12_RTA}
\end{align}
where 
\begin{align}
C = \frac{1}{\tau^{(1)}_{\text{imp}}\tau_{\text{imp}}^{(2)}} + \frac{1}{\widetilde{\tau}^{(12)}_{\text{e-e}}\tau_{\text{imp}}^{(2)}} + \frac{1}{\widetilde{\tau}^{(21)}_{\text{e-e}}\tau^{(1)}_{\text{imp}}} + \omega_{\text{c}}^{(1)}\omega_{\text{c}}^{(2)} + i \left[ \omega_{\text{c}}^{(2)}\left(\frac{1}{\tau^{(1)}_{\text{imp}}} + \frac{1}{\widetilde{\tau}^{(12)}_{\text{e-e}}} \right) - \omega_{\text{c}}^{(1)}\left(\frac{1}{\tau_{\text{imp}}^{(2)}} + \frac{1}{\widetilde{\tau}^{(21)}_{\text{e-e}}}\right) \right].
\end{align}

Therefore, we obtain the resistivity and the Hall coefficient as \cite{Gantmakher1978, Kukkonen1979},
\begin{align}
\rho^{(\text{RTA})} =& \frac{1}{e^2} \cdot \Bigg[  \left(\frac{1}{\tau^{(1)}_{\text{imp}}\tau_{\text{imp}}^{(2)}} + \frac{1}{\widetilde{\tau}^{(12)}_{\text{e-e}}\tau_{\text{imp}}^{(2)}} + \frac{1}{\widetilde{\tau}^{(21)}_{\text{e-e}}\tau^{(1)}_{\text{imp}}}\right) \left( \frac{n_1}{m_1\tau^{(2)}_{\text{imp}}} + \frac{n_2}{m_2\tau^{(1)}_{\text{imp}}} + \left( \frac{n_1 - n_2}{m_1\widetilde{\tau}^{(21)}_{\text{e-e}}} - \frac{n_1 - n_2}{m_2\widetilde{\tau}^{(12)}_{\text{e-e}}} \right) \right) \nonumber \\
& + \frac{e^2B^2}{m_1m_2} \left( \frac{n_1}{m_2 \tau^{(1)}_{\text{imp}}} +  \frac{n_2}{m_1 \tau^{(2)}_{\text{imp}}} \right) \Bigg] \nonumber \\
&\times \left[ \left( \frac{n_1}{m_1\tau^{(2)}_{\text{imp}}} + \frac{n_2}{m_2\tau^{(1)}_{\text{imp}}} + \left( \frac{n_1 - n_2}{m_1\widetilde{\tau}^{(21)}_{\text{e-e}}} - \frac{n_1 - n_2}{m_2\widetilde{\tau}^{(12)}_{\text{e-e}}} \right) \right)^2 + \left(\frac{(n_1 - n_2)|e|B}{m_1m_2} \right)^2 \right]^{-1}, \\
R_{\text{H}}^{(\text{RTA})} =& \frac{1}{|e|} \cdot \Bigg[  \frac{n_2}{m_2^2 (\tau^{(1)}_{\text{imp}})^2} - \frac{n_1}{m_1^2 (\tau^{(2)}_{\text{imp}})^2} +  2(n_2 - n_1) \left( \frac{1}{m_2^2\tau^{(1)}_{\text{imp}} \widetilde{\tau}^{(12)}_{\text{e-e}}} + \frac{1}{m_1^2\tau^{(2)}_{\text{imp}} \widetilde{\tau}^{(21)}_{\text{e-e}}} \right) \nonumber \\
& + \left( \frac{1}{m_2\widetilde{\tau}^{(12)}_{\text{e-e}}}  - \frac{1}{m_1\widetilde{\tau}^{(21)}_{\text{e-e}}}  \right)^2 (n_2 - n_1) + \frac{e^2(n_2 - n_1) B^2}{m_1^2m_2^2} \Bigg] \nonumber \\
&\times \left[ \left( \frac{n_1}{m_1\tau^{(2)}_{\text{imp}}} + \frac{n_2}{m_2\tau^{(1)}_{\text{imp}}} + \left( \frac{n_1 - n_2}{m_1\widetilde{\tau}^{(21)}_{\text{e-e}}} - \frac{n_1 - n_2}{m_2\widetilde{\tau}^{(12)}_{\text{e-e}}} \right) \right)^2 + \left(\frac{(n_1 - n_2)|e|B}{m_1m_2} \right)^2 \right]^{-1}.
\end{align} 
From these, we get Eqs.~(\ref{eq:semimetal_rho_RTA})-(\ref{eq:semimetal_resistivity_limit_B}), and (\ref{eq:semimetal_Hall_RTA})-(\ref{eq:semimetal_Hall_uncompensated_strong_ee_RTA})

Next, we consider the thermal response. From Eq.~(\ref{eq:distr_RTA}), we obtain
\begin{align}
L^{(\text{RTA})}_{22;xx} + iL^{(\text{RTA})}_{22;yx} 
=& \frac{n_1 (\braket{\xi_{1,\bm{k}}^2}_{1} - \braket{\xi_{1,\bm{k}}}_{1}^2)}{m_1} \cdot \frac{1}{1/ \widetilde{\tau}^{(1)} - i \omega^{(1)}_{\text{c}}} + \frac{n_2 (\braket{\xi_{2,\bm{k}}^2}_{2} - \braket{\xi_{2,\bm{k}}}_{2}^2)}{m_2} \cdot \frac{1}{1/ \widetilde{\tau}^{(2)} + i \omega^{(2)}_{\text{c}}}\nonumber \\
&+ \frac{1}{m_1m_2C} \Bigg\{ \frac{n_1\braket{\xi_{1,\bm{k}}}_{1}}{m_1} \left[ m_1m_2\left(\frac{1}{\tau_{\text{imp}}^{(2)}} + \frac{1}{\widetilde{\tau}^{(21)}_{\text{e-e}}} + i\omega_{\text{c}}^{(2)} \right)\braket{\xi_{1,\bm{k}} }_{1}- \frac{m_1^2}{\widetilde{\tau}^{(12)}_{\text{e-e}}} \braket{\xi_{2,\bm{k}} }_{2} \right]\nonumber \\
&+ \frac{n_2\braket{\xi_{2,\bm{k}}}_{2}}{m_2} \left[ m_1m_2 \left(\frac{1}{\tau^{(1)}_{\text{imp}}} + \frac{1}{\widetilde{\tau}^{(12)}_{\text{e-e}}} - i \omega_{\text{c}}^{(1)} \right)\braket{\xi_{2,\bm{k}} }_{2} - \frac{m_2^2}{\widetilde{\tau}^{(21)}_{\text{e-e}}} \braket{\xi_{1,\bm{k}} }_{1} \right] 
\Bigg\}. \label{eq:L_22_RTA}
\end{align}
Using the identification, 
\begin{align}
&\hat{\kappa} = \frac{1}{T} \left[ \hat{L}_{22} - \hat{L}_{21}\hat{L}^{-1}_{11} \hat{L}_{12} \right] \nonumber \\
&\leftrightarrow 
\kappa_{xx} + i\kappa_{yx} = \frac{1}{T} \left[ L_{22;xx} + iL_{22;yx} - (L_{21;xx} + iL_{21;yx} ) (L_{11;xx} + iL_{11;yx})^{-1}(L_{12;xx} + iL_{12;yx} )\right],
\end{align}
the thermal conductivity is given by
\begin{align}
&\kappa^{(\text{RTA})}_{xx} + i\kappa^{(\text{RTA})}_{yx} \nonumber \\
=& \frac{1}{T} \left[ \frac{n_1 (\braket{\xi_{1,\bm{k}}^2}_{1} - \braket{\xi_{1,\bm{k}}}_{1}^2)}{m_1} \cdot \frac{1}{1/ \widetilde{\tau}^{(1)} - i \omega^{(1)}_{\text{c}}} + \frac{n_2 (\braket{\xi_{2,\bm{k}}^2}_{2} - \braket{\xi_{2,\bm{k}}}_{2}^2)}{m_2}\cdot \frac{1}{1/ \widetilde{\tau}^{(2)} + i \omega^{(2)}_{\text{c}}} \right] \nonumber \\
&+ \frac{n_1n_2}{Tm_1m_2} \left[ \braket{\xi_{1,\bm{k}}}_{1}- \braket{\xi_{2,\bm{k}}}_{2} \right]^2 \left[ 
 \left( \frac{1}{\tau^{(2)}_{\text{imp}}} + i \omega^{(2)}_{\text{c}} \right)\frac{n_1}{m_1} + \frac{n_2 - n_1}{m_2\widetilde{\tau}^{(12)}_{\text{e-e}}} + \frac{n_1 - n_2}{m_1\widetilde{\tau}^{(21)}_{\text{e-e}}} + \left( \frac{1}{\tau^{(1)}_{\text{imp}}} - i \omega^{(1)}_{\text{c}} \right)\frac{n_2}{m_2}
\right]^{-1}. \label{eq:semimetal_th_cond_RTA}
\end{align}
The first term is the sum of the thermal conductivity of a single-carrier system. The second term is interpreted as the ambipolar contribution as discussed in the absence of the magnetic field \cite{Lee2021, Takahashi2023}. In low temperatures, $\braket{\xi_{1,\bm{k}}^2}_{1}\simeq \pi^2(k_BT)^2/3$, $\braket{\xi_{2,\bm{k}}^2}_{2} \simeq \pi^2(k_BT)^2/3$, $\braket{\xi_{1,\bm{k}}}_{1} \simeq \pi^2 (k_BT)^2/2\varepsilon_{\text{F}}$, and $\braket{\xi_{2,\bm{k}}}_{2} \simeq - \pi^2(k_BT)^2/2(\Delta - \varepsilon_{\text{F}})$. Therefore, the ambipolar contribution is not leading order here and may be neglected. In the compensated system, the imaginary part of the second term in Eq.~(\ref{eq:semimetal_th_cond_RTA}) vanishes, indicating that the ambipolar contribution does not contribute to the thermal Hall effect within the RTA. From the first term in Eq.~(\ref{eq:semimetal_th_cond_RTA}), we see a correspondence $1/\tau^{(l)}_{\text{tr},\kappa} = 1/\tau^{(1)}_{\text{imp}} + 1/\widetilde{\tau}^{(l1)}_{\text{e-e}}+ 1/\widetilde{\tau}^{(l2)}_{\text{e-e}}$ and we obtain 
\begin{align} 
\kappa^{(\text{RTA})}_{xx} + i\kappa^{(\text{RTA})}_{yx} \simeq \sum_{l = 1,2} \frac{\pi^2k_B^2Tn_{l}}{3m_{l}} \left(\frac{1}{\tau^{(l)}_{\text{tr},\kappa}} - i \eta_{l} \omega_{\text{c}}^{(l)} \right)^{-1}, \label{eq:semimetal_thermal_conductiviy_RTA}
\end{align}
which is just the sum of the Drude formula. From this, we obtain Eqs.~(\ref{eq:semimetal_WT_RTA}) and (\ref{eq:semimetal_th_Hall_RTA}). Also, the thermal conductivities for Baber scattering [Eqs.~(\ref{eq:single_th_RTA_xx}) and (\ref{eq:single_th_RTA_xx})] are obtained by setting $\tau^{(1)}_{\text{tr},\kappa} \to \tau_{\text{tr},\kappa}$.
\end{widetext}

\bibliographystyle{apsrev4-2}
\bibliography{main}

\begin{thebibliography}{76}%
\makeatletter
\providecommand \@ifxundefined [1]{%
 \@ifx{#1\undefined}
}%
\providecommand \@ifnum [1]{%
 \ifnum #1\expandafter \@firstoftwo
 \else \expandafter \@secondoftwo
 \fi
}%
\providecommand \@ifx [1]{%
 \ifx #1\expandafter \@firstoftwo
 \else \expandafter \@secondoftwo
 \fi
}%
\providecommand \natexlab [1]{#1}%
\providecommand \enquote  [1]{``#1''}%
\providecommand \bibnamefont  [1]{#1}%
\providecommand \bibfnamefont [1]{#1}%
\providecommand \citenamefont [1]{#1}%
\providecommand \href@noop [0]{\@secondoftwo}%
\providecommand \href [0]{\begingroup \@sanitize@url \@href}%
\providecommand \@href[1]{\@@startlink{#1}\@@href}%
\providecommand \@@href[1]{\endgroup#1\@@endlink}%
\providecommand \@sanitize@url [0]{\catcode `\\12\catcode `\$12\catcode `\&12\catcode `\#12\catcode `\^12\catcode `\_12\catcode `\%12\relax}%
\providecommand \@@startlink[1]{}%
\providecommand \@@endlink[0]{}%
\providecommand \url  [0]{\begingroup\@sanitize@url \@url }%
\providecommand \@url [1]{\endgroup\@href {#1}{\urlprefix }}%
\providecommand \urlprefix  [0]{URL }%
\providecommand \Eprint [0]{\href }%
\providecommand \doibase [0]{https://doi.org/}%
\providecommand \selectlanguage [0]{\@gobble}%
\providecommand \bibinfo  [0]{\@secondoftwo}%
\providecommand \bibfield  [0]{\@secondoftwo}%
\providecommand \translation [1]{[#1]}%
\providecommand \BibitemOpen [0]{}%
\providecommand \bibitemStop [0]{}%
\providecommand \bibitemNoStop [0]{.\EOS\space}%
\providecommand \EOS [0]{\spacefactor3000\relax}%
\providecommand \BibitemShut  [1]{\csname bibitem#1\endcsname}%
\let\auto@bib@innerbib\@empty
\bibitem [{\citenamefont {Ziman}(1972)}]{Ziman1972}%
  \BibitemOpen
  \bibfield  {author} {\bibinfo {author} {\bibfnamefont {J.~M.}\ \bibnamefont {Ziman}},\ }\href@noop {} {\emph {\bibinfo {title} {{Principles of the Theory of Solids}}}}\ (\bibinfo  {publisher} {Cambridge University Press},\ \bibinfo {address} {Cambridge},\ \bibinfo {year} {1972})\BibitemShut {NoStop}%
\bibitem [{\citenamefont {Ashcroft}\ and\ \citenamefont {Mermin}(1976)}]{AshcroftMermin}%
  \BibitemOpen
  \bibfield  {author} {\bibinfo {author} {\bibfnamefont {N.~W.}\ \bibnamefont {Ashcroft}}\ and\ \bibinfo {author} {\bibfnamefont {N.~D.}\ \bibnamefont {Mermin}},\ }\href@noop {} {\emph {\bibinfo {title} {{Solid State Physics}}}}\ (\bibinfo  {publisher} {Holt, Rinehart and Winston},\ \bibinfo {address} {New York},\ \bibinfo {year} {1976})\BibitemShut {NoStop}%
\bibitem [{\citenamefont {Smith}\ and\ \citenamefont {Jensen}(1989)}]{Smith_Jensen1989}%
  \BibitemOpen
  \bibfield  {author} {\bibinfo {author} {\bibfnamefont {H.}~\bibnamefont {Smith}}\ and\ \bibinfo {author} {\bibfnamefont {H.~H.}\ \bibnamefont {Jensen}},\ }\href@noop {} {\emph {\bibinfo {title} {{Transport Phenomena}}}}\ (\bibinfo  {publisher} {Clarendon Press},\ \bibinfo {address} {Oxford},\ \bibinfo {year} {1989})\BibitemShut {NoStop}%
\bibitem [{\citenamefont {Ziman}(2001)}]{Ziman2001}%
  \BibitemOpen
  \bibfield  {author} {\bibinfo {author} {\bibfnamefont {J.~M.}\ \bibnamefont {Ziman}},\ }\href@noop {} {\emph {\bibinfo {title} {{Electrons and Phonons}}}}\ (\bibinfo  {publisher} {Oxford University Press},\ \bibinfo {address} {Oxford},\ \bibinfo {year} {2001})\BibitemShut {NoStop}%
\bibitem [{\citenamefont {Gooth}\ \emph {et~al.}(2018)\citenamefont {Gooth}, \citenamefont {Menges}, \citenamefont {Kumar}, \citenamefont {S{\"u}{\ss}}, \citenamefont {Shekhar}, \citenamefont {Sun}, \citenamefont {Drechsler}, \citenamefont {Zierold}, \citenamefont {Felser},\ and\ \citenamefont {Gotsmann}}]{Gooth2018}%
  \BibitemOpen
  \bibfield  {author} {\bibinfo {author} {\bibfnamefont {J.}~\bibnamefont {Gooth}}, \bibinfo {author} {\bibfnamefont {F.}~\bibnamefont {Menges}}, \bibinfo {author} {\bibfnamefont {N.}~\bibnamefont {Kumar}}, \bibinfo {author} {\bibfnamefont {V.}~\bibnamefont {S{\"u}{\ss}}}, \bibinfo {author} {\bibfnamefont {C.}~\bibnamefont {Shekhar}}, \bibinfo {author} {\bibfnamefont {Y.}~\bibnamefont {Sun}}, \bibinfo {author} {\bibfnamefont {U.}~\bibnamefont {Drechsler}}, \bibinfo {author} {\bibfnamefont {R.}~\bibnamefont {Zierold}}, \bibinfo {author} {\bibfnamefont {C.}~\bibnamefont {Felser}},\ and\ \bibinfo {author} {\bibfnamefont {B.}~\bibnamefont {Gotsmann}},\ }\href {https://doi.org/10.1038/s41467-018-06688-y} {\bibfield  {journal} {\bibinfo  {journal} {Nat. Commun.}\ }\textbf {\bibinfo {volume} {9}},\ \bibinfo {pages} {4093} (\bibinfo {year} {2018})}\BibitemShut {NoStop}%
\bibitem [{\citenamefont {Jaoui}\ \emph {et~al.}(2018)\citenamefont {Jaoui}, \citenamefont {Fauqu{\'e}}, \citenamefont {Rischau}, \citenamefont {Subedi}, \citenamefont {Fu}, \citenamefont {Gooth}, \citenamefont {Kumar}, \citenamefont {S{\"u}{\ss}}, \citenamefont {Maslov}, \citenamefont {Felser},\ and\ \citenamefont {Behnia}}]{Jaoui2018}%
  \BibitemOpen
  \bibfield  {author} {\bibinfo {author} {\bibfnamefont {A.}~\bibnamefont {Jaoui}}, \bibinfo {author} {\bibfnamefont {B.}~\bibnamefont {Fauqu{\'e}}}, \bibinfo {author} {\bibfnamefont {C.~W.}\ \bibnamefont {Rischau}}, \bibinfo {author} {\bibfnamefont {A.}~\bibnamefont {Subedi}}, \bibinfo {author} {\bibfnamefont {C.}~\bibnamefont {Fu}}, \bibinfo {author} {\bibfnamefont {J.}~\bibnamefont {Gooth}}, \bibinfo {author} {\bibfnamefont {N.}~\bibnamefont {Kumar}}, \bibinfo {author} {\bibfnamefont {V.}~\bibnamefont {S{\"u}{\ss}}}, \bibinfo {author} {\bibfnamefont {D.~L.}\ \bibnamefont {Maslov}}, \bibinfo {author} {\bibfnamefont {C.}~\bibnamefont {Felser}},\ and\ \bibinfo {author} {\bibfnamefont {K.}~\bibnamefont {Behnia}},\ }\href {https://doi.org/10.1038/s41535-018-0136-x} {\bibfield  {journal} {\bibinfo  {journal} {npj Quantum Mater.}\ }\textbf {\bibinfo {volume} {3}},\ \bibinfo {pages} {64} (\bibinfo {year} {2018})}\BibitemShut {NoStop}%
\bibitem [{\citenamefont {Paglione}\ \emph {et~al.}(2005)\citenamefont {Paglione}, \citenamefont {Tanatar}, \citenamefont {Hawthorn}, \citenamefont {Hill}, \citenamefont {Ronning}, \citenamefont {Sutherland}, \citenamefont {Taillefer}, \citenamefont {Petrovic},\ and\ \citenamefont {Canfield}}]{Paglione2005}%
  \BibitemOpen
  \bibfield  {author} {\bibinfo {author} {\bibfnamefont {J.}~\bibnamefont {Paglione}}, \bibinfo {author} {\bibfnamefont {M.~A.}\ \bibnamefont {Tanatar}}, \bibinfo {author} {\bibfnamefont {D.~G.}\ \bibnamefont {Hawthorn}}, \bibinfo {author} {\bibfnamefont {R.~W.}\ \bibnamefont {Hill}}, \bibinfo {author} {\bibfnamefont {F.}~\bibnamefont {Ronning}}, \bibinfo {author} {\bibfnamefont {M.}~\bibnamefont {Sutherland}}, \bibinfo {author} {\bibfnamefont {L.}~\bibnamefont {Taillefer}}, \bibinfo {author} {\bibfnamefont {C.}~\bibnamefont {Petrovic}},\ and\ \bibinfo {author} {\bibfnamefont {P.~C.}\ \bibnamefont {Canfield}},\ }\href {https://doi.org/10.1103/PhysRevLett.94.216602} {\bibfield  {journal} {\bibinfo  {journal} {Phys. Rev. Lett.}\ }\textbf {\bibinfo {volume} {94}},\ \bibinfo {pages} {216602} (\bibinfo {year} {2005})}\BibitemShut {NoStop}%
\bibitem [{\citenamefont {Herring}(1967{\natexlab{a}})}]{Herring1967}%
  \BibitemOpen
  \bibfield  {author} {\bibinfo {author} {\bibfnamefont {C.}~\bibnamefont {Herring}},\ }\href {https://doi.org/10.1103/PhysRevLett.19.167} {\bibfield  {journal} {\bibinfo  {journal} {Phys. Rev. Lett.}\ }\textbf {\bibinfo {volume} {19}},\ \bibinfo {pages} {167} (\bibinfo {year} {1967}{\natexlab{a}})}\BibitemShut {NoStop}%
\bibitem [{\citenamefont {Herring}(1967{\natexlab{b}})}]{Herring1967erratum}%
  \BibitemOpen
  \bibfield  {author} {\bibinfo {author} {\bibfnamefont {C.}~\bibnamefont {Herring}},\ }\href {https://doi.org/10.1103/PhysRevLett.19.684} {\bibfield  {journal} {\bibinfo  {journal} {Phys. Rev. Lett.}\ }\textbf {\bibinfo {volume} {19}},\ \bibinfo {pages} {684} (\bibinfo {year} {1967}{\natexlab{b}})}\BibitemShut {NoStop}%
\bibitem [{\citenamefont {Bennett}\ and\ \citenamefont {Rice}(1969)}]{Bennett1969}%
  \BibitemOpen
  \bibfield  {author} {\bibinfo {author} {\bibfnamefont {A.~J.}\ \bibnamefont {Bennett}}\ and\ \bibinfo {author} {\bibfnamefont {M.~J.}\ \bibnamefont {Rice}},\ }\href {https://doi.org/10.1103/PhysRev.185.968} {\bibfield  {journal} {\bibinfo  {journal} {Phys. Rev.}\ }\textbf {\bibinfo {volume} {185}},\ \bibinfo {pages} {968} (\bibinfo {year} {1969})}\BibitemShut {NoStop}%
\bibitem [{\citenamefont {Schriempf}\ \emph {et~al.}(1969)\citenamefont {Schriempf}, \citenamefont {Schindler},\ and\ \citenamefont {Mills}}]{Schriempf1969}%
  \BibitemOpen
  \bibfield  {author} {\bibinfo {author} {\bibfnamefont {J.~T.}\ \bibnamefont {Schriempf}}, \bibinfo {author} {\bibfnamefont {A.~I.}\ \bibnamefont {Schindler}},\ and\ \bibinfo {author} {\bibfnamefont {D.~L.}\ \bibnamefont {Mills}},\ }\href {https://doi.org/10.1103/PhysRev.187.959} {\bibfield  {journal} {\bibinfo  {journal} {Phys. Rev.}\ }\textbf {\bibinfo {volume} {187}},\ \bibinfo {pages} {959} (\bibinfo {year} {1969})}\BibitemShut {NoStop}%
\bibitem [{\citenamefont {Moshe}\ and\ \citenamefont {Nathan}(1984)}]{Kaveh1984}%
  \BibitemOpen
  \bibfield  {author} {\bibinfo {author} {\bibfnamefont {K.}~\bibnamefont {Moshe}}\ and\ \bibinfo {author} {\bibfnamefont {W.}~\bibnamefont {Nathan}},\ }\href {https://doi.org/10.1080/00018738400101671} {\bibfield  {journal} {\bibinfo  {journal} {Adv. Phys.}\ }\textbf {\bibinfo {volume} {33}},\ \bibinfo {pages} {257} (\bibinfo {year} {1984})}\BibitemShut {NoStop}%
\bibitem [{\citenamefont {Schulz}\ and\ \citenamefont {Allen}(1995)}]{Schulz1995}%
  \BibitemOpen
  \bibfield  {author} {\bibinfo {author} {\bibfnamefont {W.~W.}\ \bibnamefont {Schulz}}\ and\ \bibinfo {author} {\bibfnamefont {P.~B.}\ \bibnamefont {Allen}},\ }\href {https://doi.org/10.1103/PhysRevB.52.7994} {\bibfield  {journal} {\bibinfo  {journal} {Phys. Rev. B}\ }\textbf {\bibinfo {volume} {52}},\ \bibinfo {pages} {7994} (\bibinfo {year} {1995})}\BibitemShut {NoStop}%
\bibitem [{\citenamefont {Principi}\ and\ \citenamefont {Vignale}(2015)}]{Principi2015}%
  \BibitemOpen
  \bibfield  {author} {\bibinfo {author} {\bibfnamefont {A.}~\bibnamefont {Principi}}\ and\ \bibinfo {author} {\bibfnamefont {G.}~\bibnamefont {Vignale}},\ }\href {https://doi.org/10.1103/PhysRevLett.115.056603} {\bibfield  {journal} {\bibinfo  {journal} {Phys. Rev. Lett.}\ }\textbf {\bibinfo {volume} {115}},\ \bibinfo {pages} {056603} (\bibinfo {year} {2015})}\BibitemShut {NoStop}%
\bibitem [{\citenamefont {Lucas}\ and\ \citenamefont {Das~Sarma}(2018)}]{Lucas2018}%
  \BibitemOpen
  \bibfield  {author} {\bibinfo {author} {\bibfnamefont {A.}~\bibnamefont {Lucas}}\ and\ \bibinfo {author} {\bibfnamefont {S.}~\bibnamefont {Das~Sarma}},\ }\href {https://doi.org/10.1103/PhysRevB.97.245128} {\bibfield  {journal} {\bibinfo  {journal} {Phys. Rev. B}\ }\textbf {\bibinfo {volume} {97}},\ \bibinfo {pages} {245128} (\bibinfo {year} {2018})}\BibitemShut {NoStop}%
\bibitem [{\citenamefont {Li}\ and\ \citenamefont {Maslov}(2018)}]{Li2018}%
  \BibitemOpen
  \bibfield  {author} {\bibinfo {author} {\bibfnamefont {S.}~\bibnamefont {Li}}\ and\ \bibinfo {author} {\bibfnamefont {D.~L.}\ \bibnamefont {Maslov}},\ }\href {https://doi.org/10.1103/PhysRevB.98.245134} {\bibfield  {journal} {\bibinfo  {journal} {Phys. Rev. B}\ }\textbf {\bibinfo {volume} {98}},\ \bibinfo {pages} {245134} (\bibinfo {year} {2018})}\BibitemShut {NoStop}%
\bibitem [{\citenamefont {Zarenia}\ \emph {et~al.}(2020)\citenamefont {Zarenia}, \citenamefont {Principi},\ and\ \citenamefont {Vignale}}]{Zarenia2020}%
  \BibitemOpen
  \bibfield  {author} {\bibinfo {author} {\bibfnamefont {M.}~\bibnamefont {Zarenia}}, \bibinfo {author} {\bibfnamefont {A.}~\bibnamefont {Principi}},\ and\ \bibinfo {author} {\bibfnamefont {G.}~\bibnamefont {Vignale}},\ }\href {https://doi.org/10.1103/PhysRevB.102.214304} {\bibfield  {journal} {\bibinfo  {journal} {Phys. Rev. B}\ }\textbf {\bibinfo {volume} {102}},\ \bibinfo {pages} {214304} (\bibinfo {year} {2020})}\BibitemShut {NoStop}%
\bibitem [{\citenamefont {Lee}\ \emph {et~al.}(2021)\citenamefont {Lee}, \citenamefont {Michaeli},\ and\ \citenamefont {Schwiete}}]{Lee2021}%
  \BibitemOpen
  \bibfield  {author} {\bibinfo {author} {\bibfnamefont {W.-R.}\ \bibnamefont {Lee}}, \bibinfo {author} {\bibfnamefont {K.}~\bibnamefont {Michaeli}},\ and\ \bibinfo {author} {\bibfnamefont {G.}~\bibnamefont {Schwiete}},\ }\href {https://doi.org/10.1103/PhysRevB.103.115140} {\bibfield  {journal} {\bibinfo  {journal} {Phys. Rev. B}\ }\textbf {\bibinfo {volume} {103}},\ \bibinfo {pages} {115140} (\bibinfo {year} {2021})}\BibitemShut {NoStop}%
\bibitem [{\citenamefont {Takahashi}\ \emph {et~al.}(2023)\citenamefont {Takahashi}, \citenamefont {Matsuura}, \citenamefont {Maebashi},\ and\ \citenamefont {Ogata}}]{Takahashi2023}%
  \BibitemOpen
  \bibfield  {author} {\bibinfo {author} {\bibfnamefont {K.}~\bibnamefont {Takahashi}}, \bibinfo {author} {\bibfnamefont {H.}~\bibnamefont {Matsuura}}, \bibinfo {author} {\bibfnamefont {H.}~\bibnamefont {Maebashi}},\ and\ \bibinfo {author} {\bibfnamefont {M.}~\bibnamefont {Ogata}},\ }\href {https://doi.org/10.1103/PhysRevB.107.115158} {\bibfield  {journal} {\bibinfo  {journal} {Phys. Rev. B}\ }\textbf {\bibinfo {volume} {107}},\ \bibinfo {pages} {115158} (\bibinfo {year} {2023})}\BibitemShut {NoStop}%
\bibitem [{\citenamefont {Abrikosov}\ and\ \citenamefont {Khalatnikov}(1957)}]{Abrikosov1957}%
  \BibitemOpen
  \bibfield  {author} {\bibinfo {author} {\bibfnamefont {A.~A.}\ \bibnamefont {Abrikosov}}\ and\ \bibinfo {author} {\bibfnamefont {I.~M.}\ \bibnamefont {Khalatnikov}},\ }\href {http://www.jetp.ras.ru/cgi-bin/e/index/e/5/5/p887?a=list} {\bibfield  {journal} {\bibinfo  {journal} {Sov. Phys. JETP}\ }\textbf {\bibinfo {volume} {32}},\ \bibinfo {pages} {887} (\bibinfo {year} {1957})}\BibitemShut {NoStop}%
\bibitem [{\citenamefont {Abrikosov}\ and\ \citenamefont {Khalatnikov}(1959)}]{Abrikosov1959}%
  \BibitemOpen
  \bibfield  {author} {\bibinfo {author} {\bibfnamefont {A.~A.}\ \bibnamefont {Abrikosov}}\ and\ \bibinfo {author} {\bibfnamefont {I.~M.}\ \bibnamefont {Khalatnikov}},\ }\href {https://doi.org/10.1088/0034-4885/22/1/310} {\bibfield  {journal} {\bibinfo  {journal} {Rep. Prog. Phys.}\ }\textbf {\bibinfo {volume} {22}},\ \bibinfo {pages} {329} (\bibinfo {year} {1959})}\BibitemShut {NoStop}%
\bibitem [{\citenamefont {Brooker}\ and\ \citenamefont {Sykes}(1968)}]{Brooker1968}%
  \BibitemOpen
  \bibfield  {author} {\bibinfo {author} {\bibfnamefont {G.~A.}\ \bibnamefont {Brooker}}\ and\ \bibinfo {author} {\bibfnamefont {J.}~\bibnamefont {Sykes}},\ }\href {https://doi.org/10.1103/PhysRevLett.21.279} {\bibfield  {journal} {\bibinfo  {journal} {Phys. Rev. Lett.}\ }\textbf {\bibinfo {volume} {21}},\ \bibinfo {pages} {279} (\bibinfo {year} {1968})}\BibitemShut {NoStop}%
\bibitem [{\citenamefont {Jensen}\ \emph {et~al.}(1968)\citenamefont {Jensen}, \citenamefont {Smith},\ and\ \citenamefont {Wilkins}}]{Jensen1968}%
  \BibitemOpen
  \bibfield  {author} {\bibinfo {author} {\bibfnamefont {H.~H.}\ \bibnamefont {Jensen}}, \bibinfo {author} {\bibfnamefont {H.}~\bibnamefont {Smith}},\ and\ \bibinfo {author} {\bibfnamefont {J.~W.}\ \bibnamefont {Wilkins}},\ }\href {https://doi.org/10.1016/0375-9601(68)90904-3} {\bibfield  {journal} {\bibinfo  {journal} {Phys. Lett. A}\ }\textbf {\bibinfo {volume} {27}},\ \bibinfo {pages} {532} (\bibinfo {year} {1968})}\BibitemShut {NoStop}%
\bibitem [{\citenamefont {Smith}\ and\ \citenamefont {Wilkins}(1969)}]{Smith1969}%
  \BibitemOpen
  \bibfield  {author} {\bibinfo {author} {\bibfnamefont {H.}~\bibnamefont {Smith}}\ and\ \bibinfo {author} {\bibfnamefont {J.~W.}\ \bibnamefont {Wilkins}},\ }\href {https://doi.org/10.1103/PhysRev.183.624} {\bibfield  {journal} {\bibinfo  {journal} {Phys. Rev.}\ }\textbf {\bibinfo {volume} {183}},\ \bibinfo {pages} {624} (\bibinfo {year} {1969})}\BibitemShut {NoStop}%
\bibitem [{\citenamefont {Jensen}\ \emph {et~al.}(1969)\citenamefont {Jensen}, \citenamefont {Smith},\ and\ \citenamefont {Wilkins}}]{Jensen1969}%
  \BibitemOpen
  \bibfield  {author} {\bibinfo {author} {\bibfnamefont {H.~H.}\ \bibnamefont {Jensen}}, \bibinfo {author} {\bibfnamefont {H.}~\bibnamefont {Smith}},\ and\ \bibinfo {author} {\bibfnamefont {J.~W.}\ \bibnamefont {Wilkins}},\ }\href {https://doi.org/10.1103/PhysRev.185.323} {\bibfield  {journal} {\bibinfo  {journal} {Phys. Rev.}\ }\textbf {\bibinfo {volume} {185}},\ \bibinfo {pages} {323} (\bibinfo {year} {1969})}\BibitemShut {NoStop}%
\bibitem [{\citenamefont {Sykes}\ and\ \citenamefont {Brooker}(1970)}]{Sykes1970}%
  \BibitemOpen
  \bibfield  {author} {\bibinfo {author} {\bibfnamefont {J.}~\bibnamefont {Sykes}}\ and\ \bibinfo {author} {\bibfnamefont {G.~A.}\ \bibnamefont {Brooker}},\ }\href {https://doi.org/https://doi.org/10.1016/0003-4916(70)90002-3} {\bibfield  {journal} {\bibinfo  {journal} {Ann. Phys. (N.Y.)}\ }\textbf {\bibinfo {volume} {56}},\ \bibinfo {pages} {1} (\bibinfo {year} {1970})}\BibitemShut {NoStop}%
\bibitem [{\citenamefont {Ah-Sam}\ \emph {et~al.}(1971)\citenamefont {Ah-Sam}, \citenamefont {Jensen},\ and\ \citenamefont {Smith}}]{Ah-Sam1971}%
  \BibitemOpen
  \bibfield  {author} {\bibinfo {author} {\bibfnamefont {L.~E.~G.}\ \bibnamefont {Ah-Sam}}, \bibinfo {author} {\bibfnamefont {H.~H.}\ \bibnamefont {Jensen}},\ and\ \bibinfo {author} {\bibfnamefont {H.}~\bibnamefont {Smith}},\ }\href {https://doi.org/10.1007/BF01012184} {\bibfield  {journal} {\bibinfo  {journal} {J. Stat. Phys.}\ }\textbf {\bibinfo {volume} {3}},\ \bibinfo {pages} {17} (\bibinfo {year} {1971})}\BibitemShut {NoStop}%
\bibitem [{\citenamefont {Brooker}\ and\ \citenamefont {Sykes}(1972)}]{Brooker1972}%
  \BibitemOpen
  \bibfield  {author} {\bibinfo {author} {\bibfnamefont {G.~A.}\ \bibnamefont {Brooker}}\ and\ \bibinfo {author} {\bibfnamefont {J.}~\bibnamefont {Sykes}},\ }\href {https://doi.org/10.1016/0003-4916(72)90261-8} {\bibfield  {journal} {\bibinfo  {journal} {Ann. Phys. (N.Y.)}\ }\textbf {\bibinfo {volume} {74}},\ \bibinfo {pages} {67} (\bibinfo {year} {1972})}\BibitemShut {NoStop}%
\bibitem [{\citenamefont {Egilsson}\ and\ \citenamefont {Pethick}(1977)}]{Egilsson1977}%
  \BibitemOpen
  \bibfield  {author} {\bibinfo {author} {\bibfnamefont {E.}~\bibnamefont {Egilsson}}\ and\ \bibinfo {author} {\bibfnamefont {C.~J.}\ \bibnamefont {Pethick}},\ }\href {https://doi.org/10.1007/BF00659091} {\bibfield  {journal} {\bibinfo  {journal} {J. Low Temp. Phys.}\ }\textbf {\bibinfo {volume} {29}},\ \bibinfo {pages} {99} (\bibinfo {year} {1977})}\BibitemShut {NoStop}%
\bibitem [{\citenamefont {Oliva}\ and\ \citenamefont {Ashcroft}(1982)}]{Oliva1982}%
  \BibitemOpen
  \bibfield  {author} {\bibinfo {author} {\bibfnamefont {J.}~\bibnamefont {Oliva}}\ and\ \bibinfo {author} {\bibfnamefont {N.~W.}\ \bibnamefont {Ashcroft}},\ }\href {https://doi.org/10.1103/PhysRevB.25.223} {\bibfield  {journal} {\bibinfo  {journal} {Phys. Rev. B}\ }\textbf {\bibinfo {volume} {25}},\ \bibinfo {pages} {223} (\bibinfo {year} {1982})}\BibitemShut {NoStop}%
\bibitem [{\citenamefont {Anderson}\ \emph {et~al.}(1987)\citenamefont {Anderson}, \citenamefont {Pethick},\ and\ \citenamefont {Quader}}]{Anderson1987}%
  \BibitemOpen
  \bibfield  {author} {\bibinfo {author} {\bibfnamefont {R.~H.}\ \bibnamefont {Anderson}}, \bibinfo {author} {\bibfnamefont {C.~J.}\ \bibnamefont {Pethick}},\ and\ \bibinfo {author} {\bibfnamefont {K.~F.}\ \bibnamefont {Quader}},\ }\href {https://doi.org/10.1103/PhysRevB.35.1620} {\bibfield  {journal} {\bibinfo  {journal} {Phys. Rev. B}\ }\textbf {\bibinfo {volume} {35}},\ \bibinfo {pages} {1620} (\bibinfo {year} {1987})}\BibitemShut {NoStop}%
\bibitem [{\citenamefont {Golosov}\ and\ \citenamefont {Ruckenstein}(1995)}]{Golosov1995}%
  \BibitemOpen
  \bibfield  {author} {\bibinfo {author} {\bibfnamefont {D.~I.}\ \bibnamefont {Golosov}}\ and\ \bibinfo {author} {\bibfnamefont {A.~E.}\ \bibnamefont {Ruckenstein}},\ }\href {https://doi.org/10.1103/PhysRevLett.74.1613} {\bibfield  {journal} {\bibinfo  {journal} {Phys. Rev. Lett.}\ }\textbf {\bibinfo {volume} {74}},\ \bibinfo {pages} {1613} (\bibinfo {year} {1995})}\BibitemShut {NoStop}%
\bibitem [{\citenamefont {Golosov}\ and\ \citenamefont {Ruckenstein}(1998)}]{Golosov1998}%
  \BibitemOpen
  \bibfield  {author} {\bibinfo {author} {\bibfnamefont {D.~I.}\ \bibnamefont {Golosov}}\ and\ \bibinfo {author} {\bibfnamefont {A.~E.}\ \bibnamefont {Ruckenstein}},\ }\href {https://doi.org/10.1023/A:1022693917461} {\bibfield  {journal} {\bibinfo  {journal} {J. Low Temp. Phys.}\ }\textbf {\bibinfo {volume} {112}},\ \bibinfo {pages} {265} (\bibinfo {year} {1998})}\BibitemShut {NoStop}%
\bibitem [{\citenamefont {Pethick}\ and\ \citenamefont {Schwenk}(2009)}]{Pethick2009}%
  \BibitemOpen
  \bibfield  {author} {\bibinfo {author} {\bibfnamefont {C.~J.}\ \bibnamefont {Pethick}}\ and\ \bibinfo {author} {\bibfnamefont {A.}~\bibnamefont {Schwenk}},\ }\href {https://doi.org/10.1103/PhysRevC.80.055805} {\bibfield  {journal} {\bibinfo  {journal} {Phys. Rev. C}\ }\textbf {\bibinfo {volume} {80}},\ \bibinfo {pages} {055805} (\bibinfo {year} {2009})}\BibitemShut {NoStop}%
\bibitem [{\citenamefont {Lee}\ \emph {et~al.}(2020{\natexlab{a}})\citenamefont {Lee}, \citenamefont {Finkel'stein}, \citenamefont {Michaeli},\ and\ \citenamefont {Schwiete}}]{Lee2020}%
  \BibitemOpen
  \bibfield  {author} {\bibinfo {author} {\bibfnamefont {W.-R.}\ \bibnamefont {Lee}}, \bibinfo {author} {\bibfnamefont {A.~M.}\ \bibnamefont {Finkel'stein}}, \bibinfo {author} {\bibfnamefont {K.}~\bibnamefont {Michaeli}},\ and\ \bibinfo {author} {\bibfnamefont {G.}~\bibnamefont {Schwiete}},\ }\href {https://doi.org/10.1103/PhysRevResearch.2.013148} {\bibfield  {journal} {\bibinfo  {journal} {Phys. Rev. Res.}\ }\textbf {\bibinfo {volume} {2}},\ \bibinfo {pages} {013148} (\bibinfo {year} {2020}{\natexlab{a}})}\BibitemShut {NoStop}%
\bibitem [{\citenamefont {Baber}(1937)}]{Baber1937}%
  \BibitemOpen
  \bibfield  {author} {\bibinfo {author} {\bibfnamefont {W.~G.}\ \bibnamefont {Baber}},\ }\href {https://doi.org/10.1098/rspa.1937.0027} {\bibfield  {journal} {\bibinfo  {journal} {Proc. R. Soc. London A}\ }\textbf {\bibinfo {volume} {158}},\ \bibinfo {pages} {383} (\bibinfo {year} {1937})}\BibitemShut {NoStop}%
\bibitem [{\citenamefont {Kukkonen}\ and\ \citenamefont {Maldague}(1976)}]{Kukkonen1976}%
  \BibitemOpen
  \bibfield  {author} {\bibinfo {author} {\bibfnamefont {C.~A.}\ \bibnamefont {Kukkonen}}\ and\ \bibinfo {author} {\bibfnamefont {P.~F.}\ \bibnamefont {Maldague}},\ }\href {https://doi.org/10.1103/PhysRevLett.37.782} {\bibfield  {journal} {\bibinfo  {journal} {Phys. Rev. Lett.}\ }\textbf {\bibinfo {volume} {37}},\ \bibinfo {pages} {782} (\bibinfo {year} {1976})}\BibitemShut {NoStop}%
\bibitem [{\citenamefont {Gantmakher}\ and\ \citenamefont {Levinson}(1978)}]{Gantmakher1978}%
  \BibitemOpen
  \bibfield  {author} {\bibinfo {author} {\bibfnamefont {V.~F.}\ \bibnamefont {Gantmakher}}\ and\ \bibinfo {author} {\bibfnamefont {I.~B.}\ \bibnamefont {Levinson}},\ }\href {http://jetp.ras.ru/cgi-bin/e/index/e/47/1/p133?a=list} {\bibfield  {journal} {\bibinfo  {journal} {Sov. Phys. JETP}\ }\textbf {\bibinfo {volume} {74}},\ \bibinfo {pages} {261} (\bibinfo {year} {1978})}\BibitemShut {NoStop}%
\bibitem [{\citenamefont {Maldague}\ and\ \citenamefont {Kukkonen}(1979)}]{Maldague1979}%
  \BibitemOpen
  \bibfield  {author} {\bibinfo {author} {\bibfnamefont {P.~F.}\ \bibnamefont {Maldague}}\ and\ \bibinfo {author} {\bibfnamefont {C.~A.}\ \bibnamefont {Kukkonen}},\ }\href {https://doi.org/10.1103/PhysRevB.19.6172} {\bibfield  {journal} {\bibinfo  {journal} {Phys. Rev. B}\ }\textbf {\bibinfo {volume} {19}},\ \bibinfo {pages} {6172} (\bibinfo {year} {1979})}\BibitemShut {NoStop}%
\bibitem [{\citenamefont {Long}(1971)}]{Long1971}%
  \BibitemOpen
  \bibfield  {author} {\bibinfo {author} {\bibfnamefont {J.~R.}\ \bibnamefont {Long}},\ }\href {https://doi.org/10.1103/PhysRevB.3.2476} {\bibfield  {journal} {\bibinfo  {journal} {Phys. Rev. B}\ }\textbf {\bibinfo {volume} {3}},\ \bibinfo {pages} {2476} (\bibinfo {year} {1971})}\BibitemShut {NoStop}%
\bibitem [{\citenamefont {Zhang}\ \emph {et~al.}(2000)\citenamefont {Zhang}, \citenamefont {Ong}, \citenamefont {Xu}, \citenamefont {Krishana}, \citenamefont {Gagnon},\ and\ \citenamefont {Taillefer}}]{Zhang2000}%
  \BibitemOpen
  \bibfield  {author} {\bibinfo {author} {\bibfnamefont {Y.}~\bibnamefont {Zhang}}, \bibinfo {author} {\bibfnamefont {N.~P.}\ \bibnamefont {Ong}}, \bibinfo {author} {\bibfnamefont {Z.~A.}\ \bibnamefont {Xu}}, \bibinfo {author} {\bibfnamefont {K.}~\bibnamefont {Krishana}}, \bibinfo {author} {\bibfnamefont {R.}~\bibnamefont {Gagnon}},\ and\ \bibinfo {author} {\bibfnamefont {L.}~\bibnamefont {Taillefer}},\ }\href {https://doi.org/10.1103/PhysRevLett.84.2219} {\bibfield  {journal} {\bibinfo  {journal} {Phys. Rev. Lett.}\ }\textbf {\bibinfo {volume} {84}},\ \bibinfo {pages} {2219} (\bibinfo {year} {2000})}\BibitemShut {NoStop}%
\bibitem [{\citenamefont {Li}(2002)}]{Li2002}%
  \BibitemOpen
  \bibfield  {author} {\bibinfo {author} {\bibfnamefont {M.-R.}\ \bibnamefont {Li}},\ }\href {https://doi.org/10.1103/PhysRevB.65.184515} {\bibfield  {journal} {\bibinfo  {journal} {Phys. Rev. B}\ }\textbf {\bibinfo {volume} {65}},\ \bibinfo {pages} {184515} (\bibinfo {year} {2002})}\BibitemShut {NoStop}%
\bibitem [{\citenamefont {Matusiak}\ and\ \citenamefont {Wolf}(2005)}]{Matusiak2005}%
  \BibitemOpen
  \bibfield  {author} {\bibinfo {author} {\bibfnamefont {M.}~\bibnamefont {Matusiak}}\ and\ \bibinfo {author} {\bibfnamefont {T.}~\bibnamefont {Wolf}},\ }\href {https://doi.org/10.1103/PhysRevB.72.054508} {\bibfield  {journal} {\bibinfo  {journal} {Phys. Rev. B}\ }\textbf {\bibinfo {volume} {72}},\ \bibinfo {pages} {054508} (\bibinfo {year} {2005})}\BibitemShut {NoStop}%
\bibitem [{\citenamefont {Onose}\ \emph {et~al.}(2008)\citenamefont {Onose}, \citenamefont {Shiomi},\ and\ \citenamefont {Tokura}}]{Onose2008}%
  \BibitemOpen
  \bibfield  {author} {\bibinfo {author} {\bibfnamefont {Y.}~\bibnamefont {Onose}}, \bibinfo {author} {\bibfnamefont {Y.}~\bibnamefont {Shiomi}},\ and\ \bibinfo {author} {\bibfnamefont {Y.}~\bibnamefont {Tokura}},\ }\href {https://doi.org/10.1103/PhysRevLett.100.016601} {\bibfield  {journal} {\bibinfo  {journal} {Phys. Rev. Lett.}\ }\textbf {\bibinfo {volume} {100}},\ \bibinfo {pages} {016601} (\bibinfo {year} {2008})}\BibitemShut {NoStop}%
\bibitem [{\citenamefont {Shiomi}\ \emph {et~al.}(2009)\citenamefont {Shiomi}, \citenamefont {Onose},\ and\ \citenamefont {Tokura}}]{Shiomi2009}%
  \BibitemOpen
  \bibfield  {author} {\bibinfo {author} {\bibfnamefont {Y.}~\bibnamefont {Shiomi}}, \bibinfo {author} {\bibfnamefont {Y.}~\bibnamefont {Onose}},\ and\ \bibinfo {author} {\bibfnamefont {Y.}~\bibnamefont {Tokura}},\ }\href {https://doi.org/10.1103/PhysRevB.79.100404} {\bibfield  {journal} {\bibinfo  {journal} {Phys. Rev. B}\ }\textbf {\bibinfo {volume} {79}},\ \bibinfo {pages} {100404(R)} (\bibinfo {year} {2009})}\BibitemShut {NoStop}%
\bibitem [{\citenamefont {Matusiak}\ \emph {et~al.}(2009)\citenamefont {Matusiak}, \citenamefont {Rogacki},\ and\ \citenamefont {Veal}}]{Matusiak2009}%
  \BibitemOpen
  \bibfield  {author} {\bibinfo {author} {\bibfnamefont {M.}~\bibnamefont {Matusiak}}, \bibinfo {author} {\bibfnamefont {K.}~\bibnamefont {Rogacki}},\ and\ \bibinfo {author} {\bibfnamefont {B.~W.}\ \bibnamefont {Veal}},\ }\href {https://doi.org/10.1209/0295-5075/88/47005} {\bibfield  {journal} {\bibinfo  {journal} {Europhys. Lett.}\ }\textbf {\bibinfo {volume} {88}},\ \bibinfo {pages} {47005} (\bibinfo {year} {2009})}\BibitemShut {NoStop}%
\bibitem [{\citenamefont {Shiomi}\ \emph {et~al.}(2010)\citenamefont {Shiomi}, \citenamefont {Onose},\ and\ \citenamefont {Tokura}}]{Shiomi2010}%
  \BibitemOpen
  \bibfield  {author} {\bibinfo {author} {\bibfnamefont {Y.}~\bibnamefont {Shiomi}}, \bibinfo {author} {\bibfnamefont {Y.}~\bibnamefont {Onose}},\ and\ \bibinfo {author} {\bibfnamefont {Y.}~\bibnamefont {Tokura}},\ }\href {https://doi.org/10.1103/PhysRevB.81.054414} {\bibfield  {journal} {\bibinfo  {journal} {Phys. Rev. B}\ }\textbf {\bibinfo {volume} {81}},\ \bibinfo {pages} {054414} (\bibinfo {year} {2010})}\BibitemShut {NoStop}%
\bibitem [{\citenamefont {Matusiak}\ and\ \citenamefont {Wolf}(2015)}]{Matusiak2015}%
  \BibitemOpen
  \bibfield  {author} {\bibinfo {author} {\bibfnamefont {M.}~\bibnamefont {Matusiak}}\ and\ \bibinfo {author} {\bibfnamefont {T.}~\bibnamefont {Wolf}},\ }\href {https://doi.org/10.1103/PhysRevB.92.020507} {\bibfield  {journal} {\bibinfo  {journal} {Phys. Rev. B}\ }\textbf {\bibinfo {volume} {92}},\ \bibinfo {pages} {020507(R)} (\bibinfo {year} {2015})}\BibitemShut {NoStop}%
\bibitem [{\citenamefont {Grissonnanche}\ \emph {et~al.}(2016)\citenamefont {Grissonnanche}, \citenamefont {Lalibert\'e}, \citenamefont {Dufour-Beaus\'ejour}, \citenamefont {Matusiak}, \citenamefont {Badoux}, \citenamefont {Tafti}, \citenamefont {Michon}, \citenamefont {Riopel}, \citenamefont {Cyr-Choini\`ere}, \citenamefont {Baglo}, \citenamefont {Ramshaw}, \citenamefont {Liang}, \citenamefont {Bonn}, \citenamefont {Hardy}, \citenamefont {Kr\"amer}, \citenamefont {LeBoeuf}, \citenamefont {Graf}, \citenamefont {Doiron-Leyraud},\ and\ \citenamefont {Taillefer}}]{Grissonnanche2016}%
  \BibitemOpen
  \bibfield  {author} {\bibinfo {author} {\bibfnamefont {G.}~\bibnamefont {Grissonnanche}}, \bibinfo {author} {\bibfnamefont {F.}~\bibnamefont {Lalibert\'e}}, \bibinfo {author} {\bibfnamefont {S.}~\bibnamefont {Dufour-Beaus\'ejour}}, \bibinfo {author} {\bibfnamefont {M.}~\bibnamefont {Matusiak}}, \bibinfo {author} {\bibfnamefont {S.}~\bibnamefont {Badoux}}, \bibinfo {author} {\bibfnamefont {F.~F.}\ \bibnamefont {Tafti}}, \bibinfo {author} {\bibfnamefont {B.}~\bibnamefont {Michon}}, \bibinfo {author} {\bibfnamefont {A.}~\bibnamefont {Riopel}}, \bibinfo {author} {\bibfnamefont {O.}~\bibnamefont {Cyr-Choini\`ere}}, \bibinfo {author} {\bibfnamefont {J.~C.}\ \bibnamefont {Baglo}}, \bibinfo {author} {\bibfnamefont {B.~J.}\ \bibnamefont {Ramshaw}}, \bibinfo {author} {\bibfnamefont {R.}~\bibnamefont {Liang}}, \bibinfo {author} {\bibfnamefont {D.~A.}\ \bibnamefont {Bonn}}, \bibinfo {author} {\bibfnamefont {W.~N.}\ \bibnamefont {Hardy}}, \bibinfo {author} {\bibfnamefont {S.}~\bibnamefont {Kr\"amer}}, \bibinfo {author} {\bibfnamefont {D.}~\bibnamefont {LeBoeuf}}, \bibinfo {author} {\bibfnamefont {D.}~\bibnamefont {Graf}}, \bibinfo {author} {\bibfnamefont {N.}~\bibnamefont {Doiron-Leyraud}},\ and\ \bibinfo {author} {\bibfnamefont {L.}~\bibnamefont {Taillefer}},\ }\href {https://doi.org/10.1103/PhysRevB.93.064513} {\bibfield  {journal} {\bibinfo  {journal} {Phys. Rev. B}\ }\textbf {\bibinfo {volume} {93}},\ \bibinfo {pages} {064513} (\bibinfo {year} {2016})}\BibitemShut {NoStop}%
\bibitem [{\citenamefont {Nguyen}\ \emph {et~al.}(2020)\citenamefont {Nguyen}, \citenamefont {Wagner},\ and\ \citenamefont {Simon}}]{Nguyen2020}%
  \BibitemOpen
  \bibfield  {author} {\bibinfo {author} {\bibfnamefont {D.~X.}\ \bibnamefont {Nguyen}}, \bibinfo {author} {\bibfnamefont {G.}~\bibnamefont {Wagner}},\ and\ \bibinfo {author} {\bibfnamefont {S.~H.}\ \bibnamefont {Simon}},\ }\href {https://doi.org/10.1103/PhysRevB.101.035117} {\bibfield  {journal} {\bibinfo  {journal} {Phys. Rev. B}\ }\textbf {\bibinfo {volume} {101}},\ \bibinfo {pages} {035117} (\bibinfo {year} {2020})}\BibitemShut {NoStop}%
\bibitem [{\citenamefont {Huang}\ and\ \citenamefont {Lucas}(2021)}]{Huang2021}%
  \BibitemOpen
  \bibfield  {author} {\bibinfo {author} {\bibfnamefont {X.}~\bibnamefont {Huang}}\ and\ \bibinfo {author} {\bibfnamefont {A.}~\bibnamefont {Lucas}},\ }\href {https://doi.org/10.1103/PhysRevB.103.155128} {\bibfield  {journal} {\bibinfo  {journal} {Phys. Rev. B}\ }\textbf {\bibinfo {volume} {103}},\ \bibinfo {pages} {155128} (\bibinfo {year} {2021})}\BibitemShut {NoStop}%
\bibitem [{\citenamefont {Tulipman}\ and\ \citenamefont {Berg}(2023)}]{Tulipman2023}%
  \BibitemOpen
  \bibfield  {author} {\bibinfo {author} {\bibfnamefont {E.}~\bibnamefont {Tulipman}}\ and\ \bibinfo {author} {\bibfnamefont {E.}~\bibnamefont {Berg}},\ }\href {https://doi.org/10.1038/s41535-023-00598-z} {\bibfield  {journal} {\bibinfo  {journal} {npj Quantum Mater.}\ }\textbf {\bibinfo {volume} {8}},\ \bibinfo {pages} {66} (\bibinfo {year} {2023})}\BibitemShut {NoStop}%
\bibitem [{\citenamefont {Tu}\ and\ \citenamefont {Das~Sarma}(2023)}]{Tu2023}%
  \BibitemOpen
  \bibfield  {author} {\bibinfo {author} {\bibfnamefont {Y.-T.}\ \bibnamefont {Tu}}\ and\ \bibinfo {author} {\bibfnamefont {S.}~\bibnamefont {Das~Sarma}},\ }\href {https://doi.org/10.1103/PhysRevB.108.245415} {\bibfield  {journal} {\bibinfo  {journal} {Phys. Rev. B}\ }\textbf {\bibinfo {volume} {108}},\ \bibinfo {pages} {245415} (\bibinfo {year} {2023})}\BibitemShut {NoStop}%
\bibitem [{\citenamefont {Amundsen}(1969)}]{Amundsen1969}%
  \BibitemOpen
  \bibfield  {author} {\bibinfo {author} {\bibfnamefont {T.}~\bibnamefont {Amundsen}},\ }\href {https://doi.org/10.1080/14786436908228037} {\bibfield  {journal} {\bibinfo  {journal} {Philos. Mag.}\ }\textbf {\bibinfo {volume} {20}},\ \bibinfo {pages} {687} (\bibinfo {year} {1969})}\BibitemShut {NoStop}%
\bibitem [{\citenamefont {Stephan}\ and\ \citenamefont {Maxfield}(1972)}]{Stephan1972}%
  \BibitemOpen
  \bibfield  {author} {\bibinfo {author} {\bibfnamefont {C.~H.}\ \bibnamefont {Stephan}}\ and\ \bibinfo {author} {\bibfnamefont {B.~W.}\ \bibnamefont {Maxfield}},\ }\href {https://doi.org/10.1103/PhysRevB.6.2893} {\bibfield  {journal} {\bibinfo  {journal} {Phys. Rev. B}\ }\textbf {\bibinfo {volume} {6}},\ \bibinfo {pages} {2893} (\bibinfo {year} {1972})}\BibitemShut {NoStop}%
\bibitem [{\citenamefont {Fletcher}\ and\ \citenamefont {Friedman}(1973)}]{Fletcher1973}%
  \BibitemOpen
  \bibfield  {author} {\bibinfo {author} {\bibfnamefont {R.}~\bibnamefont {Fletcher}}\ and\ \bibinfo {author} {\bibfnamefont {A.~J.}\ \bibnamefont {Friedman}},\ }\href {https://doi.org/10.1103/PhysRevB.8.5381} {\bibfield  {journal} {\bibinfo  {journal} {Phys. Rev. B}\ }\textbf {\bibinfo {volume} {8}},\ \bibinfo {pages} {5381} (\bibinfo {year} {1973})}\BibitemShut {NoStop}%
\bibitem [{\citenamefont {Fletcher}(1977)}]{Fletcher1977}%
  \BibitemOpen
  \bibfield  {author} {\bibinfo {author} {\bibfnamefont {R.}~\bibnamefont {Fletcher}},\ }\href {https://doi.org/10.1103/PhysRevB.15.3602} {\bibfield  {journal} {\bibinfo  {journal} {Phys. Rev. B}\ }\textbf {\bibinfo {volume} {15}},\ \bibinfo {pages} {3602} (\bibinfo {year} {1977})}\BibitemShut {NoStop}%
\bibitem [{\citenamefont {Sugihara}(1980)}]{Sugihara1980}%
  \BibitemOpen
  \bibfield  {author} {\bibinfo {author} {\bibfnamefont {K.}~\bibnamefont {Sugihara}},\ }\href {https://doi.org/10.1143/JPSJ.49.1098} {\bibfield  {journal} {\bibinfo  {journal} {J. Phys. Soc. Jpn.}\ }\textbf {\bibinfo {volume} {49}},\ \bibinfo {pages} {1098} (\bibinfo {year} {1980})}\BibitemShut {NoStop}%
\bibitem [{Note1()}]{Note1}%
  \BibitemOpen
  \bibinfo {note} {$R_{\protect \text {RL}} = K_{\protect \text {H}}/T$ is often referred to as Righi-Leduc coefficient \cite {Amundsen1969, Stephan1972, Fletcher1973, Fletcher1977, Sugihara1980}. It should be noted that Righi-Leduc coefficient may be differently defined as $R_{\protect \text {RL}} = 1/B \cdot \kappa _{xy}/\kappa _{xx}$ \cite {Bridgman1924, Ziman2001}. The WF law for $R_{\protect \text {RL}}$ expressed as $R_{\protect \text {RL}} = \sigma R_{\protect \text {H}}$. This is satisfied with $L = L_{\protect \text {H}} = L_{0}$ and this has been studied as well \cite {Bridgman1924}.}\BibitemShut {Stop}%
\bibitem [{\citenamefont {Sondheimer}(1948)}]{Sondheimer1948}%
  \BibitemOpen
  \bibfield  {author} {\bibinfo {author} {\bibfnamefont {E.~H.}\ \bibnamefont {Sondheimer}},\ }\href {https://doi.org/10.1098/rspa.1948.0058} {\bibfield  {journal} {\bibinfo  {journal} {Proc. R. Soc. London A}\ }\textbf {\bibinfo {volume} {193}},\ \bibinfo {pages} {484} (\bibinfo {year} {1948})}\BibitemShut {NoStop}%
\bibitem [{\citenamefont {Hurd}(1972)}]{Hurd1972}%
  \BibitemOpen
  \bibfield  {author} {\bibinfo {author} {\bibfnamefont {C.~M.}\ \bibnamefont {Hurd}},\ }\href@noop {} {\emph {\bibinfo {title} {{The Hall effect in metals and alloys}}}}\ (\bibinfo  {publisher} {Plenum},\ \bibinfo {address} {New York},\ \bibinfo {year} {1972})\BibitemShut {NoStop}%
\bibitem [{\citenamefont {Ali}\ \emph {et~al.}(2014)\citenamefont {Ali}, \citenamefont {Xiong}, \citenamefont {Flynn}, \citenamefont {Tao}, \citenamefont {Gibson}, \citenamefont {Schoop}, \citenamefont {Liang}, \citenamefont {Haldolaarachchige}, \citenamefont {Hirschberger}, \citenamefont {Ong},\ and\ \citenamefont {Cava}}]{Ali2014}%
  \BibitemOpen
  \bibfield  {author} {\bibinfo {author} {\bibfnamefont {M.~N.}\ \bibnamefont {Ali}}, \bibinfo {author} {\bibfnamefont {J.}~\bibnamefont {Xiong}}, \bibinfo {author} {\bibfnamefont {S.}~\bibnamefont {Flynn}}, \bibinfo {author} {\bibfnamefont {J.}~\bibnamefont {Tao}}, \bibinfo {author} {\bibfnamefont {Q.~D.}\ \bibnamefont {Gibson}}, \bibinfo {author} {\bibfnamefont {L.~M.}\ \bibnamefont {Schoop}}, \bibinfo {author} {\bibfnamefont {T.}~\bibnamefont {Liang}}, \bibinfo {author} {\bibfnamefont {N.}~\bibnamefont {Haldolaarachchige}}, \bibinfo {author} {\bibfnamefont {M.}~\bibnamefont {Hirschberger}}, \bibinfo {author} {\bibfnamefont {N.~P.}\ \bibnamefont {Ong}},\ and\ \bibinfo {author} {\bibfnamefont {R.~J.}\ \bibnamefont {Cava}},\ }\href {https://doi.org/10.1038/nature13763} {\bibfield  {journal} {\bibinfo  {journal} {Nature}\ }\textbf {\bibinfo {volume} {514}},\ \bibinfo {pages} {205} (\bibinfo {year} {2014})}\BibitemShut {NoStop}%
\bibitem [{\citenamefont {Wang}\ \emph {et~al.}(2017)\citenamefont {Wang}, \citenamefont {Graf}, \citenamefont {Liu}, \citenamefont {Du}, \citenamefont {Zheng}, \citenamefont {Lei},\ and\ \citenamefont {Petrovic}}]{Wang2017}%
  \BibitemOpen
  \bibfield  {author} {\bibinfo {author} {\bibfnamefont {A.}~\bibnamefont {Wang}}, \bibinfo {author} {\bibfnamefont {D.}~\bibnamefont {Graf}}, \bibinfo {author} {\bibfnamefont {Y.}~\bibnamefont {Liu}}, \bibinfo {author} {\bibfnamefont {Q.}~\bibnamefont {Du}}, \bibinfo {author} {\bibfnamefont {J.}~\bibnamefont {Zheng}}, \bibinfo {author} {\bibfnamefont {H.}~\bibnamefont {Lei}},\ and\ \bibinfo {author} {\bibfnamefont {C.}~\bibnamefont {Petrovic}},\ }\href {https://doi.org/10.1103/PhysRevB.96.121107} {\bibfield  {journal} {\bibinfo  {journal} {Phys. Rev. B}\ }\textbf {\bibinfo {volume} {96}},\ \bibinfo {pages} {121107(R)} (\bibinfo {year} {2017})}\BibitemShut {NoStop}%
\bibitem [{\citenamefont {Kumar}\ \emph {et~al.}(2017)\citenamefont {Kumar}, \citenamefont {Sun}, \citenamefont {Xu}, \citenamefont {Manna}, \citenamefont {Yao}, \citenamefont {S{\"u}ss}, \citenamefont {Leermakers}, \citenamefont {Young}, \citenamefont {F{\"o}rster}, \citenamefont {Schmidt}, \citenamefont {Borrmann}, \citenamefont {Yan}, \citenamefont {Zeitler}, \citenamefont {Shi}, \citenamefont {Felser},\ and\ \citenamefont {Shekhar}}]{Kumar2017}%
  \BibitemOpen
  \bibfield  {author} {\bibinfo {author} {\bibfnamefont {N.}~\bibnamefont {Kumar}}, \bibinfo {author} {\bibfnamefont {Y.}~\bibnamefont {Sun}}, \bibinfo {author} {\bibfnamefont {N.}~\bibnamefont {Xu}}, \bibinfo {author} {\bibfnamefont {K.}~\bibnamefont {Manna}}, \bibinfo {author} {\bibfnamefont {M.}~\bibnamefont {Yao}}, \bibinfo {author} {\bibfnamefont {V.}~\bibnamefont {S{\"u}ss}}, \bibinfo {author} {\bibfnamefont {I.}~\bibnamefont {Leermakers}}, \bibinfo {author} {\bibfnamefont {O.}~\bibnamefont {Young}}, \bibinfo {author} {\bibfnamefont {T.}~\bibnamefont {F{\"o}rster}}, \bibinfo {author} {\bibfnamefont {M.}~\bibnamefont {Schmidt}}, \bibinfo {author} {\bibfnamefont {H.}~\bibnamefont {Borrmann}}, \bibinfo {author} {\bibfnamefont {B.}~\bibnamefont {Yan}}, \bibinfo {author} {\bibfnamefont {U.}~\bibnamefont {Zeitler}}, \bibinfo {author} {\bibfnamefont {M.}~\bibnamefont {Shi}}, \bibinfo {author} {\bibfnamefont {C.}~\bibnamefont {Felser}},\ and\ \bibinfo {author} {\bibfnamefont {C.}~\bibnamefont {Shekhar}},\ }\href {https://doi.org/10.1038/s41467-017-01758-z} {\bibfield  {journal} {\bibinfo  {journal} {Nat. Commun.}\ }\textbf {\bibinfo {volume} {8}},\ \bibinfo {pages} {1642} (\bibinfo {year} {2017})}\BibitemShut {NoStop}%
\bibitem [{\citenamefont {Lee}\ \emph {et~al.}(2020{\natexlab{b}})\citenamefont {Lee}, \citenamefont {Finkel'stein},\ and\ \citenamefont {Schwiete}}]{Lee2020magneto}%
  \BibitemOpen
  \bibfield  {author} {\bibinfo {author} {\bibfnamefont {W.-R.}\ \bibnamefont {Lee}}, \bibinfo {author} {\bibfnamefont {A.~M.}\ \bibnamefont {Finkel'stein}},\ and\ \bibinfo {author} {\bibfnamefont {G.}~\bibnamefont {Schwiete}},\ }\href {https://doi.org/10.1103/PhysRevB.102.245117} {\bibfield  {journal} {\bibinfo  {journal} {Phys. Rev. B}\ }\textbf {\bibinfo {volume} {102}},\ \bibinfo {pages} {245117} (\bibinfo {year} {2020}{\natexlab{b}})}\BibitemShut {NoStop}%
\bibitem [{\citenamefont {Kukkonen}\ and\ \citenamefont {Maldague}(1979)}]{Kukkonen1979}%
  \BibitemOpen
  \bibfield  {author} {\bibinfo {author} {\bibfnamefont {C.~A.}\ \bibnamefont {Kukkonen}}\ and\ \bibinfo {author} {\bibfnamefont {P.~F.}\ \bibnamefont {Maldague}},\ }\href {https://doi.org/10.1103/PhysRevB.19.2394} {\bibfield  {journal} {\bibinfo  {journal} {Phys. Rev. B}\ }\textbf {\bibinfo {volume} {19}},\ \bibinfo {pages} {2394} (\bibinfo {year} {1979})}\BibitemShut {NoStop}%
\bibitem [{\citenamefont {Entin}\ \emph {et~al.}(2013)\citenamefont {Entin}, \citenamefont {Magarill}, \citenamefont {Olshanetsky}, \citenamefont {Kvon}, \citenamefont {Mikhailov},\ and\ \citenamefont {Dvoretsky}}]{Entin2013}%
  \BibitemOpen
  \bibfield  {author} {\bibinfo {author} {\bibfnamefont {M.~V.}\ \bibnamefont {Entin}}, \bibinfo {author} {\bibfnamefont {L.~I.}\ \bibnamefont {Magarill}}, \bibinfo {author} {\bibfnamefont {E.~B.}\ \bibnamefont {Olshanetsky}}, \bibinfo {author} {\bibfnamefont {Z.~D.}\ \bibnamefont {Kvon}}, \bibinfo {author} {\bibfnamefont {N.~N.}\ \bibnamefont {Mikhailov}},\ and\ \bibinfo {author} {\bibfnamefont {S.~A.}\ \bibnamefont {Dvoretsky}},\ }\href {https://doi.org/https://doi.org/10.1134/S1063776113130116} {\bibfield  {journal} {\bibinfo  {journal} {J. Exp. Theor. Phys.}\ }\textbf {\bibinfo {volume} {117}},\ \bibinfo {pages} {933} (\bibinfo {year} {2013})}\BibitemShut {NoStop}%
\bibitem [{\citenamefont {Lin}\ \emph {et~al.}(2015)\citenamefont {Lin}, \citenamefont {Fauqu{\'e}},\ and\ \citenamefont {Behnia}}]{Lin2015}%
  \BibitemOpen
  \bibfield  {author} {\bibinfo {author} {\bibfnamefont {X.}~\bibnamefont {Lin}}, \bibinfo {author} {\bibfnamefont {B.}~\bibnamefont {Fauqu{\'e}}},\ and\ \bibinfo {author} {\bibfnamefont {K.}~\bibnamefont {Behnia}},\ }\href {https://doi.org/10.1126/science.aaa8655} {\bibfield  {journal} {\bibinfo  {journal} {Science}\ }\textbf {\bibinfo {volume} {349}},\ \bibinfo {pages} {945} (\bibinfo {year} {2015})}\BibitemShut {NoStop}%
\bibitem [{\citenamefont {Gradshteyn}\ and\ \citenamefont {Ryzhik}(2014)}]{Table_of_Integrals}%
  \BibitemOpen
  \bibfield  {author} {\bibinfo {author} {\bibfnamefont {I.~S.}\ \bibnamefont {Gradshteyn}}\ and\ \bibinfo {author} {\bibfnamefont {I.~M.}\ \bibnamefont {Ryzhik}},\ }\href@noop {} {\emph {\bibinfo {title} {{Table of Integrals, Series, and Products}}}}\ (\bibinfo  {publisher} {Academic press},\ \bibinfo {year} {2014})\BibitemShut {NoStop}%
\bibitem [{\citenamefont {Maebashi}\ and\ \citenamefont {Fukuyama}(1997)}]{Maebashi1997}%
  \BibitemOpen
  \bibfield  {author} {\bibinfo {author} {\bibfnamefont {H.}~\bibnamefont {Maebashi}}\ and\ \bibinfo {author} {\bibfnamefont {H.}~\bibnamefont {Fukuyama}},\ }\href {https://doi.org/10.1143/JPSJ.66.3577} {\bibfield  {journal} {\bibinfo  {journal} {J. Phys. Soc. Jpn.}\ }\textbf {\bibinfo {volume} {66}},\ \bibinfo {pages} {3577} (\bibinfo {year} {1997})}\BibitemShut {NoStop}%
\bibitem [{Note2()}]{Note2}%
  \BibitemOpen
  \bibinfo {note} {To be precise, Eq.~(\ref {eq:sum_rel_time_ee}) should acquire a factor originating from geometrical factors and the discrepancy between Eq.~(\ref {eq:sum_rel_time_ee}) and the quasiparticle relaxation time around $2/\pi ^2$. However, these factors are canceled out in Eq.~(\ref {eq:semimetal_th_Hall_RTA}).}\BibitemShut {Stop}%
\bibitem [{\citenamefont {Maebashi}\ \emph {et~al.}(2017)\citenamefont {Maebashi}, \citenamefont {Ogata},\ and\ \citenamefont {Fukuyama}}]{Maebashi2017lorentz}%
  \BibitemOpen
  \bibfield  {author} {\bibinfo {author} {\bibfnamefont {H.}~\bibnamefont {Maebashi}}, \bibinfo {author} {\bibfnamefont {M.}~\bibnamefont {Ogata}},\ and\ \bibinfo {author} {\bibfnamefont {H.}~\bibnamefont {Fukuyama}},\ }\href {https://doi.org/10.7566/JPSJ.86.083702} {\bibfield  {journal} {\bibinfo  {journal} {J. Phys. Soc. Jpn.}\ }\textbf {\bibinfo {volume} {86}},\ \bibinfo {pages} {083702} (\bibinfo {year} {2017})}\BibitemShut {NoStop}%
\bibitem [{\citenamefont {Maebashi}\ \emph {et~al.}(2019)\citenamefont {Maebashi}, \citenamefont {Hirosawa}, \citenamefont {Ogata},\ and\ \citenamefont {Fukuyama}}]{Maebashi2019nuclear}%
  \BibitemOpen
  \bibfield  {author} {\bibinfo {author} {\bibfnamefont {H.}~\bibnamefont {Maebashi}}, \bibinfo {author} {\bibfnamefont {T.}~\bibnamefont {Hirosawa}}, \bibinfo {author} {\bibfnamefont {M.}~\bibnamefont {Ogata}},\ and\ \bibinfo {author} {\bibfnamefont {H.}~\bibnamefont {Fukuyama}},\ }\href {https://doi.org/10.1016/j.jpcs.2017.12.034} {\bibfield  {journal} {\bibinfo  {journal} {J. Phys. Chem. Solids}\ }\textbf {\bibinfo {volume} {128}},\ \bibinfo {pages} {138} (\bibinfo {year} {2019})}\BibitemShut {NoStop}%
\bibitem [{\citenamefont {Gusev}\ \emph {et~al.}(2018)\citenamefont {Gusev}, \citenamefont {Olshanetsky}, \citenamefont {Kvon}, \citenamefont {Magarill}, \citenamefont {Entin}, \citenamefont {Levin},\ and\ \citenamefont {Mikhailov}}]{Gusev2018}%
  \BibitemOpen
  \bibfield  {author} {\bibinfo {author} {\bibfnamefont {G.~M.}\ \bibnamefont {Gusev}}, \bibinfo {author} {\bibfnamefont {E.~B.}\ \bibnamefont {Olshanetsky}}, \bibinfo {author} {\bibfnamefont {Z.~D.}\ \bibnamefont {Kvon}}, \bibinfo {author} {\bibfnamefont {L.~I.}\ \bibnamefont {Magarill}}, \bibinfo {author} {\bibfnamefont {M.~V.}\ \bibnamefont {Entin}}, \bibinfo {author} {\bibfnamefont {A.}~\bibnamefont {Levin}},\ and\ \bibinfo {author} {\bibfnamefont {N.~N.}\ \bibnamefont {Mikhailov}},\ }\href {https://doi.org/10.1134/S0021364018120081} {\bibfield  {journal} {\bibinfo  {journal} {JETP Lett.}\ }\textbf {\bibinfo {volume} {107}},\ \bibinfo {pages} {789} (\bibinfo {year} {2018})}\BibitemShut {NoStop}%
\bibitem [{\citenamefont {Olshanetsky}\ \emph {et~al.}(2021)\citenamefont {Olshanetsky}, \citenamefont {Kvon}, \citenamefont {Gusev}, \citenamefont {Entin}, \citenamefont {Magarill},\ and\ \citenamefont {Mikhailov}}]{Olshanetsky2021}%
  \BibitemOpen
  \bibfield  {author} {\bibinfo {author} {\bibfnamefont {E.~B.}\ \bibnamefont {Olshanetsky}}, \bibinfo {author} {\bibfnamefont {Z.~D.}\ \bibnamefont {Kvon}}, \bibinfo {author} {\bibfnamefont {G.~M.}\ \bibnamefont {Gusev}}, \bibinfo {author} {\bibfnamefont {M.~V.}\ \bibnamefont {Entin}}, \bibinfo {author} {\bibfnamefont {L.~I.}\ \bibnamefont {Magarill}},\ and\ \bibinfo {author} {\bibfnamefont {N.~N.}\ \bibnamefont {Mikhailov}},\ }\href {https://doi.org/10.1063/10.0002890} {\bibfield  {journal} {\bibinfo  {journal} {Low Temp. Phys.}\ }\textbf {\bibinfo {volume} {47}},\ \bibinfo {pages} {2} (\bibinfo {year} {2021})}\BibitemShut {NoStop}%
\bibitem [{\citenamefont {Bridgman}(1924)}]{Bridgman1924}%
  \BibitemOpen
  \bibfield  {author} {\bibinfo {author} {\bibfnamefont {P.~W.}\ \bibnamefont {Bridgman}},\ }\href {https://doi.org/10.1103/PhysRev.24.644} {\bibfield  {journal} {\bibinfo  {journal} {Phys. Rev.}\ }\textbf {\bibinfo {volume} {24}},\ \bibinfo {pages} {644} (\bibinfo {year} {1924})}\BibitemShut {NoStop}%
\end{thebibliography}%

\end{document}